\documentclass[a4paper,12pt,twoside,openright]{book}
\usepackage{color}

\usepackage[T1]{fontenc} 
\usepackage{lmodern}

\usepackage[hang,flushmargin]{footmisc} 

\usepackage{bm}
\usepackage[skip=10pt,font=footnotesize,bf]{caption}
\usepackage{url}
\usepackage{amsmath,amssymb}
\usepackage[breaklinks,colorlinks,urlcolor=blue,citecolor=blue,linkcolor=red]{hyperref}
\usepackage{epsfig,graphicx,verbatim,xspace,multirow,mathtools}
\usepackage{booktabs}
\usepackage[final]{pdfpages}
\usepackage{booktabs}
\usepackage{float}
\usepackage{cite}
\usepackage{rotating}
\usepackage[super]{nth}
\usepackage{lmodern}
\usepackage{tabto}

\usepackage{wasysym}

\usepackage[british,spanish]{babel}
\usepackage{epigraph}
\usepackage{fancyhdr}
\usepackage{indentfirst}
\usepackage{epigraph}
\usepackage[utf8]{inputenc}
\usepackage{afterpage}
\UseRawInputEncoding

\usepackage{setspace}
\onehalfspace
\singlespace
\spacing{1.2}

\usepackage[final]{pdfpages}

\def\eqref#1{{(\ref{#1})}}

\usepackage[hyperpageref]{backref} 
\usepackage{tabu}

\renewcommand*{\backref}[1]{}  
\newcommand{\be}{\begin{equation}}
\newcommand{\ee}{\end{equation}}
\newcommand{\bea}{\begin{eqnarray}}

\newcommand{\eea}{\end{eqnarray}}  
\renewcommand*{\backrefalt}[4]{
\ifcase #1 
No cited.
\or
{Cited on page} #2.
\else
{Cited on page} #2.
\fi}

\makeatletter
\def \cleardoublepage {\clearpage \if@twoside
\ifodd \c@page
\else
\null\thispagestyle{empty}\clearpage
\fi
\fi}
\makeatother

\usepackage{fancyhdr}
\renewcommand{\headrulewidth}{0.0pt}

\fancypagestyle{plain}{%
\fancyhead{} 
\fancyfoot[C]{\thepage}
\renewcommand{\headrulewidth}{0pt} 
}

\begin{document}


\setlength{\unitlength}{1cm} 
\thispagestyle{empty}
\begin{center}

\begin{center}
\includegraphics[scale=0.08]{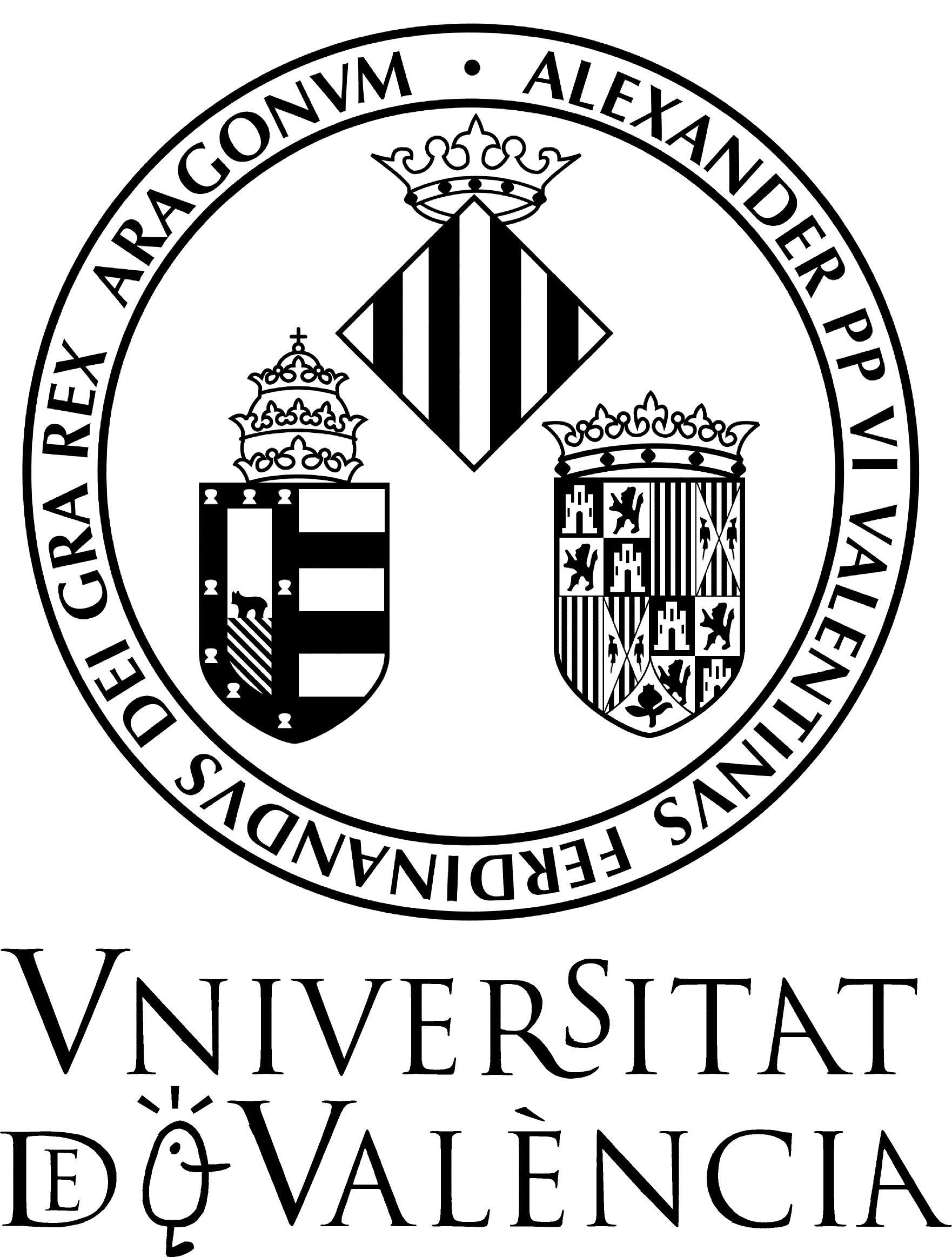}
\end{center}

\vspace{1cm}
\textbf{\Large{Dark Matter Phenomenology: Sterile Neutrino Portal and Gravitational Portal in Extra-Dimensions}\\ \vspace{1cm}
}
\large PhD Thesis

\vspace{0.75cm}

{\Large \bf{Miguel García Folgado}} \\[2ex]
{\small IFIC - Universitat de Val\`{e}ncia - CSIC\\
 Departamento de F\'isica Te\'orica \\
 Programa de Doctorado en F\'isica }\\[2ex]

\textbf{\small Under the supervision of}\\[2ex]

 {\large \textbf{Andrea Donini}} \\
 {\large \textbf{Nuria Rius Dionis}}\\
 {\large \textbf{Roberto Ruiz de Austri Bazan}}

\vspace{0.5cm}
{\large \bf{Valencia,  Marzo 2021}}

\end{center}


\begin{titlepage}
\cleardoublepage
\thispagestyle{empty}

\thispagestyle{empty}

\noindent \textbf{Andrea Donini}, Científico titular del Consejo Superior de Investigaciones Científicas, \vspace{0.2cm}

\noindent \textbf{Nuria Rius Dionis}, Catedrática del departamento de Física Teórica de la Universidad de Valencia, y \vspace{0.2cm}

\noindent \textbf{Roberto Ruiz de Austri Bazan}, Científico titular del Consejo Superior de Investigaciones Científicas, \vspace{0.2cm}

\noindent \textbf{Certifican}:\\[2ex]
\noindent Que la presente memoria, \textbf{Dark Matter Phenomenology: Sterile Neutrino Portal and Gravitational Portal in Extra-Dimensions} ha sido realizada bajo su direcci\'on en el Instituto de F\'isica Corpuscular, centro mixto de la Universidad de Valencia y del CSIC, por \textbf{Miguel Garc\'ia Folgado}, y constituye su Tesis para optar al grado de Doctor en Ciencias Físicas.\\[2ex]
\noindent Y para que as\'i conste, en cumplimiento de la legislaci\'on vigente, presenta en el Departamento de F\'isica Te\'orica de la Universidad de Valencia la referida Tesis Doctoral, y firman el presente certificado.\\

\hspace{0.75cm} 

\noindent Valencia, a 10 de Diciembre de 2020,\\

\hspace{0.75cm} 

\noindent Andrea Donini  \hspace{1.5cm} Nuria Rius Dionis \hspace{1.5cm} Roberto Ruiz de Austri \\
\noindent \hphantom{} \hspace{10.5cm} Baz\'an \hphantom{}

\newpage
\thispagestyle{empty}
$$ $$
\newpage
\thispagestyle{empty}

\end{titlepage}


\cleardoublepage
$\,$\\
\textbf{\large Comité Evaluador} \\

\vspace*{1.5cm} 
$\,$\\
\textbf{Tribunal titular} \\
\\
Dr. Alberto Casas González \hspace*{21pt} Universidad Autónoma de Madrid \\
\\
Dr. Matthew McCullough \hspace*{32pt} University of Cambridge/CERN \\
\\
Dr. Verónica Sanz González \hspace*{21pt} Universitat de València \\

\vspace*{1cm} 
$\,$\\
\textbf{Tribunal suplente} \\
\\
Dr. Arcadi Santamaría Luna \hspace*{17pt} Universitat de València \\
\\
Dr. David García Cerdeño \hspace*{29pt} Universidad Autónoma de Madrid \\
\\
Dr. Geraldine Servant \hspace*{51pt} Deutsches Elektronen Synchrotron \\
\\

\vspace*{1cm} 

\cleardoublepage


\cleardoublepage
\vspace*{4cm} 
\hspace*{200pt} \\
\hspace*{200pt} \textit{A mi familia, amigos y Andrea.} \\
\hspace*{200pt} \textit{Sin vuestro apoyo incondicional} \\
\hspace*{200pt} \textit{esta tesis no existiría} \\
\vspace*{4cm} 
\cleardoublepage


\cleardoublepage
\vspace*{4cm} 
\hspace*{200pt} \\
\hspace*{200pt} \textit{It's a dangerous business} \\
\hspace*{200pt} \textit{going out your door. You step} \\
\hspace*{200pt} \textit{onto the road, and if you don't} \\
\hspace*{200pt} \textit{keep your feet, there's no knowing} \\
\hspace*{200pt} \textit{where you might be swept off to.}\\
\hspace*{200pt} \\
\hspace*{200pt} J. R. R. Tolkien, \\
\hspace*{200pt} The Lord of the Rings \\
\vspace*{4cm} 
\cleardoublepage


\selectlanguage{british}

\frontmatter
\fancyfoot[C]{\thepage}

\chapter*{List of Publications}
\addcontentsline{toc}{chapter}{List of Publications}  

This PhD thesis is based on the following publications: \\
\begin{itemize}
\item \emph{{Probing the sterile neutrino portal to Dark Matter with $\gamma$ rays}}~\cite{Folgado:2018qlv},\\
Miguel G. Folgado, Germán A. Gómez-Vargas, Nuria Rius and Roberto Ruiz De Austri.\\  
 \href{https://doi.org/10.1088/1475-7516/2018/08/002}{\emph{JCAP}  {\bf 1808} (2018) 002}, [\href{http://arxiv.org/abs/arXiv:1803.08934}{{\tt arXiv:1803.08934}}]. 
 
 \item \emph{{Gravity-mediated Scalar Dark Matter in Warped Extra-Dimensions}}~\cite{Folgado:2019sgz},\\
Miguel G. Folgado, Andrea Donini and Nuria Rius.\\  
 \href{https://doi.org/10.1007/JHEP01(2020)161}{\emph{JHEP}  {\bf 01} (2020) 161}, [\href{http://arxiv.org/abs/arXiv:1907.04340}{{\tt arXiv:1907.04340}}]. 
 
 \item \emph{{Gravity-mediated Dark Matter in Clockwork/Linear Dilaton Extra-Dimensions}}~\cite{Folgado:2019gie},\\
Miguel G. Folgado, Andrea Donini and Nuria Rius.\\  
 \href{https://doi.org/10.1007/JHEP04(2020)036}{\emph{JHEP}  {\bf 04} (2020) 036}, [\href{http://arxiv.org/abs/arXiv:1912.02689}{{\tt arXiv:1912.02689}}]. 
 
\item \emph{{Kaluza-Klein FIMP Dark Matter in Warped Extra-Dimensions}}~\cite{Bernal:2020fvw},\\
Nicolas Bernal, Andrea Donini, Miguel G. Folgado and Nuria Rius.\\  
 \href{https://doi.org/10.1007/JHEP09(2020)142}{\emph{JHEP}  {\bf 09} (2020) 142}, [\href{https://arxiv.org/abs/2004.14403}{{\tt arXiv:2004.14403}}]. 
 
 \end{itemize}

\newpage

Other works not included in this thesis are:

\begin{itemize}

\item \emph{{On the interpretation of non-resonant phenomena at colliders}}~\cite{Folgado:2020vjb},\\
Miguel G. Folgado and Veronica Sanz.\\  
 \href{https://doi.org/10.1155/2021/2573471}{\emph{Adv.High Energy Phys. }{\bf 2021} (2021) 2573471} [\href{https://arxiv.org/abs/2005.06492}{{\tt arXiv:2005.06492}}]. 
 
\item \emph{{Spin-dependence of Gravity-mediated Dark Matter in Warped Extra-Dimensions}}~\cite{Folgado:2020utn},\\
Miguel G. Folgado, Andrea Donini and Nuria Rius.\\  
 \href{https://doi.org/10.1140/epjc/s10052-021-08989-x}{\emph{Eur.Phys.J.C} {\bf 81} (2021) 197} [\href{https://arxiv.org/abs/2006.02239}{{\tt arXiv:2006.02239}}]. 

 \item \emph{{Exploring the political pulse of a country using data science tools}}~\cite{folgado2020exploring},\\
Miguel G. Folgado, Veronica Sanz.\\ 
\href{}{} [\href{https://arxiv.org/abs/2011.10264}{{\tt arXiv:2011.10264}}].

\item \emph{{Kaluza-Klein FIMP Dark Matter in Clockwork/Linear Dilaton Extra-Dimensions}}~ \cite{Bernal:2020yqg},\\
Nicolas Bernal, Andrea Donini, Miguel G. Folgado and Nuria Rius.\\ 
\href{http://dx.doi.org/10.1007/JHEP04(2021)061}{\emph{JHEP} {\bf 04} (2021) 061} [\href{https://arxiv.org/abs/2012.10453}{{\tt arXiv:2012.10453}}].

\end{itemize}


\chapter*{Abbreviations}
\markboth{Abbreviations}{Abbreviations}
\addcontentsline{toc}{chapter}{Abbreviations}  
\def\arraystretch{1.5}
\begin{tabular}{ll}
$\Lambda$CDM &  The standard cosmological model   \\
AdS & Anti-de-sitter \\
ATLAS & Spin dependent\\
ALPs & Axion Like Particles\\
BBN & Big Bang nucleosintesis\\
BE & Bose-Einstein\\
BSM & Beyond the standard model\\
CC & Electroweak Charged Currents\\
CDM & Cold dark matter\\
CERN & Conseil Europ\'een pour la Recherche Nucl\'eaire\\
CFT & conformal field theories \\
CMB & Cosmic Microwave Background\\
CMS & Spin dependent\\
CNNS & coherent neutrino-nucleus scattering\\
CKM & Cabibbo-Kobayashi-Maskawa matrix\\
COBE & Cosmic Background Explorer \\
CP & Charge-conjugate Parity \\
CPT & Charge-conjugate Parity time\\
\end{tabular}

\begin{tabular}{ll}
CW/LD & Clockwork/Linear Dilaton\\
DD & Direct Detection\\
DRU & Differential rate unit\\
dSphs & dwarf spheroidal galaxies \\
ED & Extra-Dimensions\\
EW & Electroweak\\
EWSB & Electroweak symmetry breaking\\
FD & Fermi-Dirac\\
FIMP & Feebly Interactive Massive Particle\\
FLRW & Friedman-Lema\^itre-Robertson-Walker Metric\\
GC & Galactic Center\\
GCE & Galactic Center $\gamma$-ray Excess\\
GIM & Glashow-Iliopoulos-Maiani mechanism\\
HDM & Hot dark matter\\
ID & Indirect Detection\\
KK & Kaluza-Klein \\
LED & Large Extra-Dimensions\\
LHC & Large hadron collider\\
LSB & low surface brightness\\
NC & Electroweak Neutral Currents\\
NFW & Navarro, Frenk and White\\
PMNS & Pontecorvo-Maki-Nakagawa-Sakata\\
QED & Quantum Electrodynamics Detection\\
QCD & Quantum Chromodynamics\\
\end{tabular}

\begin{tabular}{ll}
RS & Randall-Sundrum\\
SD & Spin dependent\\
SI & Spin independent\\
SIDM & Self-interactive dark matter\\
SM  & Standard Model\\
SSB & Spontaneous Symmetry Breaking \\
SUSY & Supersymmetry\\
UED & Universal Extra-Dimensions\\
VEV & Vacuum Expectation Value\\
WDM & Warm dark matter\\
WIMP & Weakly Interactive Massive Particle\\
WMAP & Wilkinson Microwave Anisotropy Probe\\
\end{tabular}


\addcontentsline{toc}{chapter}{List of Figures}  
\hypersetup{linkcolor = black}
\listoffigures
\hypersetup{linkcolor = red}

\addcontentsline{toc}{chapter}{List of Tables}  
\hypersetup{linkcolor = black}
\listoftables
\hypersetup{linkcolor = red}

\sloppy 

\chapter*{Preface}
\addcontentsline{toc}{chapter}{Preface} 
\vspace{-0.3cm}
\selectlanguage{british}
The Standard Model of Fundamental Interactions (SM) represents one of the most precise theories in physics. Among the predictions of the SM we find, for instance, the \textit{anomalous magnetic moment} of the electron $a_e = 0.001159652181643(764)$ \cite{Aoyama:2012wj,Aoyama:2014sxa}. This prediction agrees with the experimental results to more than ten significant digits, the most accurate prediction in the history of physics. However, nowadays we have several evidences that the SM only explains $5 \%$ of the matter content of the Universe. The other $95 \%$ are composed by the so-called Dark Energy and Dark Matter. As their names suggest, the nature of these two components of the energy/matter content of the Universe is still unclear and represents one of the most important challenges for the particle physicists. 
In this Thesis we have focused in the study of the phenomenology of one of these mysterious components of the Universe, the Dark Matter. Although we have many evidences of its existence, this new type of matter has not been detected yet. As a consequence, the landscape of the models that can explain the Dark Matter properties is huge. 
In the present work we propose and study several Dark Matter models, setting limits by using experimental results. 

This Thesis is organized in two parts: introduction (Part \ref{sec:introduction}) and scientific research (Part \ref{sec:papers}). First, in Part \ref{sec:introduction} we provide an overview of the current status of the Dark Matter physics: Chapter \ref{sec:SM} explains the fundamental properties of the Standard Model, showing its different open problems. In Chapter \ref{sec:Cosmo} we review the standard cosmological
model $\Lambda$CDM. Chapter \ref{sec:DM} summarizes the fundamental properties and evidences of Dark Matter. Chapter \ref{sec:termo} deals with the tools needed to understand the thermal evolution of cold relics, which plays a central role in this Thesis. In Chapter \ref{sec:Detection} we review the Dark Matter experimental landscape, focusing in Direct Detection and Indirect Detection. Chapter \ref{sec:ED} discusses the fundamental tools to understand extra-dimensional models.
To conclude, in Chapter \ref{sec:summary} we summarize the most important results of the publications that compose this Thesis.
In Part \ref{sec:papers} we present a collection of the publications done during the research.

\hypersetup{urlcolor=black}
\fancyfoot[C]{\thepage}
\selectlanguage{spanish}
\chapter*{Agradecimientos}
\addcontentsline{toc}{chapter}{Agradecimientos}  

Si esta Tesis existe es gracias al apoyo y a la ayuda de toda la gente que me rodea. Me gustaría hacer unos agradecimientos que estén a la altura de todo lo que habéis hecho por mí, pero creo que eso va a ser más difícil que escribir cualquiera de los capítulos que componen este libro. Aunque aún no sé muy bien cómo escribir esta parte, si tengo muy claro por quien empezar. Este trabajo no hubiese sido posible sin la dedicación y la ayuda incondicional de mis directores, Andrea, Nuria y Roberto. Investigar junto a vosotros durante estos cuatro años ha sido todo un placer. Sería imposible cuantificar la gran cantidad de cosas que me habéis enseñado durante este tiempo.

Roberto, te agradezco enormemente tu guía durante las primeras etapas de la Tesis. Gracias al tiempo que trabajamos juntos desarrollé todas mis habilidades con la programación, las cuales valoro muchísimo. Tu dedicación al trabajo siempre me ha parecido impresionante, espero que se me haya quedado algo de tu productividad.

Andrea, aunque solo hemos trabajado juntos durante la segunda mitad del doctorado, has sido un pilar clave en esta Tesis. Desde que empezamos a colaborar has estado pendiente de mí, tus consejos han sido vitales para el desarrollo de todos nuestros proyectos. Tu meticulosa forma de abordar los problemas siempre me ha fascinado, cada vez que hablamos, tu pasión por la física teórica prácticamente se palpa en el ambiente. Espero que se me haya pegado algo de tu forma de trabajar, porque la verdad es que me encanta.

Nuria, tu tiempo y dedicación durante el desarrollo de la Tesis han significado mucho para mí. Siempre con la puerta de tu despacho abierta, dispuesta a hablar de física sea la hora que sea. Me ha encantado la forma en que me has guiado a lo largo de estos años, apoyándome en todo momento y dándome los conocimientos necesarios para llevar a cabo nuestros proyectos, a la vez que me dotabas de autonomía suficiente para desarrollar las ideas por mí mismo. Quiero pensar que el doctorado me ha hecho una persona más independiente y autosuficiente, y eso te lo debo todo a ti, muchas gracias por todo.

Además de a mis directores, me gustaría agradecer también a Nicolás Bernal, con quien he tenido el placer de colaborar en un par de proyectos y de quien he aprendido muchísimo sobre Materia Oscura FIMP, y a Verónica Sanz, que me ha enseñado lo apasionante que puede ser la investigación en temas más allá de la física.

Mis directores no son los únicos que me han ayudado a escribir y corregir esta Tesis, estoy infinitamente agradecido a Andrea, Eva, Iván (desde que te marchaste a Londres se echan tanto de menos nuestros partidos de padel y frontón y nuestras cervezas... \textexclamdown Vuelve pronto!), Pablo, Pau y Stefan por echarme una mano con las correcciones ortográficas de los capítulos en inglés, así como a mi hermana, por ayudarme con la ortografía de la parte en español. \textexclamdown Muchas gracias a todos! No se que haría sin vosotros (tengo que buscar alguna forma de compensaros esto...).

Aparte de la gente que ha participado activamente en mi investigación, quiero agradecer al resto de miembros del grupo SOM: Olga Mena, Pilar Hernández y Carlos Peña. Muchas gracias por todas las charlas, meetings y discusiones en general. Ha sido genial compartir estos años con vosotros, viviendo vuestra pasión por la física de partículas. Además de por lo mucho que he disfrutado en el grupo, siempre estaré en deuda con vosotros por financiarme el doctorado y por darme la oportunidad de visitar Fermilab y la Universidad de Tokyo (y por los muchos congresos a los que he podido asistir gracias a vosotros), estancias que me han encantado y que han significado mucho para mi a nivel personal. Por supuesto, también estoy muy agradecido de haber compartido esta experiencia con lo miembros más nuevos del grupo, Dani Figueroa y Jacobo López. Muchas gracias Jacobo por haberme dado estabilidad económica durante este último año, gracias a ti he podido finalizar esta Tesis. Finalmente, y no menos importante, quiero agradecer también a los doctorandos y postdocs del grupo: Andrea Caputo, David, Fer, Hector, Jordi, Miguel Escudero, Pablo, Sam, Stefan y Victor. Esta experiencia no hubiese sido lo mismo sin vosotros.

Del mismo modo que agradezco a mi grupo la oportunidad que me han dado permitiéndome hacer las estancias y asistir a diversos congresos, quiero dar las gracias a las universidades que me acogieron en estos eventos. En especial quiero darle las gracias a Pedro Machado, por hacerme sentir en Fermilab como en mi propia casa, y a toda la IPMU en general, por la gran acogida que me dieron en Tokyo. Aparte de en lo profesional, estas dos estancias no hubiesen sido nada a nivel personal de no ser por Brais, con quien viví un montón de aventuras y comí miles de hamburguesas en EEUU (conseguiste contagiarme tu pasión por esta joya gastronómica), y por Pablo, con quien recorrí todo Japón y visité lugares que jamás olvidaré (de no ser por tu iniciativa y tus grandes ansias de viaje la estancia no hubiese sido ni la mitad de lo que fue).

La investigación no ha sido la única labor que he desempeñado durante estos años, también he podido experimentar la docencia, la cual me ha sorprendido muy gratamente. Hasta entrar en el doctorado nunca me planteé siquiera que dar clase pudiese gustarme, pero ha resultado ser una de las actividades que más he disfrutado. Quiero agradecer a Arcadi Santamaría y a Mariam Tórtola, con quienes he tenido el placer de compartir esta experiencia, por toda la autonomía que me han dado y lo mucho que me han enseñado de este bello arte. Así mismo, quiero agradecer a todos los alumnos a los que he dado clase durante este tiempo, no se cuánto habrán aprendido ellos de mí, pero espero que sea comparable a lo mucho que he aprendido yo de ellos.

Durante estos cuatro años he compartido el IFIC con gente genial. En especial me gustaría darle las gracias a Avelino por las muchas oportunidades que me ha dado para hablar y presentar mis trabajos en el journal club, la calidez del hogar siempre es el mejor lugar para practicar charlas y debatir sobre física.

A nivel personal, los años de doctorado me han dejado un montón de experiencias que no cambiaría por nada. Me alegro muchísimo de haber compartido este tiempo con Juan (nuestro año viviendo juntos fue una pasada), Pablo y Stefan, los días en el IFIC no hubiesen sido lo mismo sin nuestras meriendas y almuerzos. Al igual que no hubiesen sido lo mismo sin las quedadas con vosotros y con Brais, Inma y Lydia (nunca me canso de vivir aventuras contigo, ni lo haré jamas), sois lo mejor que me ha dado el doctorado. Nuestras escapadas a Moixent durante los meses post cuarentena fueron geniales.

Esta Tesis no solo cierra el doctorado, para mi representa el final de una etapa que comenzó hace ya diez años, cuando empecé la carrera. Dicen que los años universitarios se recuerdan siempre como la mejor etapa de la vida, y en mi caso se ha cumplido totalmente. Los amigos que hice durante aquellos años han estado siempre a mi lado, hemos vivido infinidad de cosas juntos y no los cambiaría absolutamente por nada, si aquellos años fueron tan fantasticos es gracias a vosotros. Primero me gustaría agradecer a mis amigos de física, Andrea Gonzalez, Andrés, Iván, Lydia, Marina, Pau y Román, con quienes he vivido viajes, paellas, carnavales, fiestas y, en resumen, un montón de aventuras desde que nos conocimos, y espero seguir haciéndolo siempre, nunca me canso de estar con vosotros. A mis amigas de bioquímica, Eva, Marcela y Carla, hicisteis que aquellas noches de estudio fuesen geniales (en general no muy productivas, pero inmejorables). Y, finalmente, a Laura, tus consejos siempre me han ayudado a seguir adelante. Mi vida como investigador empezó con todos vosotros, no se me ocurre un comienzo mejor.

Ademas de a mi gente de la uni, quiero agradecer a mis amigos del barrio por estar siempre ahí, tanto en los buenos como en los malos momentos. Abraham, Raul y Rover, no podría imaginarme la vida sin vosotros, de hecho hace tanto que estamos juntos que casi ni recuerdo como era todo antes de conoceros. Siempre me he considerado una persona muy afortunada, y es gracias a vosotros. Alba y Victor, aunque nos conocemos de hace menos tiempo también sois vitales para mi, por muy malo que haya sido el día, cuando estoy con vosotros conseguís que se me olvide. Finalmente, me alegro mucho de haberte conocido Bea, siempre consigues que una tarde/noche cualquiera sea genial, por supuesto con un buen café o una cerveza de por medio.

Quiero agradecer también a toda mi familia por apoyarme incondicionalmente y por haber estado a mi lado desde que tengo memoria. En general, le doy las gracias a mis abuelas, abuelos, tías, tíos, primas y primos por todo lo que hemos vivido juntos y por todo el amor y el cariño que me habéis transmitido a lo largo de toda mi vida.
A mi madre, por cuidarme desde que tengo uso de razón. Siempre has apostado por mi, a pesar de que al principio me costase encontrar mi camino.
A mi padre, no solo por el afecto que me has mostrado durante toda mi vida, también por transmitirme todas tus inquietudes intelectuales sobre astrofísica y cosmología. Todas las historias que me contabas de pequeño sobre la teoría de la relatividad y sobre el tiempo hicieron mella en mi y han acabado convirtiéndome en quien soy.
Y finalmente, y para nada menos importante, todo lo contrario, a mi hermana, por su alegría y por aguantarme y quererme siempre tal como soy, con todos mis (muchos) defectos y virtudes.

Por último, quiero darle las gracias a Andrea: desde que nos conocimos has estado siempre ahí, soportando todas mis tonterias, mostrándome siempre todo tu apoyo y amor. Me encanta como eres, la persona más dulce que he conocido, aunque fuerte y dura cuando es necesario (sin ti no se cuánto tiempo hubiese tardado en escribir esta Tesis). No puedo imaginarme la vida sin nuestros paseos, nuestras series y pelis, nuestras cenas y, en definitiva, sin ti.

Después de leer esta parte como cincuenta veces creo que estoy contento con como ha quedado, así que ya, sin más dilación, \textexclamdown os dedico esta Tesis!
\selectlanguage{british}
\hypersetup{urlcolor=blue}
\fancyfoot[C]{} 
\hypersetup{linkcolor=black} 
\tableofcontents
\definecolor{linkcolour}{rgb}{0.85,0.15,0.15}
\hypersetup{linkcolor=linkcolour}

\mainmatter 
\renewcommand{\headrulewidth}{0.5pt} 



\part{Introduction}\label{sec:introduction}\thispagestyle{empty}
\renewcommand{\headrulewidth}{0.4pt}

\lhead[{\bfseries \thepage}]{ \rightmark}
\rhead[ Chapter \thechapter. \leftmark]{\bfseries \thepage}
\chapter{Standard Model of Particles: A Brief Review}
\label{sec:SM}
Humankind have always tried to understand the great mysteries of the Universe, as well as those of the matter that surrounds us. The Greeks were the first to try to model nature by postulating that all forms of matter can be understood starting from four fundamental elements: water, earth, fire and air. It took thousands of years to refine this description. In the 17th century the first definition of a chemical element was made and after two centuries (1869) the periodic table of the elements was published, which order them according to their chemical properties, with a number of elements very similar to that we know today.

\begin{figure}[htbp]
\centering
\includegraphics[width=80mm]{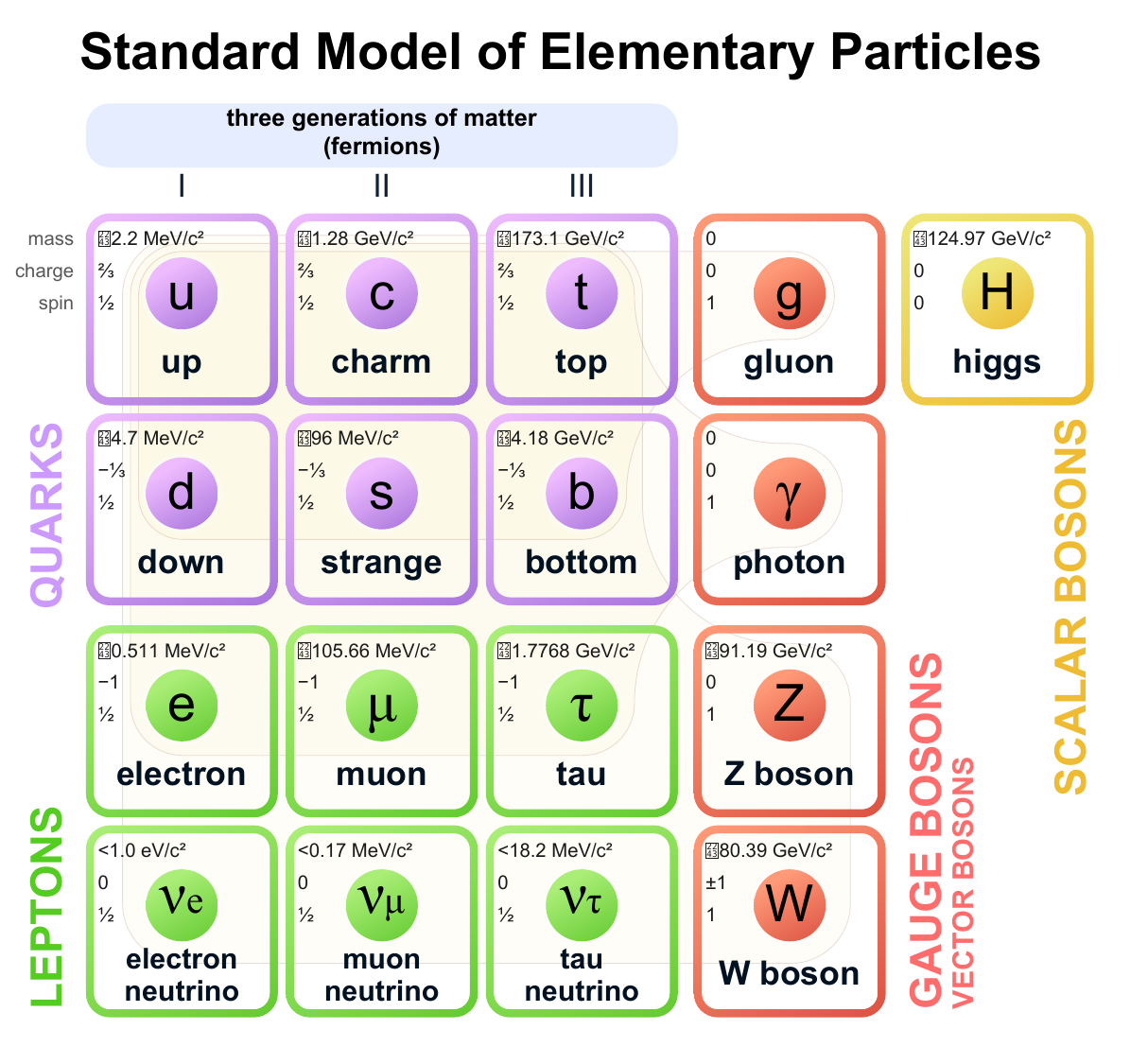}
\caption[Standard model of particles]{\it Standard model of particles: Purple, green and red particles represents, respectively, quarks, leptons and gauge bosons. The yellow particle represents the Higgs boson. Image taken from Ref.~\cite{citaSM}.}
 \label{fig:SM_particle_content}
\end{figure}

The high number of elements that were known at the beginning of the 20th century led us to think that there had to be a more elementary underlying structure that we did not understand. It was finally Niels Bohr who first proposed the current atomic theory \cite{Bohr:1913:CAMa,Bohr:1913:CAMb,Bohr:1913:CAMc}, in which matter was explained by electrons, protons, and subsequently neutrons. This simplified theory was refined over the years, giving birth to in the Standard Model of Fundamental Interactions (SM).

The model accurately describes nature at the microscopic level using a total of twelve elementary particles that constitute matter at the fundamental level and three force fields. The gravitational field is excluded from this description since, to this day, the principles of the quantum world and the theory of General Relativity have not been reconciled.

\section{Particle Content}

The Standard Model of particles \cite{PhysRevLett.19.1264,GLASHOW1961579,Salam:1968rm,PhysRevLett.13.321,PhysRev.145.1156,PhysRevLett.13.585,PhysRev.155.1554,PhysRev.139.B1006,PhysRevD.2.1285,FRITZSCH1973365,PhysRevLett.13.508} is a relativistic quantum field theory with gauge symmetry $SU(3)_C \times SU(2)_L \times U(1)_Y$ (in the next section we provide a quick justification for the choice of these groups). The model describes with great precision the strong, weak and electromagnetic interactions through the exchange of different spin-1 fields, which constitute the gauge bosons of the theory. The symmetry group $SU(3)_C$ is the one associated with strong interactions, while $SU(2)_L \times U(1)_Y$ is the symmetry group of the Electroweak Theory, that unifies electromagnetic and weak interactions.

\begin{table}
\begin{sideways}
\centering
\begin{tabular}{c c c c c c}
    \hline
    \bfseries Particle & \bfseries Discovered at & \bfseries Mass & \bfseries charge & \bfseries spin & \bfseries lifetime [s]  \\ 
    \hline
     Electron $e$ & Cavendish Laboratory (1897) \cite{Thomson:1897cm} & $510.9989461(31)$ keV & $-1$ & $1/2$ & Stable \\
     muon $\mu$ & Caltech (1937) \cite{Street:1937me} & $105.6583745(24)$ MeV & $-1$ & $1/2$ &  $2.2\times 10^{-6}$  \\
     tau $\tau$ & SLAC (1976) \cite{Perl:1975bf} & $1776.86(12)$ MeV & $-1$ &  $1/2$ & $2.9\times 10^{-13}$ \\
     neutrino $\nu_e$ & Savannah River Plant (1956) \cite{Reines:1956rs} & $< 2$ eV & $0$ & $1/2$ & Stable  \\
     neutrino $\nu_\mu$ & Brookhaven (1962) \cite{Danby:1962nd} & $< 2$ eV & $0$ & $1/2$ & Stable \\
     neutrino $\nu_\tau$ & Fermilab (2000) \cite{Kodama:2000mp} & $< 2$ eV & $0$ & $1/2$ & Stable \\
     quark $u$ & SLAC (1968) \cite{Aubert:1974js,Augustin:1974xw} & $2.2(5)$ MeV & $2/3$ & $1/2$ & Stable \\
     quark $d$ & SLAC (1968) \cite{Aubert:1974js,Augustin:1974xw} & $4.7(5)$ MeV& $-1/3$ & $1/2$ & Stable \\
     quark $c$ & Brookhaven \& SLAC (1974) \cite{Bloom:1969kc,Breidenbach:1969kd} & $1.275(35)$ GeV & $2/3$ & $1/2$ & $1.1\times 10^{-12}$ \\
     quark $s$ & SLAC (1968) \cite{Aubert:1974js,Augustin:1974xw} & $95(9)$ MeV & $-1/3$ & $1/2$ & $1.24\times 10^{-8}$ \\
     quark $t$ & Fermilab (1995) \cite{D0:1995jca} & $173.0(4)$ GeV & $2/3$ & $1/2$ & $4.6\times 10^{-25}$ \\
     quark $b$ & Fermilab (1977) \cite{Herb:1977ek} & $4.18(4)$ GeV & $-1/3$ & $1/2$ & $1.3\times 10^{-12}$ \\
     photon $\gamma$ & Washington University (1923) \cite{Compton:1923zz} & $<10^{-18}$ eV & $0$ & $1$ & Stable \\
     gluon $G$ & DESY (1979) \cite{Berger:1978rr} & 0 & $0$ & $1$ & Stable \\
     $W^{\pm}$ boson & CERN (1983) \cite{Banner:1983jy,Arnison:1983rp} & $80.379(12)$ GeV  & $\pm 1$ & $1$ & $3.2\times 10^{-25}$ \\
     $Z$ boson & CERN (1983) \cite{Bagnaia:1983zx,Arnison:1983mk} & $91.1876(21)$ GeV & $0$ & $1$ & $2.6\times 10^{-25}$ \\
     Higgs boson $H$ & CERN (2012)  \cite{Aad_2012,Chatrchyan_2012} & $125.18(16)$ GeV  & $0$ & $0$ & $>5.1\times 10^{-23}$ \\     
    \bottomrule
\end{tabular}
\end{sideways}
\caption[Properties of SM particles.]{Properties of SM particles. Idea taken from \cite{CottinBuracchio:2276806}. The different data has been extracted from \cite{PhysRevD.98.030001}. Up, down and strange quark masses are estimates of so-called \textit{current quark masses}. On the other hand, charm and beauty quark masses are the \textit{running} masses, while the top quark mass comes from direct measurements.}
\label{propiedades_particulas}
\end{table}

The fundamental constituents of matter are fermions, described by the fermionic sector of the Standard Model. It is made up of quarks and leptons, both types of particles separated into three flavour families. Quarks are charged under $SU(3)_C$. As a consequence, each quark appears in the model in three different colours. Leptons can be separated into two kinds of particles: charged leptons and neutrinos (the SM predicts zero mass for the neutrinos). The Standard Model is a chiral gauge theory, in the sense that it treats differently particles with right- and left-handed chiralities, grouping the right-handed in singlets and the left-handed in doublets under the symmetry group $SU(2)_L$. The properties of all these particles are collected in the Review of Particle Physics \cite{PhysRevD.98.030001}, published and reviewed annually by the Particle Data Group (PDG). To date, no evidence of the existence of right-handed neutrinos has been found, so that they are not included in the model:
\bea
\label{familia_1}
\nth{1} \, \text{Family}: L_{1} \equiv \left( \begin{array}{c}
\nu_e  \\
e^{-}  \end{array} \right)_{L} \, ; \, e_1 \equiv e^{-}_{_R} \, ; \, 
Q_1 \equiv \left( \begin{array}{c}
u  \\
d  \end{array} \right)_{L} \, ; \, U_1 \equiv u_R \, ; \, D_1 \equiv d_R \, , \nonumber \\
\label{familia_2}
\nth{2} \, \text{Family}: L_{2}\equiv\left( \begin{array}{c}
\nu_{\mu}  \\
\mu^{-}  \end{array} \right)_{L} \, ; \, e_2 \equiv \mu^{-}_{_R} \, ; \,
Q_2 \equiv\left( \begin{array}{c}
c  \\
s  \end{array} \right)_{L} \, ; \, U_2 \equiv c_R \, ; \, D_2 \equiv s_R \, , \nonumber \\
\label{familia_3}
\nth{3} \, \text{Family}: L_{3}\equiv\left( \begin{array}{c}
\nu_{\tau}  \\
\tau^{-}  \end{array} \right)_{L} \, ; \, e_3 \equiv \tau^{-}_{_R} \, ; \, 
Q_3 \equiv\left( \begin{array}{c}
t  \\
b  \end{array} \right)_{L} \, ; \, U_3 \equiv t_R \, ; \, D_3 \equiv b_R \, . \nonumber
\eea

The different interactions are mediated by spin-1 particles, the so-called gauge bosons. The electromagnetic and strong interactions are mediated by the photon $\gamma$ and the gluon $G^a$ (with $a = 1, ..., 8$), respectively, and have an infinite range of interaction, due to the zero mass of the mediators. The weak interactions are mediated by the massive $W^{\pm}$ and $Z$ bosons; due to the mass of the mediators the weak interaction is a short-range force. Eventually, the only spin-0 particle in the model is the Higgs boson, the most recently discovered components of the SM. The interaction of the Higgs field with the rest of the particles explains the mass generation in the SM. 
The different properties of particles and mediators of the SM are collected in Tab.~\ref{propiedades_particulas}.

\section{Electroweak Unification: The Election of $SU(2)_L \times U(1)_Y$}
Having understood the components of the Standard Model that we observe in experiments, we will try to establish the theoretical framework in which the interactions between the different fermions take place. In order to do this, we must first talk about the choice of the symmetry group. 

The description of the weak interactions was one of the great problems of the second half of the 20th century. At that moment, only one gauge theory was known: Quantum Electrodynamics (QED) \cite{10.1143/PTP.1.27,10.1143/ptp/2.3.101,10.1143/ptp/3.2.101,PhysRev.74.224,PhysRev.74.1439,Feynman:1948ur,Feynman:1949zx} that describes the electromagnetic interactions. The structure of QED is very simple, as the gauge group is $U(1)$, the only mediator is the photon. Its simplicity allows to point out clearly two important characteristics of every gauge theory: on the one hand, the interaction is composed by a gauge field times a fermionic current; on the other hand, the associated charge in QED is the symmetry group generator. However, the observation of the parity violation \cite{1957PhRv..105.1413W,1957PhRv..105.1415G} makes weak interactions totally different from QED.

The election of $SU(2)$ to describe a gauge theory of the weak interaction seems logic: experimentally, we observe three gauge bosons ($W^\pm, Z^0$) and $SU(2)$ has three generators. The weak interaction between charged leptons and neutrons would be described by the Lagrangian
\be
\mathcal{L} \propto  (J_\mu W^\mu + \text{h.c}) \, ,
\ee
with
\be
J_\mu \equiv \bar{\nu}_e \gamma_\mu (1 - \gamma_5) \, e \, ,
\ee
where $J_\mu$ is the weak current and $\gamma_\mu$  and $\gamma_5$ are Dirac matrices\footnote{There are many books where it is possible to find a complete description of the algebra of these matrices \cite{Peskin:1995ev,Mandl:1985bg}, known as Clifford Algebra.}. The gauge bosons in this group are denoted as $W_i^\mu = \left(W_1^\mu,W_2^\mu,W_3^\mu\right)$. This description has several problems, starting with the fact that the term with $W_3$ is not electrically neutral. On the other hand, the three charges of this Lagrangian $\left(T_1, T_2, T_3\right)$ do not form a closed algebra.

$SU(2)$ can not explain the weak interaction since $m_Z \neq m_W$, but $SU(2) \times U(1)$ can explain it\footnote{Historically, the election of the group was not clear until the observation of the Z boson mass. If the mass of the three mediators would have been equal, $m_Z = m_W$, the gauge group $O(3)$ \cite{Georgi:1974sy} could have explained the weak interaction. However, the discovery that $m_Z$ and $m_W$ are related via the Weinberg angle was the key to understand that $SU(2) \times U(1)$ was the correct gauge group.} (at the same time that unifies it with the electromagnetic interaction!). The $U(1)$ group of this new theory is different to the $U(1)$ of QED. The conserved charge in this case is not the electrical charge $Q$. The gauge boson of this new $U(1)$ symmetry group is denoted as $B^\mu$. In order to obtain zero electrical charge for all Lagrangian terms, at the same time that it mixes charged leptons and neutrinos, a complicate structure that differentiate between left- and right-handed fields is needed. The left-handed fields are charged under the $SU(2)$ while the right-handed fields are neutral under this group (for this reason the group is labelled as $SU(2)_L$). This charge is the so-called \textit{weak isospin}, $T_3$. On the other hand, the charge of the new $U(1)$ group $Y$ receives the name of \textit{hypercharge} and is related with $Q$ and $T_3$ by $Y = Q - T_3$.

The unification of QED and the weak interactions receives the name of the \textit{Electroweak Theory} \cite{PhysRevLett.19.1264,GLASHOW1961579,Salam:1968rm}. 

\section{Quantum Chromodynamics (QCD): The Strong Interaction Gauge Group}

The gauge theory that describes the strong interactions is called Quantum Chromodynamics (QCD). The fundamental structure of QCD is similar to the QED structure, both are vector theories: left- and right-handed representations are the same.
Such as in the QED case, QCD presents a conserved charge called \textit{colour}. 
The particles that have colour charge in the SM are the quarks while QCD mediators are the gluons, which contrary to the photon also have color.
Despite the similarities, QCD has two main properties that makes it totally different to QED. On the one hand, the theory presents color confinement: it is impossible to observe free particles with colour charge. On the other hand, the theory has asymptotic freedom: discovered by David Gross \cite{PhysRevLett.30.1343} and Frank Wilczek \cite{PhysRevLett.30.1346} in 1973, this property can be described as a reduction in the strength of interactions between quarks and gluons, going from low to high-energy. This two properties makes Quantum Chromodynamics one of the most complex theories in particle physics, as at low energies the relevant degrees of freedom are not quarks and gluons (that are confined) but colourless mesons and baryons. A complete description of this theory can be found in Ref.~\cite{Pich:1999yz}.

Until the present day, the Electroweak Theory and QCD have not been properly unified. However, both theories can be grouped under the simple group:
\be
G = SU(3)_C \times SU(2)_L \times U(1)_Y ,
\label{SM_gauge_group}
\ee
the gauge symmetry group of the Standard Model.
The different charges of the right- and left-handed fermionic fields under the electroweak symmetry group are summarized in Tab.~\ref{cargas_particulas}.
\begin{table}
\centering
\begin{tabular}{c c c }
    \hline
    \bfseries Particle Name & \bfseries Field & \bfseries $SU(3)_C \times SU(2)_L \times U(1)_Y$   \\ 
    \hline
     Quarks left-handed & $Q_\alpha$ & $(3,2,1/6)$  \\
     Quarks right-handed & $U_\alpha$ & $(3,1,1/6)$  \\
    	  & $D_\alpha$ & $(3,1,-1/3)$  \\
     Lepton left-handed & $L_\alpha$ & $(1,2,-1/2)$  \\
     Lepton right-handed & $e_\alpha$ & $(1,1,-1)$  \\
    \bottomrule
\end{tabular}
\caption[Charge of the different SM fermionic components under the gauge fields of the SM $SU(3)_C \times SU(2)_L \times U(1)_Y$.]{Charge of the different SM fermionic components under the gauge fields of the SM $SU(3)_C \times SU(2)_L \times U(1)_Y$. In all fields $\alpha = 1,\, 2,\, 3$ represents the family. }
\label{cargas_particulas}
\end{table}

\section{Standard Model Lagrangian without masses}
In the previous section we have established the gauge symmetry group of the theory. The gauge fields of $SU(2)_L \times U(1)_Y$ are $W_i^\mu = \left(W_1^\mu,W_2^\mu,W_3^\mu\right)$ and $B^\mu$, while the QCD gauge boson is the gluon $G^\mu_a$, where ($a = 1, ..., 8$). It is very common to distinguish between two sectors to describe the SM Lagrangian: the gauge sector, that describes the interactions between the different gauge fields, and the fermionic sector, that describes the matter Lagrangian.

The Lagrangian that describes the gauge sector can be written as:
\be
\mathcal{L}_{\text{kin,gauge}} =-\dfrac{1}{4} G_{\mu \nu}^a G^{\mu \nu}_a - \dfrac{1}{4} W_{\mu \nu}^i W^{\mu \nu}_i - \dfrac{1}{4} B_{\mu \nu} B^{\mu \nu} \, ,
\label{lagrangiano_v1}
\ee 
where, from the gauge fields, the following tensors have been defined
\bea
\left \{
\begin{array}{lll}
G_{\mu \nu}^a &=& \partial_\mu G_\nu^a - \partial_\nu G_\mu^a + g_s f^{abc}G_\mu^b G_\nu^c \hspace{1.5cm} a, b, c =1, ..., 8 \, ; \nonumber \\
W_{\mu \nu}^i &=& \partial_\mu W_\nu^i - \partial_\nu W_\mu^i + g \epsilon^{ijk}W_\mu^j W_\nu^k \hspace{1.5cm} i, j, k =1, 2, 3 \, ; \nonumber \\
B_{\mu \nu} &=& \partial_\mu B_\nu - \partial_\nu B_\mu \, ,
\label{tensores_SM}
\end{array}
\right .
\\
\, 
\eea
being ($g_s,g,g'$) the different couplings of each symmetry group of the SM, $SU(3)_C$, $SU(2)_L$ and $U(1)_Y$, respectively. In the above expressions $f^{abc}$ and $\epsilon^{ijk}$ are the antisymmetric structure constants of the $SU(3)_C$ and $SU(2)_L$ gauge groups and are defined through the commutators of the different group generators
\bea
\left \{
\begin{array}{lll}
\left[\lambda_a , \lambda_b\right] &=& i f^{abc} \lambda_c \, , \\
\left[\sigma_i , \sigma_j\right] &=& 2 i \epsilon^{ijk} \sigma_k \, ,
\end{array}
\right .
\eea
where $\lambda$ and $\sigma$ are the Gell-Mann and Pauli matrices.

The Lagrangian that describes the fermionic content of the Standard Model can be written as
\bea
\mathcal{L}_{\text{kin,fermions}} &=& i \sum_\alpha \, \left( \bar{Q}_\alpha \gamma^\mu \mathcal{D}_\mu Q_\alpha \, + \, \bar{U}_\alpha \gamma^\mu \mathcal{D}_\mu U_\alpha \, + \, \bar{D}_\alpha \gamma^\mu \mathcal{D}_\mu D_\alpha \right. \, \nonumber \\ 
&+& \, \left. \bar{L}_\alpha \gamma^\mu \mathcal{D}_\mu L_\alpha \, + \, \bar{e}_\alpha \gamma^\mu \mathcal{D}_\mu e_\alpha \, \right) \, ,
\label{lagran_ferm}
\eea
where the sum is over the three flavour families. $\mathcal{D}_\mu$ is the covariant derivative that preserves the gauge invariance of the Lagrangian and $\gamma^\mu$ are the Dirac matrices. The covariant derivative can be written as
\bea
\mathcal{D}_\mu \equiv \partial_\mu - ig_s\dfrac{\lambda_a}{2}G^a_\mu - i g \dfrac{\sigma_i}{2}W^i_\mu - i g'YB_\mu \, ,
\label{derivadas_covariantes}
\eea
where the interaction with a given gauge boson arises only if the matter field is charged under the corresponding group.

With the Lagrangian described in this section, the particle content of the SM is fixed. The problem with this description is that the SM does not accept the existence of masses for any field. On the one hand, left- and right-handed fermions are different with respect to the $SU(2)_L$ gauge group. This fact makes it impossible to write gauge invariant mass terms for the fermions in the Lagrangian. On the other hand, any mass term for the gauge fields is not gauge invariant. As a consequence, all gauge and fermionic fields are massless in the described theory. 
If the weak theory did not exist, this issues would not affect the description of QED and QCD: in both theories, left- and right-handed representations of the fermions fields are the same and the mediators (gluons and photons) are massless. However, the $SU(2)_L$ description of the weak interaction and the fact that the range of this interaction is finite (which implies massive mediators) points out a problem of the Lagrangian in Eq.~\ref{lagrangiano_v1} and \ref{lagran_ferm}.

In order to understand the problem with the fermion masses we must remember the structure of the electroweak interaction and the different treatment of the right-handed (singlets of $SU(2)_L$) and left-handed (doublets of $SU(2)_L$) fields. The mass terms usually have the form
\be
m_\psi \bar{\psi} \psi = m_\psi \left(\psi_L^\dagger \psi_R + \psi_R^\dagger \psi_L\right) \, .
\label{mass_term}
\ee
As a consequence of the different representation of left- and right-handed fields under the $SU(2)_L$ group, Eq.~\ref{mass_term} can not be part of the Lagrangian because it explicitly breaks $SU(2)_L$ invariance. For this reason, in Eq.~\ref{lagran_ferm} these kind of terms are not present.

\section{The Higgs Mechanism}

The scalar sector was the last stone added to the SM at the turn of the beginning of the 21st century. The gauge bosons of the Electroweak Theory have mass. However, gauge invariance forbids explicit mass terms in the Lagrangian. The solution to this problem was proposed by Peter Higgs, Robert Brout, Francois Englert, Gerald Guralnik, Carl Richard Hagen and Tom Kibble in 1962 and developed in what is currently known as the Higgs Mechanism \cite{PhysRevLett.13.321,PhysRev.145.1156,PhysRevLett.13.585,PhysRev.155.1554}.

The Higgs Mechanism is based on the idea of the Spontaneous Symmetry Braking (SSB). The symmetry that we have to break is the electroweak symmetry, that has four generators, or four gauge bosons. The final four bosons have to be ($W_\mu^\pm,Z^0_\mu$) and $A_\mu$ (the gauge field of QED, the photon), of which three have non-zero mass. On the one hand, the Electroweak Theory mixes the weak neutral currents with the hypercharge one; on the other hand, we know that QED has only one generator. In other words, the Higgs Mechanism must break the electroweak symmetry to QED:
\be
\label{SSB_esquema}
SU(2)_L \times U(1)_Y \, \longrightarrow \, U(1)_{\text{EM}} \, .
\ee 
The idea of the mechanism consists in to add a new scalar field with a non-zero vacuum value that breaks the symmetry, giving masses to the fermions and gauge fields. The question now is, how must be the structure of this new field?

Before starting to describe the Higgs Mechanism it is important to understand the meaning of SSB. According to the Noether theorem \cite{1971TTSP....1..186N}, each Lagrangian symmetry implies a conserved charge. This theorem was proposed for classical mechanics and is totally valid in quantum mechanics and quantum field theory. However, there are two different ways to realize the theorem in Nature. On the one hand, the most common one is to assume that the vacuum is symmetric under the associated transformation (if $Q$ represents the charge operator, then $Q|0\rangle = 0$). This quantization mechanism is known as Wigner-Weyl quantization \cite{1927ZPhy...46....1W} and the symmetries that describes are called \textit{exact symmetries}. On the other hand, if the vacuum state is not symmetric under some Lagrangian symmetry ($Q|0\rangle \neq 0$) we say that the symmetry is spontaneously broken. This case is known as Nambu-Goldstone quantization \cite{1960PhRv..117..648N,Goldstone:1961eq} and its most relevant consequence is the prediction of massless bosons associated with the broken symmetry, known as \textit{Goldstone bosons}\footnote{The prediction is known as \textit{Goldstone theorem}.}. More concretely, for each broken generator of the theory a new Goldstone boson appears. Fig.~\ref{SSB_esquema} shows a representation of the implications of this kind of symmetries. Despite the fact that the potential is symmetric under a certain transformation, its vacuum state is not (it presents a non-zero expectation value). 

The Higgs Mechanism breaks three generators of the original Electroweak Theory. In order to break the Electroweak Theory as in Eq.~\ref{SSB_esquema}, according to the Goldstone theorem, three massless real scalar fields appear. These fields are \textit{eaten} by the gauge bosons of the Electroweak Theory, becoming on its longitudinal degrees of freedom and providing masses for the particles of the SM. One of this new real scalar fields takes a non-zero \textit{vacuum expectation value} (VEV), providing the structure for the mass terms of the Lagrangian. This field must be electrically neutral. The reason is easy: the electromagnetism is an exact symmetry of the vacuum and, as a consequence, the field that takes the VEV cannot be charged under $U(1)_{\text{EM}}$.


\begin{figure}[htbp]
\centering
\includegraphics[width=80mm]{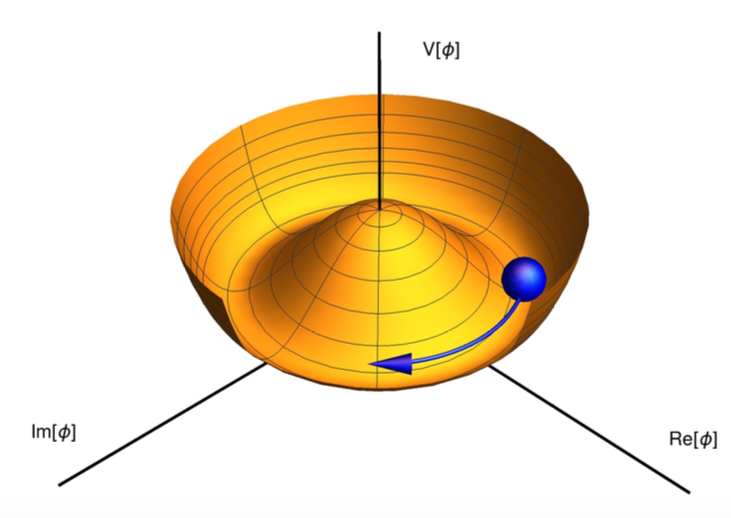}
\caption[Representation of the Spontaneous symmetry Breaking]{\it Representation of the Spontaneous symmetry Breaking.}
 \label{fig:goldstone_theorem}
\end{figure}

As the theory does not allow terms like Eq.~\ref{mass_term} for the fermions, it is necessary that the new field couples to the left-handed doublets of $SU(2)_L$ to generate this kind of mass terms. The minimal candidate that fulfills all requirements is
\be
\Phi = \left( \begin{array}{c}
\Phi^+  \\
\Phi^0  \end{array} \right) = 
\left( \begin{array}{c}
\Phi^+  \\
\dfrac{1}{\sqrt{2}}(v + \phi_1 + i\phi_2)  \end{array} \right) \, ,
\ee
where the $\Phi^+$ is a complex scalar field, $\phi_1$ and $\phi_2$ are real scalar fields and $v$ is the VEV of the Higgs field. The charges of this field under the gauge groups of the SM $SU(3)_C \times SU(2)_L \times U(1)_Y$ are $(1,2,1/2)$. The scalar sector of the Lagrangian of the SM takes the form
\be
\mathcal{L}_\text{scalar} = (\mathcal{D}_\mu \Phi )^\dagger (\mathcal{D}^\mu \Phi )^\dagger - \mu_\Phi^2 \Phi^\dagger \Phi - \lambda_4 (\Phi^\dagger \Phi)^2  \, .
\label{Lagrangiano_Higgs}
\ee
If the mass of this new field is imaginary ($\mu_\Phi^2 < 0$) there will be a VEV different from zero
\be
\langle \Phi \rangle = \left( \begin{array}{c}
0  \\
v/\sqrt{2}  \end{array} \right) =
 \left( \begin{array}{c}
0  \\[8pt]
\sqrt{\dfrac{-\mu_\Phi^2}{2 \lambda_4}}  \end{array} \right)
\ee

A clever way to break the symmetry is to use the Kibble parametrization \cite{PhysRev.155.1554}
\be
\Phi = \exp \left( i \, \dfrac{\vec{\sigma}}{2} \cdot \dfrac{\vec{\theta}}{v} \right) \left( \begin{array}{c}
0  \\[8pt]
\dfrac{H + v}{\sqrt{2}}  \end{array} \right) \, ,
\label{kibble_prametrization_Higgs_field}
\ee
where $\vec{\sigma} = (\sigma_1,\sigma_2,\sigma_3)$ are the Pauli matrices, $\vec{\theta} = (\theta_1,\theta_2,\theta_3)$ are the three real fields (Goldstone bosons) that will be absorbed by gauge bosons after the SSB and $H$ the massive field responsible for the SSB (Higgs boson). Applying the corresponding gauge transformation to Eq.~\ref{kibble_prametrization_Higgs_field} the scalar doublet takes the form
\be
\Phi =\left( \begin{array}{c}
0  \\[8pt]
\dfrac{H + v}{\sqrt{2}}  \end{array} \right) \, .
\ee

In July of 2012, the hypothesis of the Higgs Mechanism was confirmed with the discovery at CERN, simultaneously by the LHC experiments ATLAS \cite{Aad_2012} and CMS \cite{Chatrchyan_2012}, of a new particle with mass $m_H = 125.3 \pm 0.4$ GeV that coincides in properties with the boson mediator of the Higgs field (a spin-0 particle with positive parity). The data from CDF and D0 collaborations of the Tevatron experiment at Fermilab confirmed the discovery \cite{Aaltonen_2012}.

\subsection{Gauge Bosons Masses}

The mass terms of the gauge fields come from the derivative terms of Eq.~\ref{Lagrangiano_Higgs}. If we consider the electroweak structure of Eq.~\ref{derivadas_covariantes} the derivative term of the Higgs potential is:
\bea
(\mathcal{D}_\mu \Phi^\dagger)(\mathcal{D}^{\mu} \Phi) &=& \left| \left( i g \dfrac{\vec{\sigma}}{2}\vec{W}_\mu \, + \, \dfrac{ig'}{2}B_\mu\right) \Phi \right|^2 \, \nonumber \\
&=&  (v + H)^2 \left( \dfrac{g^2}{4} W^+_\mu W^{\mu -} + \dfrac{g^2}{8\, \text{cos}^2(\theta_W)} Z_\mu Z^\mu \right) \, , \nonumber \\
\,
\label{terminos_masas_bosones_weak}
\eea
where the mass states are given by
\bea
W^\pm &\equiv& \dfrac{1}{\sqrt{2}} \left( W_1^\mu \mp i W_2^\mu  \right)
\eea
and
\bea
\left( \begin{array}{c}
A^\mu  \\
Z^\mu  \end{array} \right) 
&\equiv & 
\left( \begin{array}{cc}
\cos (\theta_W) & \sin (\theta_W)  \\
-\sin (\theta_W) & \cos (\theta_W)  \end{array} \right)
\left( \begin{array}{c}
B^\mu  \\
W_3^\mu  \end{array} \right)  \, ,
\eea
where $A^{\mu}$ represents the gauge field of the photon while ($W^{\pm}_\mu,Z_\mu$) are the weak bosons, that acquire masses after the SSB thanks to the VEV of the Higgs field. 

The mixing angle between ($B^\mu, W_3^\mu$) and ($A^\mu, Z^\mu$) is defined as a combination of the $g$ and $g'$ couplings to the gauge groups
\be
\cos (\theta_W) = g/\sqrt{g^2 + g'^2} \, ,
\ee
and the masses of the gauge bosons are given by the different quadratic terms in Eq.~\ref{terminos_masas_bosones_weak}:
\bea
\left \{
\begin{array}{lll}
m_W &=& \dfrac{1}{2} g v \, , \\[8pt]
m_Z &=& \dfrac{1}{2} v \sqrt{g^2 + g'^2} \, , \\
m_A &=& 0 \, .
\end{array}
\right .
\label{masas_gauge}
\eea
The difference $m_W \neq m_Z$ was the key to understand that $SU(2)\times U(1)$ is the gauge group of the weak interaction. This fact ends the discussions about the possible description of the weak theory using $O(3)$.

\subsection{Vacuum Expectation Value and Higgs Boson Mass}
The mass of the Higgs boson $H$ is given by the non-derivative terms in Eq.~\ref{Lagrangiano_Higgs}:
\be
m_H = \sqrt{2 \lambda_4} \, v \, .
\label{higgs_mass}
\ee
The question now is, how to compute the VEV? The value of this constant is computed using the muon decay channel $\mu \rightarrow \nu_\mu \, + \, e^-  \, + \, \bar{\nu}_e$. 
\begin{figure}[htbp]
\centering
\includegraphics[width=50mm]{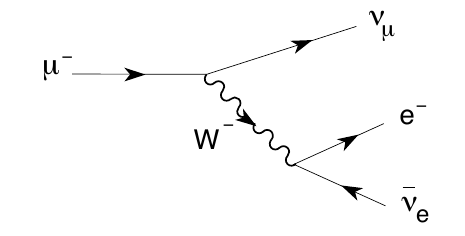}
\caption[Muon decay channel $\mu \rightarrow \nu_\mu \, + \, e^-  \, + \, \bar{\nu}_e$]{\it Muon decay channel $\mu \rightarrow \nu_\mu \, + \, e^-  \, + \, \bar{\nu}_e$. The prediction of the effective weak theory of this decay, compared with the result of the Electroweak Theory, was used to calculate the VEV. }
 \label{fig:epocas}
\end{figure}
The prediction of the Electroweak Theory is proportional to $g^2/(8m_W^2)$, while the prediction of the effective weak currents (where the gauge boson has been integrated out) is $G_F/\sqrt{2}$, where $G_F=1.17\times10^{-5} \, \text{GeV}$ represents the Fermi constant. Both prediction must be the same and, as a consequence,
\be
v = \dfrac{1}{\sqrt{\sqrt{2}G_F}} \backsimeq 246 \, \text{GeV}  \, .
\label{vev_value}
\ee

\subsection{Fermion Masses}

As already commented, the SM does not allow mass terms for the fermion particles. The reason is the difference between the left- and right-handed fields, doublets and singlets of $SU(2)$, respectively. One can analyse the problem using the hypercharge. All terms in the Lagrangian must have zero hypercharge. However, fermionic mass terms have non-zero hypercharge
\bea
\left \{
\begin{array}{lll}
Y(\bar{Q}_1 D_1) = -1/2 \, , \\
Y(\bar{D}_1 Q_1) = 1/2 \, . \\
\end{array}
\right .
\label{hypercargas_varias}
\eea

Now, in order to give mass to the gauge bosons $Z$ and $W^\pm$ we have introduced a new scalar field. The hypercharge of the Higgs field is irrelevant to give mass to the gauge bosons as only the combination $g' \, Y$ is relevant. However, we can choose $Y(\Phi) = 1/2$ to solve at the same time the fermion mass problem. Therefore, terms like $\bar{Q}_i \Phi D_i$ or $\bar{Q}_i \tilde{\Phi} U_i$ (where $\tilde{\Phi} = i \sigma_2 \Phi^\ast$) have zero hypercharge and can be part of the Lagrangian. After the SSB, when the Higgs field takes a VEV, mass terms are generated automatically in the Lagrangian
\bea
\bar{Q}_1 \Phi D_1 \, + \, \text{h.c.}  &\xrightarrow[\text{SSB}]{\text{}} (\bar{u}_1 \, \, \,\, \, \bar{d}_1)_L \left( \begin{array}{c}
0 \nonumber  \\
v/\sqrt{2}  \end{array} \right) D_1 \, + \, \text{h.c.} \, \\
 &= \dfrac{v}{\sqrt{2}} \, (d_L^\dagger d_R + d_R^\dagger d_L) = \dfrac{v}{\sqrt{2}} \, \bar{d} d \, .
\eea
Once $Y(\Phi)$ is fixed to $1/2$, terms that mix leptons and quarks are not allowed because the hypercharge continues to be different to zero. 

One problem of the SM is that we have three different flavour families of fields. This makes it more difficult to write the mass terms. Terms that mix leptons and quarks have non-zero hypercharge and are forbidden, but terms that mix quarks from different flavour families can be part of the Lagrangian, and the same happens for the leptonic terms. As a consequence, the most general Lagrangian that we can build is
\be
\mathcal{L}_Y =  - \bar{L}_\alpha \, Y^L_{\alpha \beta} \,  \Phi \,  e_\beta - \bar{Q}_\alpha \,  Y^d_{\alpha \beta} \,  \Phi \,  D_\beta - \bar{Q}_\alpha \,  Y^u_{\alpha\, \beta}  \, \tilde{\Phi} \,  U_\beta + \, \text{h.c.} \, ,
\label{Lagrangiano_yukawa}
\ee 
where $Y_{\alpha \beta}$ are the ($3\times 3$) Yukawa matrices.

After the SSB, the Yukawa Lagrangian takes the form:
\be
\mathcal{L}_Y =  -\dfrac{ H + v }{\sqrt{2}} \left( \bar{e}_{L\alpha} \, Y^L_{\alpha \beta} \,  e_{R\beta} + \bar{u}_{L\alpha} \, Y^u_{\alpha \beta} \,  u_{R\beta} + \bar{d}_{L\alpha} \, Y^d_{\alpha \beta} \,  d_{R\beta} + \, \text{h.c.} \right) \, .
\label{Lagrangiano_yukawa_after_SSB}
\ee 
The Yukawa matrices are not diagonal in general. In order to obtain the mass and interaction eigenstates it is necessary to diagonalise them. First, a redefinition of the fields is needed. For instance, for the leptonic left-handed doublet, $L \rightarrow \mathcal{U}_L \, L$. At the end, we need five matrices belonging to global $SU(3)$ flavour group, one per field in Tab.~\ref{cargas_particulas}: $\left(\mathcal{U}_L, \mathcal{U}_Q, \mathcal{U}_e, \mathcal{U}_u, \mathcal{U}_d\right)$. Now, we can fix these matrices in order to diagonalise the different Yukawa matrices. For the charged leptons it is an easy task: $\mathcal{M}^L \equiv \mathcal{U}_L^\dagger Y^L \mathcal{U}_e$, we can always find two $SU(3)$ matrices that diagonalise the Yukawa matrix $Y^L$.

The case of the quarks is more complicate. We can find two matrices to diagonalise $Y^u$ matrix $\mathcal{M}^u \equiv \mathcal{U}_Q^\dagger Y^u \mathcal{U}_u$. The problem appears when we try to do the same for the $Y^d$ matrix, $\tilde{\mathcal{M}}^d \equiv \mathcal{U}_Q^\dagger Y^d \mathcal{U}_d$: the $\mathcal{U}_Q$ matrix has already been fixed to diagonalise $Y^u$ and, as a consequence, it is impossible to diagonalise simultaneously the two Yukawa matrices of the quarks. However, we can chose $\mathcal{U}_d$ in order to get $\tilde{\mathcal{M}}_d = \tilde{\mathcal{M}}_d^\dagger$. 

\subsection{Cabibbo-Kobayashi-Maskawa (CKM) matrix}

While the zero mass of the neutrinos always allows to transform the leptonic fields to obtain a diagonal basis, in the quark sector up and down types fields are massive. As a consequence, it is impossible to diagonalise both kind of fields simultaneously keeping the Lagrangian invariant. However, it is possible to introduce a rotation matrix to obtain $Y^d$ diagonal keeping the gauge invariance. After the diagonalization of $Y^L$ and $Y^u$, Eq.~\ref{Lagrangiano_yukawa_after_SSB} can be written as
\be
\mathcal{L}_Y =  -\left( 1 + \dfrac{H}{v} \right)  \left( \bar{e}_{L\alpha} \, \mathcal{M}^L_{\alpha \beta} \,  e_{R\beta} + \bar{u}_{L\alpha} \, \mathcal{M}^u_{\alpha \beta} \,  u_{R\beta} + \bar{d}_{L\alpha} \, \tilde{\mathcal{M}}^d_{\alpha \beta} \,  d_{R\beta} + \, \text{h.c.} \right) \, ,
\label{Lagrangiano_yukawa_diagonal}
\ee 
where it has been reabsorbed a factor $v/\sqrt{2}$ in the definition of the matrices. The diagonal matrices are $\mathcal{M}^u = \text{diag}(m_u, m_c, m_t)$ and $\mathcal{M}^L = \text{diag}(m_e, m_\mu, m_\tau)$, while $\tilde{\mathcal{M}}^d$ is an hermitian matrix. One can always find a $V$ matrix that diagonalise $\tilde{\mathcal{M}}^d$. If we assume the presence of this $V$ matrix 
we can write $\tilde{\mathcal{M}}^d = V \mathcal{M}^d V^\dagger$, where $\mathcal{M}^d = \text{diag}(m_d, m_s, m_b)$. In order to keep the invariance of the Lagrangian it is necessary to rotate the $d$-type fields 
\bea
\left \{
\begin{array}{lll}
d_R &\rightarrow & V \, d_R \, , \\
d_L &\rightarrow & V \, d_L \, .
\end{array}
\right .
\eea
With this new rotation, the $d$-type fields are diagonals at the same time that we keep diagonal the other two Yukawa matrices. The question now is, what implications have this rotation? 

The interaction between $W^\pm$, $Z$ and $A$ gauge bosons and the matter fields are determined by the Neutral Currents (NC) and Charged Currents (CC) Lagrangians. The rotations of the $d-$type fields imply that the rotation matrix $V$ appears in the CC Lagrangian. The final interaction is given by
\be
\mathcal{L}_{CC} = \dfrac{g}{\sqrt{2}} W^+_\mu \left( \bar{u}_{L \alpha} \, \gamma^\mu \, V_{\alpha  \beta} \, d_{L \beta} + \bar{\nu}_{L \alpha} \, \gamma^\mu \, e_{L \alpha}  \right) + \, \text{h.c.} \, .
\ee
The rotation matrix $V_{\alpha \beta}$ receives the name of Cabibbo-Kobayashi-Maskawa (CKM) matrix \cite{Cabibbo:1963yz, Kobayashi:1973fv} and it is the source of the flavour changing in CC processes.

The NC Lagrangian does not mix the quark flavour, it can be written in compact form
\be
\mathcal{L}_{NC} = - \, e \, A_\mu \sum_f \, Q^f \, \bar{f} \, \gamma^\mu \, f \, - \,  \dfrac{e}{\sin (\theta_W) \cos (\theta_W)} \, Z_\mu \, \sum_f \, \bar{f} \, (v_f - a_f \, \gamma_5) \, f \, ,
\ee
where $e$ is the electron charge and the sum is over all the physical fermions. The different $a_f$ and $v_f$ constants depend on each fermion:
\bea
\left \{
\begin{array}{lll}
a_f &=& \dfrac{1}{2} T^f_3 \, , \\[8pt]
v_f &=& \dfrac{1}{2} T^f_3 \left[ 1 - 4 |Q^f| \sin^2 (\theta_W)  \right] \, ,
\end{array}
\right .
\label{masas_gauge}
\eea
where $T_3^f$ and $Q^f$ are the isospin and the electrical charge of the fermion $f$. Notice that in this case, rotating $d \rightarrow V d$ does not introduces any matrix in NC processes, as they cancel in terms such as $d\, \Gamma d$ (being $\Gamma$ some combination of Dirac matrices).

\section{Open Problems of the Standard Model}
\label{sec:openproblems}

The Standard Model represents one of the most relevant achievements in physics. It took many years to understand how the microscopic world works. But this is not the end of the story as the SM presents various problems that until the present day have no solution. Examples of open problems in particle physics are the neutrino masses, the hierarchy problem or the strong CP problem. In the rest of this Section we described briefly these topics.

In addition to these problems, there are strong cosmological evidences of the existence of a new kind of matter that does not have the same interaction rules that the particles of the SM. This new kind of matter receives the name of \textit{Dark Matter} (DM). Its phenomenology is still a mystery and it is the main topic of this thesis. The nature and properties of DM will be studied in Chapter \ref{sec:DM}.

\subsection{The Hierarchy Problem}
\label{sec:hierarchyproblem}

There are some hints for the existence of physics beyond the SM (BSM). However, it is unclear at which scale this new physics enters the game. According to the Higgs Mechanism, the new physical scale must be close to the electroweak scale: the technical reason is that the Higgs mass is quadratically sensitive to high scales. If we analyse the Higgs potential (Eq.~\ref{Lagrangiano_Higgs}), the first order quantum corrections to the mass parameter $\mu_\Phi$ are given by
\be
\delta \mu_\Phi^2 = \dfrac{\Lambda^2}{32 \, \pi^2} \left[ -6\,Y_t^2 + \dfrac{1}{4}\,(9\,g^2 + 3\,g'^2) + 6\lambda_4 \right] \, ,
\label{Higgs_mass_correction}
\ee
where $\Lambda$ is the cutoff of the theory, $Y_t$ the top Yukawa coupling and $g$ and $g'$ the electroweak couplings. Since we know the value of the VEV of the Higgs field (Eq.~\ref{vev_value}) and the Higgs mass, we can calculate the value of $\lambda_4 = 0.13 $ using Eq.~\ref{higgs_mass}. If the scale of the new physics is close to the EW scale, the hierarchy problem is not a real problem. However, the landscape of the current experiments in high-energy physics makes us think that these scales are not as close as it should. If the new scale is $\Lambda \gg 1 \, \text{TeV}$ the Eq.~\ref{Higgs_mass_correction} implies $\delta \mu_\Phi^2 \gg \mu_\Phi^2$. If quantum corrections are much larger than the experimentally measured value of $\mu_\Phi$, then extremely large, cancellations should be at work. This fact is known as hierarchy problem, a complete revision about this topic can be found in Ref.~\cite{Csaki:2018muy}.
Different models to try to solve the hierarchy problem have been proposed in the last decades. The most popular are Supersymmetry (a review can be found in Ref.~\cite{Haber:2017aci}), technicolor \cite{1976PhRvD..13..974W,1979PhRvD..19.1277W,1979PhRvD..20.2619S}, composite Higgs \cite{Kaplan:1983fs} and warped extra-dimensions \cite{Randall:1999ee,Randall:1999vf}.

\subsection{Strong CP problem}
\label{sec:strongcpproblem}

Quantum Chromodynamics predicts the existence of processes with CP violation. However, no violation of the CP-symmetry is observed experimentally. There are no theoretical reasons to preserve this symmetry and, as a consequence, this represents a  fine tuning problem.

The absence of any observed violation in strong interactions is a problem because the QCD Lagrangian presents natural terms that break the CP simmetry \cite{Peccei:1977hh}:
\be
\mathcal{L} \supset - \dfrac{n_f g_s^2 \theta_{CP}}{32 \pi^2} \, G_{a\mu \nu} \widetilde{G}^{a\mu \nu}  \, ,
\ee
where $\theta_{CP}$ is the vacuum phase and $n_f$ the number of flavours. This term comes directly from the vacuum QCD structure and it would be absent in presence of massless quarks. The phase is related to the value of the neutral dipole moment \cite{Crewther:1979pi}, whose current limits \cite{Pospelov_1999,Guo_2015} implies that $|\theta_{CP}| < 10^{-10}$.

Different solutions have been proposed to solve the problem, the most popular among them being the one proposed by Roberto D. Peccei and Helen R. Quinn in Ref.~\cite{PhysRevLett.38.1440}, introducing a new symmetry $U(1)_{\text{PC}}$. This new symmetry is spontaneously broken generating the Weinberg-Wilczek axion \cite{Wilczek:1977pj,Weinberg:1977ma} (the Goldstone boson of the broken PQ symmetry). In this scenario, the $\theta$-phase is related with the VEV of a new field and its small value is the consequence of the symmetry breaking at high scales. The value of $\theta$ in this approach is determined by irrelevant operators \cite{Nelson:1983zb,Barr:1984qx}. 

Different reviews about the strong CP problem and its possible solutions can be found in Refs.~\cite{Hook:2018dlk,Kaloper:2017fsa}.

\subsection{Neutrino Masses: The Seesaw Mechanism}
\label{sec:neutrinomasses}

In the Standard Model the neutrinos are massless, but nowadays it is experimentally shown that they have a non-zero mass. In 1957 Bruno Pontecorvo predicted the existence of neutrino oscillations \cite{Pontecorvo:1957cp}, as a consequence of the difference between the interaction (weak) and mass eigenstates. This effect implies non-zero mass for the neutrinos and ever since Pontecorvo predicted its existence, several experiments searched for it and studied their effect \cite{Hirata:1990xa,Hirata:1992ku,Fukuda:1998mi,Cleveland:1998nv,Hampel:1998xg,Abdurashitov:1999zd,Fukuda:2001nj,Ahmad:2002jz,Altmann:2005ix,Ahn:2006zza,Michael:2006rx,Abe:2008aa,Abe:2011sj,Abe:2011fz,An:2012eh,Ahn:2012nd,Abe:2014ugx}.

There are different mechanisms to generate neutrino masses, for instance add a new right-handed neutrino ($N$). However, when we add a mass term for the neutrinos with a new state $N$, singlet under the SM symmetry group, we have the same problem as for the quarks: we need a new matrix to diagonalize charged leptons and neutrino mass matrices simultaneously. The relation between the mass and weak eigenstates can be fixed using the unitary Pontecorvo-Maki-Nakagawa-Sakata (PMNS) matrix $U_{\alpha i}$ \cite{Pontecorvo:1957qd,Maki:1962mu}:
\be
\nu_\alpha = \sum_i U_{\alpha i} \nu_i \, ,
\ee
where $\alpha = e, \mu, \tau$ represent the weak eigenstates while $i = 1, 2, 3$ are the mass eigenstates. The PMNS matrix can be parametrized with three mixing angles $(\theta_{12},\theta_{23},\theta_{13})$ and a CP-violating Dirac phase $\delta$. The mixing angles are usually refereed to as solar, atmospheric and reactor angles, respectively, because at the kind of experiment where they were measured for the first time. On the other hand, the oscillation lengths are $(\Delta m_{12}, \Delta m_{23}, \Delta m_{13})$. The most recent values of all of these parameters can be found in Ref.~\cite{Esteban:2016qun}.

Until this point we only talked about the mixing and oscillation parameters; but what is the mass scale of the neutrinos? The KATRIN experiment puts the upper bound at $\sum m_\nu \lesssim 2.7 \, \text{eV}$ at $95\%$ Ref.~\cite{Aker:2019uuj}. Extending the SM to add neutrino masses is not a complicated task: it is enough to introduce three new fields $N$ that represent the right-handed neutrinos\footnote{Actually, two new fields are sufficient to explain present observations; albeit, with the consequence that the lightest neutrino should be massless.}. In that case, an extra term would appear in Eq.~\ref{Lagrangiano_yukawa} giving mass to the neutrinos via the Higgs Mechanism, as in the case of quarks and charged leptons. The question now is, if the mechanism to give mass to the neutrinos is the Higgs Mechanism, why the neutrinos masses are so different from the rest of the fundamental particle masses?

The right-handed neutrinos $N$ have a special property that make them different from the rest of the SM particles: they are singlets under all SM gauge groups. This allows the neutrino to be its own antiparticle! While the usual fermions receive the name of Dirac particles, this kind of particles receives the name of Majorana particles \cite{Majorana:1937vz}. This means that we could add a new extra term to the Lagrangian:
\be
\mathcal{L} = - \bar{L}_i \, Y^\nu_{ij} \Phi N_j - \dfrac{1}{2} N^T_i C^{-1} M^R_{ij} N_j + \, \text{h.c.} \, ,
\ee
where $M^R$ is the $3 \times 3$ right-handed neutrino Majorana mass matrix, $Y^\nu$ the Yukawa matrix of the neutrinos and $C$ the charge conjugate operator. The \textit{usual} mass of the neutrinos, the so-called Dirac mass, is given by the Yukawa couplings
\be
m_D =\dfrac{v}{\sqrt{2}} \, Y^\nu \, .
\ee
It is important to keep in mind that we have three flavour families and, as a consequence, $m_D$ is not a parameter, but a $3 \times 3$ matrix.

After SSB, the mass matrix of the neutrinos takes the form:
\be
\mathcal{M}_\nu = \left( \begin{array}{cc}
0 & m_D  \\
m_D^T & M^R \end{array} \right) \, ,
\ee
As an example, consider the $2\times 2$ case (i.e. for one generator only).
In the limit $|m_D| \gg |M^R|$ there is a large hierarchy between the eigenvalues, that are given by $m_\nu \backsimeq m_D^2/M^R$ and $m_N \backsimeq M^R$, and approximately corresponds to the eigenvectors of $\nu_1 \backsim \nu_L$ and $\nu_2 \backsim N$. In this way, we could have a \textit{natural} explanation why neutrinos are much lighter than other fermions, even if their Yukawa couplings (and, thus, $m_D$) are similar. This mechanism receives the name of \textit{seesaw mechanism} type I \cite{GellMann:1980vs,PhysRevLett.44.912,Yanagida:1979as,Glashow1111,MINKOWSKI1977421}, as the larger $M_N$ the smaller $m_\nu$. This is only one of the different seesaw mechanisms able to provide mass to the neutrinos. However, other variants of this mechanism do not need the existence of the right-handed neutrinos.

Complete reviews of different seesaw models can be fond in Ref.~\cite{Grossman:2003eb}.


\chapter{Introduction to Cosmology: The Homogeneous Universe}
\label{sec:Cosmo}

\section{An Expanding Universe: The FLRW Metric}

Developing a theory related to the matter of the Universe, regardless of what type the matter is, implies a deep knowledge of the shape of the Universe on large scales. The science that is investigating this is called \textit{cosmology}. Although the word cosmology was used for the first time in 1656 in Thomas Blount's \textit{Glossographia} \cite{blount1681glossographia}, its origins began long ago.
Already the ancient Greeks tried to explain the position and nature of the astronomical objects they observed. At that time notable authors such as Aristoteles and Claudius Ptolemy developed the geocentric model, which placed the Earth as the center of the Universe. Many centuries later, Nicol\'as Copernicus (1473-1543) developed the heliocentric model, which was strongly supported by Galileo Galilei (1564-1642), laying the first foundations for our current astronomical models. However, the modern cosmology was born during the first half of the 20th century with the discovery of the expansion of the Universe. In 1929 Edwin Hubble found the first evidence of the expansion of the Universe \cite{Hubble168}. He observed that all distant galaxies and astronomical objects were moving away from us as
\be
z = \dfrac{ \lambda_{\text{observed}}-\lambda_{\text{emitted}} }{\lambda_{\text{emited}}} \backsimeq H_0 \, d_L \, .
\ee
This expression is called the \textit{Hubble Law} and establishes a relationship between the \textit{luminosity distance}\footnote{Defined as $d_L = 10^{(m-M)/5 + 1}$, where $M$ is the absolute magnitude while $m$ the apparent magnitude of an astronomical object. The luminosity distance is usually measured in parsecs (pc).} $d_L$ of some astronomical object with its \textit{redshift}\footnote{The redshift $z$ is the difference between the observed and the emitted wavelength of the astronomical body.} $z$. At first order, the relation is linear and only depends on the Hubble constant \cite{Aghanim:2018eyx}
\be
H_0 = 100 \, h \,  \text{km}\, \text{s}^{-1}\, \text{Mpc}^{-1} = 67.66 \pm 0.42 \, \text{km} \, \text{s}^{-1} \, \text{Mpc}^{-1} \, ,
\label{Hubble_cte}
\ee
where $h$ is the reduced Hubble constant.

\begin{figure}[htbp]
\centering
\includegraphics[width=120mm]{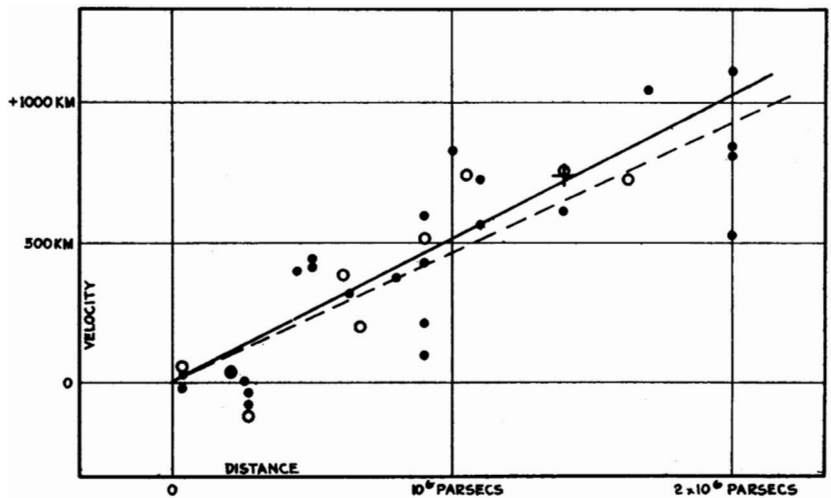}
\caption[Results obtained by Edwin Hubble comparing the measurements about the radial velocity of the galaxy with the red shift of 22 different astronomical clusters]{\it Original figure of \cite{Hubble168} showing the results obtained by Edwin Hubble comparing the measurements about the radial velocity of the galaxy with the redshift of 22 different astronomical clusters. The results shown in this figure represent the first proof of the expansion of the Universe.}
 \label{fig:hubble}
\end{figure}
The results obtained by Hubble are shown in Fig.~\ref{fig:hubble}. The original results of Hubble's work analysed 22 different galaxies. Nowadays, we have data of thousands of galaxies and the most part of these shows $z>0$. This fact is considered an irrefutable proof of the expansion of the Universe. The expansion of the Universe and the assumption that we live in an isotropic and homogeneous Universe\footnote{This is called \textit{cosmological principle} and is backed up by strong evidences \cite{Aghanim:2018eyx,Akrami:2018vks}.} lead us to the Big Bang model. 

Nowadays, our understanding of the evolution of the Universe is based on the Friedman-Lema\^itre-Robertson-Walker (FLRW) cosmological model, that describes an isotropic, homogeneous and expanding Universe \cite{Friedman:1922kd, 10.1112/plms/s2-42.1.90,10.1093/mnras/91.5.483, Robertson:1935zz} with metric
\be
\label{metric_FLRW}
ds^2 = g_{\mu \nu}dx^\mu dx^\nu = dt^2 - a(t)^2 \left[ \dfrac{dr^2}{1-kr^2} + r^2 d\theta^2 + r^2 \sin^2 \theta d\phi \right] \, ,
\ee
where $(t,r,\theta, \phi)$ are the comoving coordinates and $a(t)$ is the cosmic scale factor. The curvature of the space-time is given by $k$ and can be $+1$, $-1$ or $0$ describing an open, close and flat space-time, respectively.

In order to quantify the expansion of the Universe it is necessary to study the variation of the scale factor $a(t)$. The most convenient way to perform this study is to analyse the so-called expansion rate or Hubble parameter, defined as
\be
H \equiv \dfrac{\dot{a}}{a} \, ,
\label{Hubble_parameter}
\ee
where $\dot{a} = da/dt$. The current value of the Hubble parameter is the Hubble constant, $H_0$, defined in Eq.~\ref{Hubble_cte}.

\section{Einstein Field Equations}

To understand the evolution of the Universe through the FLRW metric a deep knowledge of General Relativity and the Einstein gravitational field equations is necessary. First proposed in 1915 by Albert Einstein in Ref.~\cite{Einstein:1916vd}, the gravitational field equations take the form:
\be
G_{\mu\nu} = \dfrac{8\, \pi \,G}{c^4} \, T_{\mu \nu} + \Lambda\, g_{\mu \nu} \, ,
\label{einstein_equation}
\ee
where $T_{\mu \nu}$ is known as the \textit{energy-momentum} tensor and represents the energy flux and momentum of a matter distribution, $\Lambda$ is the cosmological constant and $G_{\mu \nu}$ is the unique divergence free tensor which can be built with linear combinations of the space-time metric and its first and second derivatives
\be
G_{\mu \nu} \equiv R_{\mu \nu} - \dfrac{1}{2} g_{\mu \nu} R \, .
\label{tensor_einstein}
\ee

The Einstein field equations form a system of ten coupled differential equations and describe the evolution of the space-time metric tensor $g_{\mu \nu}$ under the influence of the $T_{\mu \nu}$ tensor, and vice-versa.
To understand Eq.~\ref{einstein_equation}, a deep knowledge of the different elements of the differential geometry is needed (see, for instance, Ref.~\cite{Weinberg:1972kfs}):
\bea
\left \{
\begin{array}{llll}
\Gamma^{\mu}_{\alpha \beta} &=& \dfrac{1}{2}g^{\mu \nu}\left( -\dfrac{\partial g_{\alpha \beta}}{\partial x^\nu} + \dfrac{\partial g_{\nu \alpha}}{\partial x^\beta} + \dfrac{\partial g_{\nu \beta}}{\partial x^\alpha} \right) \quad \quad &\text{Cristoffel Symbols} \, , \nonumber \\  [8pt]
	R^{\alpha}_{\beta \gamma \sigma} &=& \dfrac{\partial \Gamma^{\alpha}_{\beta \sigma}}{\partial x^{\gamma}}-\dfrac{\partial \Gamma^{\alpha}_{\gamma \sigma}}{\partial x^{\beta}}+ \Gamma^{\mu}_{\beta \sigma} \Gamma^{\alpha}_{\gamma \mu}- \Gamma^{\mu}_{\gamma \sigma} \Gamma^{\alpha}_{\beta \mu} \quad \quad &\text{Riemann Tensor} \, , \\[8pt]
	R_{\sigma \nu} &=& R^{\rho}_{\ \sigma \rho \nu} \quad \quad &\text{Ricci Tensor} \, , \\[8pt]
	\mathcal{R} &=& R^{\mu}_{\mu} \quad \quad &\text{Scalar Curvature} \, .
\label{cosas_varias_einstein}
\end{array}
\right . \\
\eea
In General Relativity $g_{\mu \nu}$ plays a fundamental role: each solution of the Einstein field equations is characterized by its respective metric, which is defined by the energy density of the Universe.
The existence of the last term of Eq.~\ref{einstein_equation} has been a topic of debate since Einstein postulated it to give a solution of his equations that predicted a static Universe. In the original formulation of the FLRW model, $\Lambda$ is supposed to be absent (to get a constant expansion of the Universe). Current cosmology rescued it as a possible explanation of the observed accelerated expansion of the Universe at recent times \cite{Peebles:2002gy}.

\section{Dynamics of the Universe}

The structure of the Universe is fixed by Eq.~\ref{einstein_equation}. In order to solve this equation it is necessary to know the form of the energy-momentum tensor $T_{\mu \nu}$. Under the assumption of homogeneity and isotropy, the content of the primordial Universe can be described as a perfect fluid, and the energy-momentum tensor can be written as:
\be
T^{\mu}_{\, \, \, \nu} = p \, g^{\mu}_{\, \, \, \nu} + (\rho + p) \, U_\nu U^\mu \equiv \text{diag}(\rho,-p,-p,-p) \, ,
\label{energy_momentum_tensor}
\ee
where $U_\mu \equiv dX^\mu/d\tau$ is the four-velocity of the fluid, $\rho$ the energy density and $p$ the pressure. The energy-momentum conservation principle $dU = -p \, dV$, where $U = \rho \, V$ is the total energy of the fluid and $V \propto a^3$ the volume, directly implies 
\be
\label{ecuacion_2}
\dfrac{d \rho}{d p} + 3 \dfrac{\dot{a}}{a}(\rho + p) = 0 \, .
\ee
Eq.~\ref{ecuacion_2} allows to obtain the relation between the energy density $\rho$ and the scale factor $a$ when the relation between the energy density and the pressure\footnote{The relation between the pressure and the energy density is called \textit{equation of state}.} $p$ is known. Most cosmological fluids can be described by a simple time-independent equation of state, where the energy and the pressure are proportional, $p = \omega \rho$, being $\omega$ an arbitrary constant. In these cases the energy density can be expressed as $\rho \propto a^{-3(1+\omega)}$. 

In order to describe the evolution of the Universe, it is necessary to understand the different components that contribute to the energy-momentum tensor. It is possible to distinguish three different components of the content of the Universe: \textit{matter}, \textit{radiation} and \textit{Dark Energy}. The nature of the first two components is easy to explain: the cosmological definition of matter says that it includes all the different non-relativistic matter species, while radiation includes the relativistic particles.

The third component, the Dark Energy, is a kind of unexplained energy with negative pressure that is necessary to understand our current knowledge about the evolution of the Universe.
Quantum field theory predicts the existence of a vacuum energy with negative pressure \cite{Zeldovich:1968ehl}. This energy can be calculated for the energy-momentum tensor as $T_{\mu \nu}^{\text{vac}} = \rho_{\text{vac}} g_{\mu \nu}$. The problem with this explanation of the Dark Energy nature lies in the fact that there is a large discrepancy between the observed and the calculated energy density
\be
\dfrac{\rho_{\text{vac}}}{\rho_{\text{obs}}} \backsim 10^{120} \, .
\ee
This huge discrepancy receives the name of \textit{vacuum catastrophe}\footnote{Also called sometimes \textit{cosmological constant problem}.}. The nature of this component of the Universe is still unclear. The scientific community agrees that it could be related to the cosmological constant, but alternative ideas could also work. A detailed description of the current status of the problem can be found in Ref.~\cite{Martin:2012bt}.

It is possible to distinguish between three different epochs in the evolution of the Universe, depending on whether matter, radiation, or Dark Energy dominates. 
\bea
\left \{
\begin{array}{lll}
     p = \dfrac{1}{3}\rho \, \,  &\Rightarrow  \, \,  \rho \propto a^{-4}  \quad \quad &\text{Radiation Epoch} \, , \\  
	 p = 0  \, \,  &\Rightarrow  \, \,  \rho \propto a^{-3} \quad \quad &\text{Matter Epoch} \, , \\
	 p = -\rho \, \,  &\Rightarrow  \, \,  \rho \propto \text{const.} \quad \quad &\text{Dark Energy Epoch} \, . \\  
\end{array}
\right .
\eea

\begin{figure}[htbp]
\centering
\includegraphics[width=120mm]{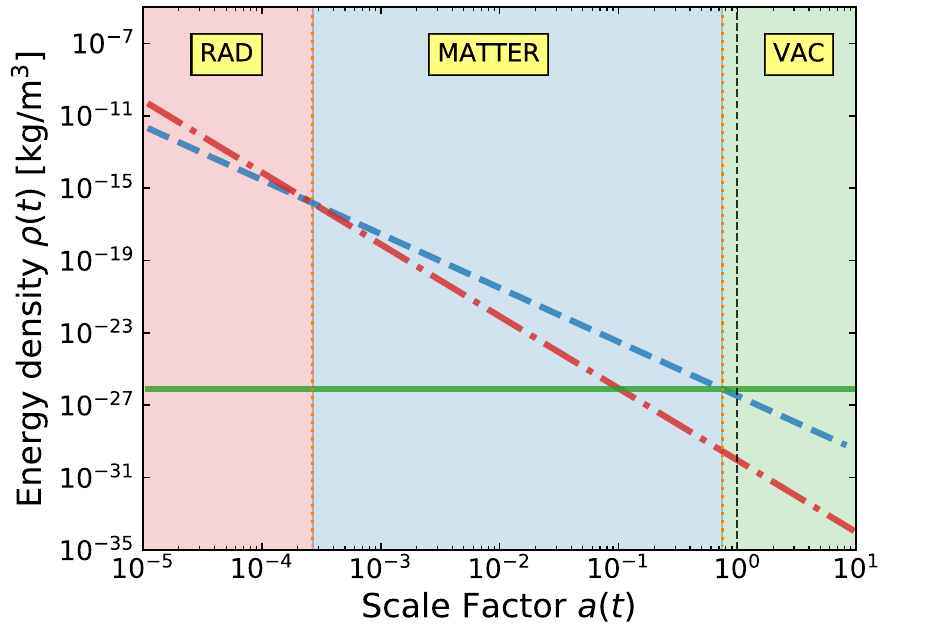}
\caption[Thermal history of the Universe]{\it Different epochs of the Universe, depending on which contribution dominates. The red dot-dashed, blue dashed and green solid lines represent the different contributions of radiation, matter and Dark Energy (or vacuum energy), respectively. The black dashed vertical line shows the present moment of the Universe.}
 \label{fig:epocas}
\end{figure}

Fig.~\ref{fig:epocas} shows the different contributions to the total energy density of the different components of the Universe. The red dot-dashed, blue dashed and green solid lines show, respectively, the radiation, matter and Dark Energy contributions. In the early Universe, most parts of the components were relativistic; this era is dominated by the radiation contribution. In the \textit{adolescent} Universe the SM particles, except photons and neutrinos, are non-relativistic, the matter contribution dominates the total energy density. The present moment of the Universe (black-dashed line in Fig.~\ref{fig:epocas}) is close to the point at which the vacuum contribution begins to dominate over the matter contribution ($z \backsimeq 0.55$). This fact receives the name of \textit{coincidence problem} \cite{Velten:2014nra} and its possible anthropic implications have been studied by different authors (see, for instance, Refs.~\cite{Barreira:2011qi,Fedrow:2013rwa}).

\section{The Friedman Equations}

To analyse the evolution of the scale factor it is necessary to simplify the different terms of Eq.~\ref{einstein_equation} using the definitions of Eq.~\ref{cosas_varias_einstein} with the metric of Eq.~\ref{metric_FLRW} and the form of the energy-momentum tensor that, under the assumption of homogeneity and isotropy, takes the form of Eq.~\ref{energy_momentum_tensor}. The resulting expressions receive the name of Friedman equations
\bea
\label{FriedmannEquations}
\left \{
\begin{array}{lll}
      \left( \dfrac{\dot{a}}{a} \right)^2 &=& \dfrac{8\pi G}{3} \rho - \dfrac{k}{a^2} \, , \\[8pt]
	  \left( \dfrac{\ddot{a}}{a} \right) &=& -\dfrac{4 \pi G}{3} (\rho + 3 p)  \, ,
\end{array}
\right .
\eea
where $\rho$ and $p$ can be understood as the sum of all contributions to the energy density and pressure in the Universe. The first Friedman equation is usually written in terms of the Hubble parameter (Eq.~\ref{Hubble_parameter})
\be
H^2 = \dfrac{8 \pi G}{3} \rho - \dfrac{k}{a^2} \, .
\label{first_FE_guay}
\ee

The flat space case ($k=0$) in Eq.~\ref{first_FE_guay} defines the critical case
\be
\rho_{\text{crit}} \equiv \dfrac{3H^2}{8 \pi G} \, ,
\ee    
that can be estimated today using Eq.~\ref{Hubble_cte} obtaining $\rho_{\text{crit}, 0} \equiv \left. \rho_{\text{crit}} \right\rvert_{H = H_0} = 1.9 \times 10^{-29} \, h^2 \, \text{g} \, \text{cm}^{-3}$. The critical density is used to define dimensionless density parameters 
\be
\Omega \equiv \dfrac{\rho}{\rho_{\text{crit}}} \, .
\ee
This is very convenient because the energy densities of the different components of the Universe have enormous values.
Since the density parameter $\Omega$ is related with $k$, which describes the curvature of space-time, its value allows to analyse the geometry of the Universe
\bea
\left \{
\begin{array}{lll}
     \Omega > 1 \quad \quad &\text{Closed} \, , \\  
	\Omega = 0 \quad \quad &\text{Flat} \, , \\
	\Omega < 1 \quad \quad &\text{Open} \, . \\  
\end{array}
\right .
\eea

The first Friedmann equation (Eq.~\ref{first_FE_guay}) can be written in terms of $\Omega$ and the Hubble constant
\be
H^2 = H_0^2 \left[ \Omega_\text{r} \, \left( \dfrac{a_0}{a} \right)^4 + \Omega_\text{m} \, \left( \dfrac{a_0}{a} \right)^3 + \Omega_k \,  \left(\dfrac{a_0}{a} \right)^2 + \Omega_\Lambda   \right] \, ,
\label{expresion_de_H_cuadrado}
\ee
where $\Omega_\text{r}$, $\Omega_\text{m}$, $\Omega_k$ and $\Omega_\Lambda$ denotes the density parameters of radiation, matter, curvature and vacuum in the present epoch, respectively. In Eq.~\ref{expresion_de_H_cuadrado} we define the curvature density parameter in the present epoch as $\Omega_k \equiv -k/(a_0 H_0)$. The expression is written in terms of $H_0$ and $a_0$, where $a_0$ represents the scale factor today. It is very common in cosmology to take the normalization for the scale factor $a_0 \equiv 1$. With this normalization, the above expression becomes
\be
H^2 = H_0^2 \left( \Omega_\text{r} \, a^{-4} + \Omega_\text{m} \, a^{-3} + \Omega_k \, a^{-2} + \Omega_\Lambda  \right) \, .
\ee
The question now is, what is the value of these parameters?

\section{Cosmology in the present days: $\Lambda$CDM model}
\label{Cosmology_in_the_present_days}

\begin{figure}
\centering
\includegraphics[width=130mm]{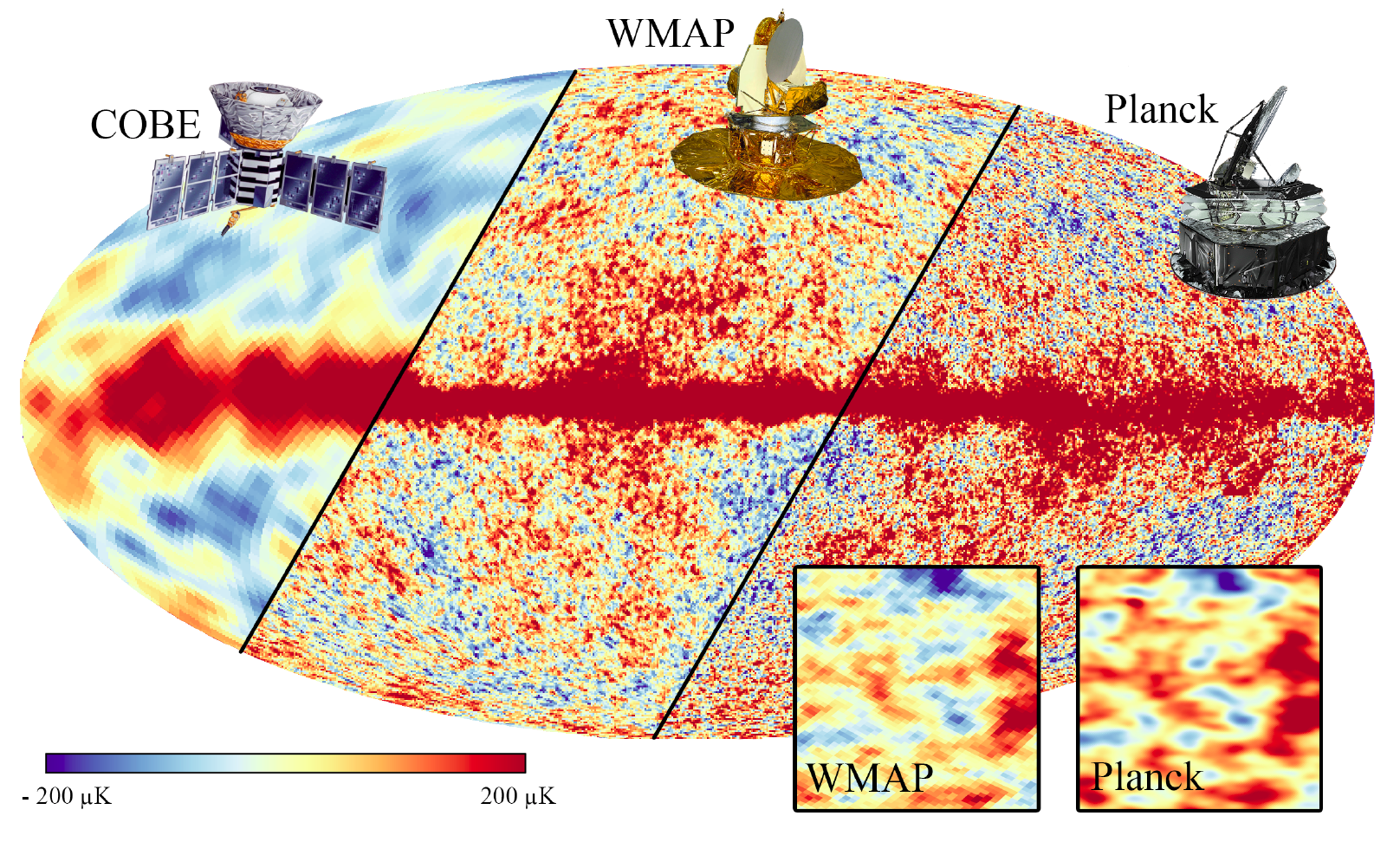}
\caption[Cosmic microwave background (CMB)]{\it Evolution along the last 30 years of the measurements of the anisotropies in the temperature of the Cosmic Microwave Background. From the left to the right the figure shows the data from COBE (Cosmic Background Explorer) \cite{Bennett:1996ce},  WMAP (Wilkinson Microwave Anisotropy Probe) \cite{Hinshaw_2013} and Planck \cite{Aghanim:2018eyx}. Image taken from \cite{Vogelsberger_ISAPP_2017}.}
 \label{fig:CMB}
\end{figure}

In 1964 Arno Penzias and Robert Woodrow Wilson \cite{Penzias:1965wn} discovered the \textit{Cosmic Microwave Background} (CMB), a noise, apparently isotropic, in the form of electromagnetic radiation that populates the Universe. Since then, several experiments measured the CMB finding small temperature anisotropies (the evolution of our knowledge about the CMB can be observed in Fig.~\ref{fig:CMB}) such as the case of COBE (Cosmic Background Explorer) \cite{Bennett:1996ce},  WMAP (Wilkinson Microwave Anisotropy Probe) \cite{Hinshaw_2013} or Planck \cite{Aghanim:2018eyx}, the latter being the most accurate measurement today. The CMB discovery confirmed a key prediction of the Big Bang cosmology. Since that moment, the scientific community accepted that the Universe started in a hot and dense state and has been expanding ever since.

The current cosmological model includes a non-vanishing cosmological constant $\Lambda$, that represent the Dark Energy or vacuum component of the Universe. As for matter, it assumes that most part of the matter is non-barionic and is mostly composed of \textit{Cold Dark matter}\footnote{See Sect.~\ref{sec:Hotwarmcold} for more details.} (CDM). The evidence of this fact will be commented in Chapter \ref{sec:DM}. Respect to the curvature, the model assumes that the Universe is practically flat at large scale. The name of this model that accepts the existence of two new, and unexplained, components of the energy density receives the name of $\Lambda$CDM model.

\begin{table}
\centering
\begin{tabular}{c c}
    \hline
    \multicolumn{2}{c}{\bfseries Cosmological Parameters Planck 2018} \\ 
    \hline
     Expansion & $h = 0.677 \pm 0.004 $\\
     Barionic Matter & $\Omega_\text{b} h^2 = 0.02242 \pm 0.00014$ \\
     Dark Matter & $\Omega_\text{DM} h^2 = 0.1193 \pm 0.0009$ \\
     Dark Energy & $\Omega_\Lambda h^2 = 0.689 \pm 0.006$ \\
     Radiation & $\Omega_\text{r} h^2 = (9.2 \pm 0.4 ) \times 10^{-5}$ \\
     Curvature & $\Omega_k h^2 = -0.004 \pm 0.015$ \\
    \bottomrule
\end{tabular}
\caption[Cosmological parameters]{Cosmological parameters published by Planck \cite{Aghanim:2018eyx}. These values represent the conclusion of the Experiment.}
\label{tabla_cosmo_param}
\end{table}

The $\Lambda$CDM model is a parametrization of the cosmological measurements. The accuracy of the model depends on the precision of the astrophysical experiments that estimate its parameters. 
Tab.~\ref{tabla_cosmo_param} shows the most recent measurements taken by the Planck collaboration of the cosmological parameters. These results show that the most part of the Universe being Dark Energy ($\backsim$ 69\%) and Cold Dark Matter ($\backsim$ 26\%) while the baryonic matter only represents $\backsim$ 5\% of the total energy content. The Dark Energy is still a complete mystery today: the most accepted theory is that is related to the cosmological constant of the Einstein field equations. On the other hand, what is this Dark Matter? This $\backsim$ 26\% of the content of the Universe is the main topic of this Thesis.
\chapter{About the Nature of Dark Matter}
\label{sec:DM}

As it was explained in Sect.~\ref{Cosmology_in_the_present_days}, there are unequivocal evidences that point out that the \textit{baryonic} matter (where baryonic in cosmology includes not only baryons, but also all of the SM particles) represents the $\backsim$ 5\% of the energy density of the Universe, while Dark Matter constitutes the $\backsim$ 26\%. The implication of this fact is absolutely strong: the SM of fundamental interactions described in Chapter \ref{sec:SM} only explains a minuscule portion of the matter of the Universe, the rest is still a mystery. Along this Thesis we try to bring some light over the DM enigma. In order to perform this task it is necessary to understand the nature of this new kind of matter. What are the evidences of DM? is it possible to observe these elusive particles? which is its the nature?

\section{Dark Matter Evidences}
\label{sec:evidences}

The first observational evidence of the existence of DM date from the early 1930's when Fritz Zwicky measured the velocity dispersion of several galaxies of the Coma Cluster. Zwicky concluded that a bigger amount of matter than the visible one was necessary to keep the galaxy cluster together\footnote{More precise estimations were made after the first Zwicky observation, using the virial theorem \cite{1976ApJ...204..668F}.} \cite{Zwicky:1933gu,1937ApJ....86..217Z}.   Previously to Zwicky, other observations suggesting missing mass in our galaxy were made by Jacobus Cornelius Kapteyn (1922) \cite{Kapteyn:1922zz} and by Jan Hendrik Oort (1932) \cite{1932BAN6249O}. 

\begin{figure}[htbp]
\centering
\includegraphics[width=100mm]{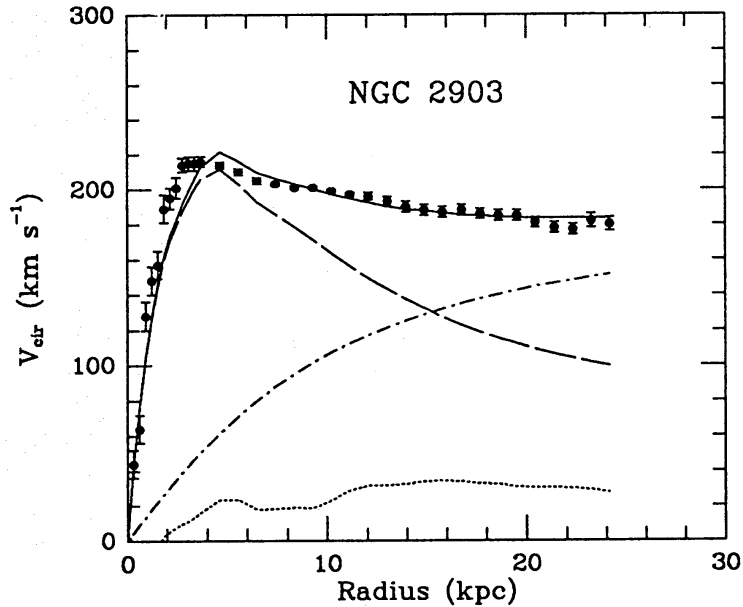}
\caption[Galaxy rotation curve of NGC 2903.]{\it Galaxy rotation curve of NGC 2903 \cite{Begeman:1991iy}. The solid line represents the data fit of the observed galaxy rotation curve while the dashed, dotted and dash-dotted represent, respectively, the rotation curves of the individual components: the visible components, the gas and the dark halo.}
 \label{fig:galaxyrotation}
\end{figure}

In the 1960's and 1970's the first astrophysical Dark Matter studies were made. Vera Cooper Rubin, Kent Ford and Ken Freeman measured the velocity rotation curve of different spiral galaxies \cite{1970ApJ...159..379R,1970ApJ...160..811F}. In these works they concluded that the velocity rotation curve of the spiral galaxies display an anomalous behaviour contrary to the galaxies luminosity measurements. According to our knowledge about the relation between the luminosity and the mass of the galaxy, if the only kind of matter in it is baryonic, the rotational velocity should follow the dash line in Fig.~\ref{fig:galaxyrotation}. Conversely, as we can understand from the data points in the Figure, the velocity remains almost constant. Since then, many measurements of the velocity rotation curves of several galaxies have been done (see, for instance, Refs.~\cite{Sofue:2000jx,10.1093/mnras/278.1.27}). Nowadays, there are strong evidences that the $95$\% of the matter content of almost every galaxies is DM.

Galaxy rotation curves were the first solid proof, and probably the most famous, of the DM existence, but are not the only one. Several evidences of the DM content in the Universe have been discovered since Rubin, Ford and Freeman researches, including the fact that the mass of the galaxy clusters is in agreement with the $\Lambda$CDM model, supporting the DM theories \cite{2011ARA&A..49..409A}.

One of the ways to estimate the mass of any astronomical body is the \textit{gravitational lensing}. This method uses light that arrives at the Earth emitted by galaxies, clusters, quasar and other astronomical objects. In most cases, these objects are not located close to the Earth, as a consequence, it is quite common the presence of some astronomical bodies along the emitted light path to the Earth. When the light goes through these astronomical objects, according to General Relativity, the gravitational field distorts its propagation. This distortion receives the name of gravitational lensing. The measurement of this effect allows the mass of galaxies, clusters and other astronomical bodies between us and the light source to be estimated. The gravitational lensing measurements of different astronomical objects point to DM predominance in almost every galaxies and clusters \cite{Taylor:1998uk,Wu:1998ju,Natarajan:2017sbo,Refregier:2003ct}.

\begin{figure}[htbp]
\centering
\includegraphics[width=100mm]{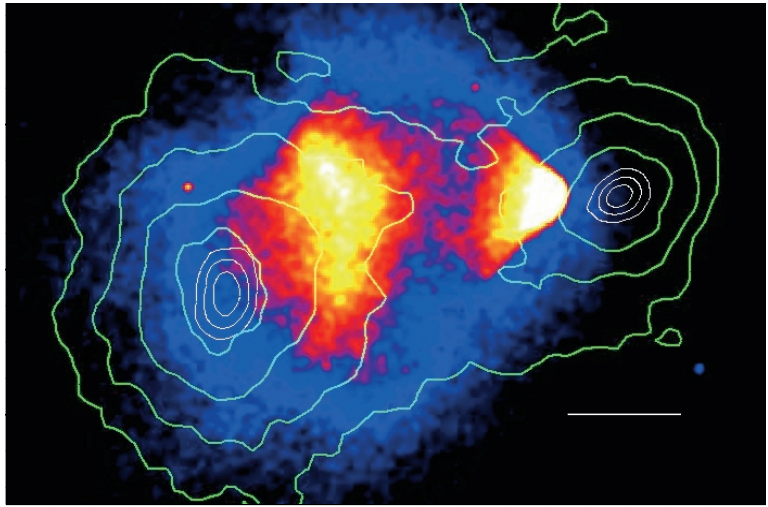}
\caption[Bullet Cluster]{\it Image taken from \cite{Clowe:2006eq}. It shows the gas distribution of the Bullet Cluster. Contour lines depict gravitational equipotential lines which indicate the DM location.}
 \label{fig:bullet_cluster}
\end{figure}

The \textit{Bullet Cluster} is probably the best example of how gravitational lensing proofs the existence of DM. The Bullet Cluster or 1E 0657-56\footnote{The Bullet Cluster is composed by two colliding clusters. Was discovered in 1995 by \textit{Chandra} X-ray \cite{Tucker:1998tp}.} has displaced its center of mass with respect to the observed baryonic center of mass. DM models can easily explain this effect. Other alternatives would require a modification of General Relativity \cite{Clowe:2003tk,Markevitch:2003at}. 

All the evidences illustrated above are astrophysical proofs, but there are several cosmological indications of the existence of DM in agreement with these evidences. The Friedmann equations and General Relativity describe a homogeneous Universe. As a consequence, the galaxies, stars and the rest of the astronomical bodies were originated by small density perturbations after the Big Bang. If had only existed baryonic matter in the early Universe, the presence of galaxies and clusters would not be possible today. In that hypothetical case, the evolution of the primordial density perturbations would not have been sufficient \cite{1993sfu..book.....P,Peebles:2017bzw,Frenk:2012ph}. 

On the other hand, the temperature anisotropies measured in the CMB by COBE, WMAP and Planck \cite{Bennett:1996ce,Hinshaw_2013,Akrami:2018vks} absolutely agree with a Universe made of $69\%$ Dark Energy and $31\%$ matter.

In summary, nowadays there are several irrefutable evidences of the existence of DM. It is true that no direct or indirect detection of DM has been until today, but the exceptional predictive power of the $\Lambda$CDM cosmological model represents an excellent proof that our Universe is mostly dark.

\section{Properties of Dark Matter}
\label{sec:Properties}

As it was explained in Sect.~\ref{sec:evidences}, there are several cosmological and astrophysical evidences of the existence of DM. It is true that, to until now, Dark Matter observations have not been done. As a consequence, the interactions and properties of DM are still unknown. However, the different proofs that we have about its existence allow us to predict some of its properties.

\begin{figure}[htbp]
\centering
\includegraphics[width=150mm]{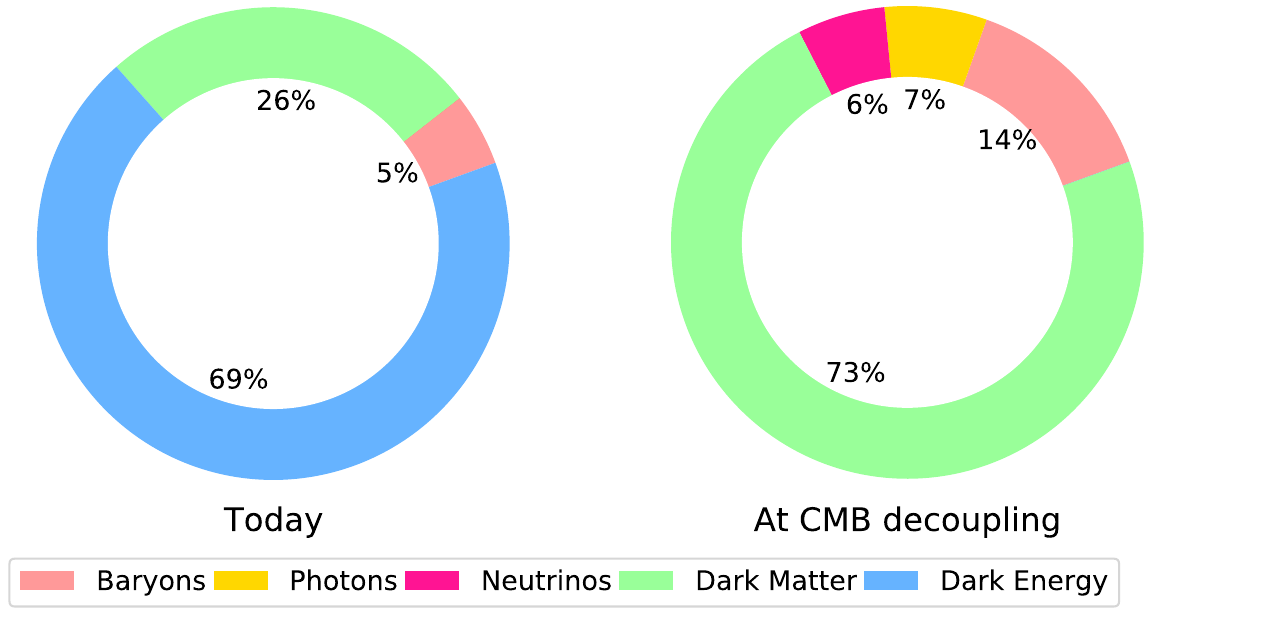}
\caption[Energy and matter content of the Universe]{\it Energy and matter content of the Universe in two different ages. Left: today; Right: at the CMB decoupling. }
 \label{fig:abundancia_DM}
\end{figure}

The abundance of DM along the evolution of the Universe is well known: Fig.~\ref{fig:abundancia_DM} shows the matter and energy content of the Universe today (left) and after the CMB decoupling (right). Nevertheless, the abundance of DM is not the unique property that it is possible to predict with the current data. In this section we explain the mostly accepted DM properties by the scientific community.

\subsection{DM - SM interactions}
\label{sec:DMinteraction}

Assuming that Dark Matter exists, the first question we must ask is: how does it interact with the rest of the particles? We have clear proofs that DM interacts, at least, through one of the four fundamental forces, the gravitational one. This fact is indisputable: all evidences of the existence of Dark Matter are related to gravitation. Now, what about the other three? 

In 1990 strongly interacting DM was proposed \cite{PhysRevD.41.3594}; nevertheless, not many years later this option was totally ruled out. The implications of the existence of this Dark Matter type are so strong that even in the Earth heat flow it would be detected \cite{Mack:2007xj}.

Another Dark Matter theory proposal assumes that DM has electrical charge \cite{DERUJULA1990173}. However, non-detection of DM and other reasons set strong limits on DM particles with an electric charge, practically ruling out this option. \cite{Taoso:2007qk}. Consequently, the most accepted hypothesis is that DM is a singlet under the color and electromagnetic SM gauge groups. However, some physicists have speculated about the possibility of having DM composed of particles with a fractional electrical charge, also known as milli-charged particles \cite{Holdom:164785,HOLDOM1991329,Abel_2008,Mass__2006,Abel_2004,Batell_2006}. These kind of DM candidates may have effects in the CMB, setting strong bounds \cite{Dubovsky_2004}. Besides the CMB, there are other sources of bounds for this type of particles (different constraints are summarized in Ref.~\cite{Davidson_2000}).

Regarding the last of the 4 forces, the weak interactions of DM with SM neutrinos are analysed in several works \cite{B_hm_2001,Wilkinson_2014,Arg_elles_2017,Olivares_Del_Campo_2018}. The elusive nature of neutrinos makes it difficult to constrain these kind of interactions. Weakly interacting DM will be reviewed in Sect.~\ref{sec:wimps}. 

\subsection{Dark Matter self-interactions}
\label{sec:DMselfinteractions}

In Sect.~\ref{sec:DMinteraction} it were examined the different DM interactions with SM particles. However, what happens with the self-interaction of the DM? The self-interactions of DM have been a subject of debate for many years. Theories with self-interactive Dark Matter (SIDM) were proposed at the end of the last century \cite{Spergel:1999mh}, motivated by the problems generated by the most popular kind of DM, the Cold Dark Matter\footnote{See Sect.~\ref{sec:Hotwarmcold}.}.

However, since DM must explain observations such as the bullet cluster, in order to keep General Relativity, strong bounds are imposed on SIDM \cite{Dave_2001,Feng_2009,Agrawal_2017,Randall_2008,Kahlhoefer_2013}:
\be
\sigma/m \lesssim 10^{-24} \, \text{cm}^2/\text{GeV} .
\ee

\subsection{Dark Matter stability}
\label{sec:DMstability}

If there is a clear property of Dark Matter in which everybody agrees is the DM lifetime. In order to reproduce the current observations of the Dark Matter abundance, any candidate must have a lifetime larger than the age of the Universe, $t_0 = 13.8 \, \text{Gyr}$ \cite{Aghanim:2018eyx}. 
Nowadays, DM is part of the content of the Universe as a relic density. If the lifetime condition is not satisfy, Dark Matter would have started to decay after the decoupling moment; therefore, there would be nothing today.

\section{Hot, Warm and Cold Dark Matter}
\label{sec:Hotwarmcold}

Since Dark Matter is the dominant matter component, the formation of the different structures observed nowadays in the Cosmos is fixed by the random movement of DM in the early times. The DM velocity in the primordial Universe is a function of the distance travelled by the DM particles due to their random motion. The name of this distance is \textit{free streaming length}, $\lambda_{\text{FS}}$. According to $\lambda_{\text{FS}}$, DM can be classified into three groups: 
if $\lambda_{\text{FS}}$ is much smaller than a typical protogalaxy size ($\diameter \backsim 300 \, \text{pc}$), DM is \textit{cold}; if it is much larger \textit{hot} and finally if it is comparable \textit{warm} \cite{1984ApJ...285L..39V,1985ApJ...299..583U}. In the middle of the 1990's theories of mixed DM became popular, nevertheless today are ruled out. In Fig.~\ref{fig:abundancia_DM} are represented the structures predicted by the three DM types. 

\begin{figure}[htbp]
\centering
\includegraphics[width=130mm]{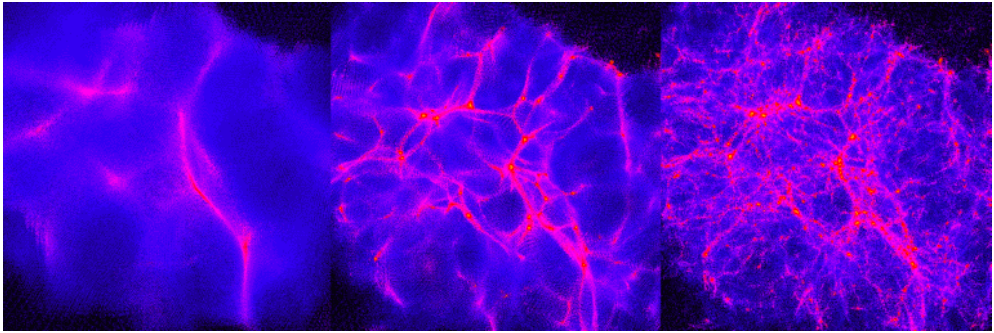}
\caption[Examples of structure formation of hot, warm and cold Dark Matter.]{\it Examples of structure formation with hot (left), warm (middle) and cold (right) Dark Matter. Simulation made by Ben Moore, Zurich University \cite{citahotwarmcold}.}
 \label{fig:abundancia_DM}
\end{figure}

Hot Dark Matter (HDM) refers to particles that move with velocity close to the speed of light, like SM neutrinos. The main property of the HDM is that the DM species are relativistic at the time of the structure formation, this implies large damping scales\footnote{Photons and baryons are imperfectly coupled and, as a consequence, a series of anisotropy damping are produced in small scale, this effect is the so-called \textit{Silk damping} \cite{1968ApJ...151..459S}. Collision-less species that move from areas of higher density to areas of lower density also produce this kind of effects \cite{1983ApJ...274..443B}.}. Nowadays, the HDM is disfavoured by N-body simulations since the Universe predicted by this Dark Matter type is incompatible with the current observations of the structure formation. For a complete description about the HDM problems see Ref.~\cite{primack2001happened}.

Cold Dark Matter (CDM) was proposed in 1982 in Refs.~\cite{1982ApJ...263L...1P,1982PhRvL..48.1636B,1982Natur.299...37B} (the details of the theory were developed in Ref.~\cite{1984Natur.311..517B}). Nowadays, CDM is the most accepted DM model. Its predictions are in agreement with a great number of observations, such as the abundance of clusters at $z\lesssim 1$ and the galaxy-galaxy correlation function. However, in the last years, several discrepancies have been found in CDM scenarios. For example, the CDM models usually predict more \textit{Dwarf Spheroidal Galaxies}\footnote{Low-luminosity galaxies with older stellar population.} (dSphs) than observed ones \cite{Klypin_1999,1999ApJ...524L..19M}. In addition to this problem, N-body CDM simulations predict rotation curves for low surface brightness galaxies\footnote{Low surface brightness galaxies are a diffuse kind of galaxies with a surface brightness that is one magnitude lower than the ambient night sky.} not compatible with the observations \cite{Flores:1994gz,McGaugh_1998,Gentile:2005tu,Gentile_2004}. A complete review of CDM can be found in Ref.~\cite{Primack:2001ia}.

In order to alleviate the Cold Dark Matter problems, Warm Dark Matter was proposed (WDM). The larger $\lambda_{\text{FS}}$ of WDM with respect to the CDM ones suppresses the formation of small structures, solving the Dwarf Spheroidal Galaxies problem. In the WDM case, the current DM abundance can be obtained for $\lambda_{\text{FS}} \backsim 0.3 \, \text{Mpc}$ \cite{Viel:2006kd}. The WDM inhibit the formation of small DM halos at high redshift, that are needed in the star formation processes. This fact, and the observations of the so called \textit{Lyman-$\alpha$ forest}\footnote{Discovered in 1970 by Roger Lynds with the observations of the quasar 4C 05.34 \cite{2006ApJS..163...80M}, the Lyman-$\alpha$ forest is a series of absorption lines in electron transition of the neutral hydrogen atom.}, set bounds to the WDM mass.

\section{Dark Matter distribution in the Galaxy}
\label{sec:distr}

In previous sections, all properties and proofs of the existence of DM have been explained, as well as the amount of DM that populates our Universe. But how is the DM distributed? is it possible to predict the density profile of DM in our galaxy? The answer is yes! 

There are several models that describe the distribution of DM along the Milky Way. The distribution is given by a Dark Matter halo profile model, that relates, for each point, the DM density with the distance between this point and the Galactic Center (GC).

\begin{table}
\centering
\begin{tabular}{c c c}
    \hline
    \bfseries Profile Name & \bfseries Predicted density $\rho(r)$ & Ref.   \\
    \hline
     NFW & $\dfrac{\rho_s}{\eta} \left(1+\eta \right)^{-2}$ & \cite{Navarro:1995iw} \\[8pt]
     Einasto & $\rho_s \, \exp \left( -\dfrac{2}{\alpha} \left[ \left( \eta \right)^\alpha  - 1 \right] \right)$ & \cite{Graham:2005xx,Navarro:2008kc}  \\[8pt]
     Isothermal & $\dfrac{\rho_s}{1 + \eta^2}$ & \cite{Bahcall:1980fb,Begeman:1991iy} \\[8pt]
     Burkert & $\dfrac{\rho_s}{\left(1 + \eta\right)\left(1 + \eta^2\right)}$ & \cite{Burkert:1995yz} \\[8pt]
     Moore & $\rho_s \, \eta^{-1.16}\left( 1 + \eta\right)^{-1.84}$ & \cite{Diemand:2004wh} \\[8pt]
    \bottomrule
\end{tabular}
\caption[Dark Matter halo profiles.]{List of the most common Dark Matter halo profiles. We have defined $\eta = r/r_s$ to alleviate the notation. In all profiles there are two parameters that it is necessary to determine with observations: $r_s$, that represents a typical scale radius, and $\rho_s$, a typical scale density. The Einasto profile presents an extra parameter, $\alpha$, that varies from simulation to simulation.}
\label{DMprofiles}
\end{table}

Tab.~\ref{DMprofiles} summarizes the most common DM density profiles in the literature. The most common one is the Navarro, Frenk and White (NFW) profile, motivated by N-body simulations. However, recent simulations favour the Einasto profile over the NFW \cite{2004MNRAS.349.1039N,Retana_Montenegro_2012}. Other models, such as the Isothermal or the Burkert profiles, seem more motivated by the observations of galactic rotation curves. All profiles showed in Tab.~\ref{DMprofiles} assume spherical symmetry\footnote{There are strong evidences in N-body simulations to assume spherical symmetry in the DM halo profiles \cite{Jing:2002np}.}. A complete discussion about the advantages and disadvantages of different DM density profiles can be found in Ref.\cite{2010AdAst2010E...5D}. 

All models present two free parameters\footnote{The Einasto profile needs an extra parameter, $\alpha$. This shape parameter varies from simulation to simulation.} $(r_s, \rho_s)$ that must be determined using astrophysical observations of the Milky Way. The two fundamental measurements used to fit these free parameters are the DM density at the Sun location respect to the Galactic Center\footnote{Recent measurements determined $R_\odot\ = 8.33$ kpc \cite{Bovy_2009,Gillessen:2008qv}, in any case, the most extended value for the distance GC-Sun is still $R_\odot\ = 8.5$ kpc \cite{Kerr:1986hz}.}, $\rho_\odot\ = 0.3 \pm 0.1 \, \text{GeV}/\text{cm}^3$ \cite{Read:2014qva}\footnote{Measurements of the Sloan Digital Sky Survey estimate $\rho_\odot = 0.46 \, \text{GeV}/\text{cm}^3$ \cite{Sivertsson:2017rkp}. However, the most extended value is still $\rho_\odot\ = 0.3 \, \text{GeV}/\text{cm}^3$.}, and the DM contained in 60 kpc, estimated as $M_{60} = 4.7 \times 10^{11} M_\odot\ $ \cite{Przybilla_2010,Sakamoto:2002zr,Xue:2008se}.

\begin{table}
\centering
\begin{tabular}{c c c c}
    \hline
    \bfseries Profile & \bfseries $\alpha$ & \bfseries $r_s \, [\text{kpc}]$ & \bfseries $\rho_s \, [\text{GeV}/\text{cm}^3]$  \\ 
    \hline
     NFW & $-$ & $24.42$ & $0.184$ \\
     Einasto & $0.17$ & $28.44$ & $0.033$ \\
     EinastoB & $0.11$ & $35.24$ & $0.021$ \\
     Isothermal & $-$ & $4.38$ & $1.387$ \\
     Burkert & $-$ & $12.67$ & $0.712$ \\
     Moore & $-$ & $30.28$ & $0.105$ \\
    \bottomrule
\end{tabular}
\caption[Fitted parameter of the Dark Matter halo profiles.]{Fitted parameter of the Dark Matter halo profiles.}
\label{DMprofiles_fit}
\end{table}

\begin{figure}[htbp]
\centering
\includegraphics[width=120mm]{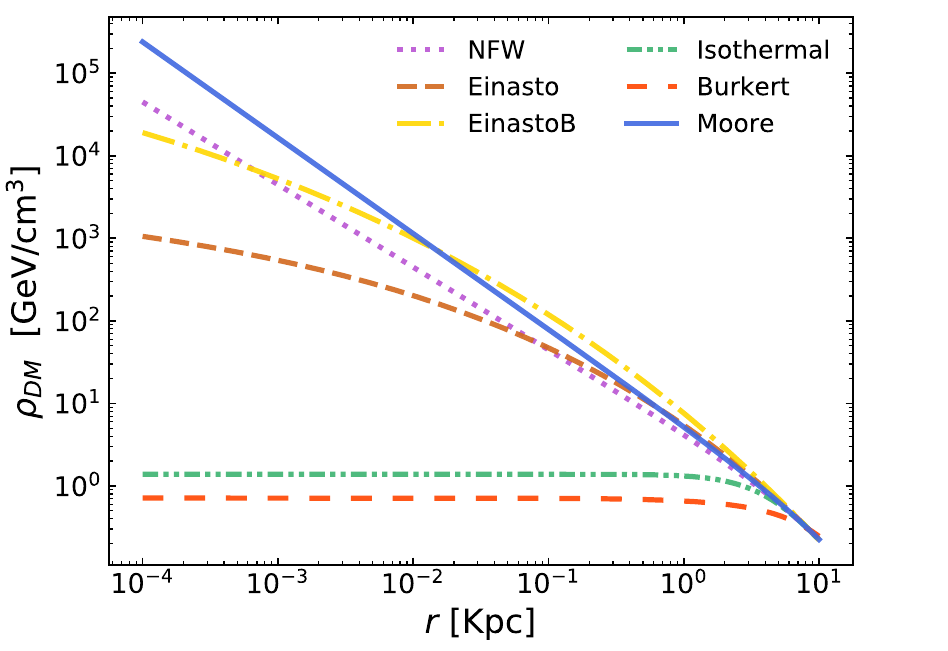}
\caption[DM halo profile models]{\it DM density as a function of the radius to the GC for different DM halo profile models.}
 \label{fig:profiles_DM}
\end{figure}

Tab.~\ref{DMprofiles_fit} shows the values of the free parameters of the DM halo profile models, which have been taken from Ref.~\cite{Cirelli:2010xx}. The Einasto and EinastoB models have the same dependence with the distance to the GC, nevertheless they are completely different in terms of particle inclusion. While in the first one the baryons are not considered, in the second one all SM is present. Fig.~\ref{fig:profiles_DM} shows the DM density as a function of the distance $r$ for the different DM halo profile models.

\section{Candidates}
\label{sec:candidates}

Up to this time the evidences and general properties of DM have been explained. The next step is to analyse the possible Dark Matter candidates that would fit the observations. The DM candidates landscape is huge; here we will make a summary of those that are, or have been, most popular. For a complete review about the DM candidates see Ref.~\cite{Baer:2014eja}.


\subsection{MACHOs}

One of the first studied cases was the possibility that the DM was baryonic matter. In this hypothesis DM would consist of small astronomical inert bodies that receive the name of MACHOs\footnote{Massive Astrophysical Compact Halo Objects. This term was coined by the astrophysicist Kim Griest.} \cite{1986ApJ...304....1P}. Nowadays, it is known that this kind of DM involves several problems. The current bounds are derived from the microlensing observations causing the exclusion of masses below the solar mass, $M_\odot\ $ \cite{Tisserand_2007,Alcock_2000}. Besides, since the MACHOs were produced after the BBN, its existence should leave a mark on the abundance of baryons that has not been observed \cite{Fields:1999ar}.

\subsection{Weakly interactive massive particles (WIMPs)}
\label{sec:wimps}

One of the most studied Dark Matter candidates is the weakly interactive massive particles (WIMPs). Firstly Proposed by Benjamin W. Lee and Steven Weinberg \cite{PhysRevLett.39.165} and studied later in several researches, this kind of particles interact very weakly with the rest of the particles of the SM. In the WIMP paradigm the DM particles were in thermal equilibrium with the SM in the early Universe. When the rate of the interactions between the DM and the SM particles became smaller than the expansion rate of the Universe, the WIMP particles decoupled from the thermal bath leaving a relic abundance that can be observed nowadays\footnote{This process receives the name of \textit{freeze-out}.}. If the WIMP particles are in the GeV-TeV mass range, the interaction scale to obtain the current DM abundance of the Universe is just the electroweak scale \cite{Kolb:1990vq,PhysRevLett.39.165,PhysRevD.43.3191,Gondolo:1990dk}. This fact, that receives the name of \textit{WIMP miracle}, has motivated the study of these particles during the last 40 years.
For instance, in Refs.~\cite{Folgado:2018qlv,Folgado:2019sgz,Folgado:2019gie,Folgado:2020utn} (included in Part \ref{sec:papers}) we have analysed different scenarios where the DM is a WIMP particle. As WIMPs are the main DM candidate studied in the this Thesis, a detailed description of the processes needed to generate the DM abundance in this scenario is provided in Sect.~\ref{Sec:freezeout}. Several examples of theories that predict the existence of stable particles at the electroweak scale that can be interpreted as WIMP particles are: SUSY\cite{MARTIN_1998,Jungman_1996,Bertone_2005,Bergstr_m_2012,Roszkowski_2018}, UED \cite{2001PhRvD..64c5002A} or little-Higgs theories\footnote{In all cases the stable particle is consequence of a conserved symmetry.} \cite{PhysRevD.10.539,PhysRevLett.29.1698,PhysRevD.12.508}.

Until the present day, WIMP searches have been unsuccessful. As a consequence, the possible cross-section of WIMP DM with the SM particles in the mass range $m_{\text{DM}} \in [1,1000]$ GeV is significantly constrained. However, great efforts are being made by the experimental community in this area. Nowadays, there are three fundamental strategies in order to detect WIMP Dark Matter: Direct Detection (DD), Indirect Detection (ID) and collider searches. The DD consists in the detection of DM-nucleus scattering processes. Some DD experiments are, for instance, Xenon1T \cite{PhysRevLett.121.111302} or PandaX-II \cite{PhysRevLett.119.181302}. On the other hand, ID experiments try to observe the SM particles that results from the annihilation and decay of particles in the cosmic ray fluxes. Different examples of ID techniques include the detection of $\gamma$-rays (such as the Fermi-LAT experiment \cite{Fermi-LAT:2016uux,Ackermann:2015zua}) or the detection of charged particles (such as AMS-02 \cite{Tomassetti:2015lva}). The detection techniques of the WIMP DM are described in detail in Chapter \ref{sec:Detection}. For interesting recent reviews about this topic see Refs.~\cite{Baer:2014eja,Klasen:2015uma,Adhikari:2016bei,Alexander:2016aln,Tulin:2017ara}.

\subsection{Feebly interactive massive particles (FIMPs)}
\label{sec:FIMP_intro}

\begin{figure}[htbp]
\centering
\includegraphics[width=120mm]{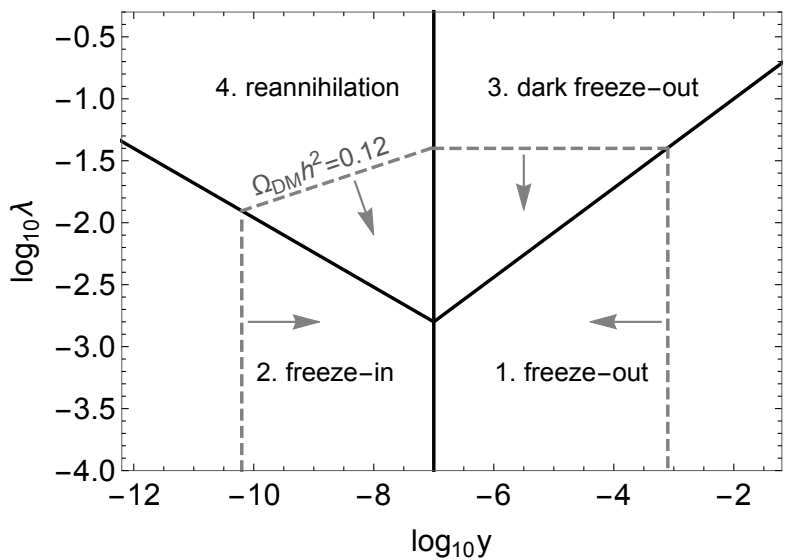}
\caption[Freeze-out and Freeze-in couplings to obtain the correct DM relic abundance]{\it Values of the DM-visible sector coupling ($y$) and the DM self-interaction coupling ($\lambda$) to obtain the correct Dark Matter relic abundance. Image taken from Ref.~\cite{Bernal:2017kxu}.}
 \label{fig:FOFIcouplings}
\end{figure}

In order to produce the light elements and the observed structure of the CMB, the SM particles must have been in thermal equilibrium in the early Universe. However, DM may or may not have been part of the same heat bath that the SM. If DM never was in thermal equilibrium with the rest of the particles, the observed DM abundance could be generated via the \textit{freeze-in} mechanism\footnote{In Sect.~\ref{Sec:freezein} all the details of the freeze-in production mechanism are explained.} \cite{McDonald:2001vt,Choi:2005vq,Kusenko:2006rh,Petraki:2007gq,Hall:2009bx}. Fig.~\ref{fig:FOFIcouplings} shows the values of the DM coupling to SM particles ($y$) and the DM self-coupling ($\lambda$) with which it can be obtained the correct relic abundance. While in the WIMP scenario to figure the correct relic abundance the needed interaction scale is the electroweak scale, in this new paradigm the interaction scale is much weaker because the DM particles never reached thermal equilibrium with the SM particles. The name of this new DM candidate is Feebly Interacting Massive Particles (FIMPs) \cite{Hall:2009bx}. In Ref.~\cite{Bernal:2020fvw} we have considered a FIMP candidate to solve the DM problem.

The detection of FIMP particles is difficult. Since the interaction scale between the SM and the DM candidate is $\text{log}_{10} (y) \in  [-10,-7]$, the DD experiments can not impose limits to the scattering cross-section. On the other hand, the signature of the mediators can be searched in the LHC, setting different limits. A summary of the different detection techniques and signals of FIMP Dark Matter can be found in Ref.~\cite{Bernal:2017kxu}.


\subsection{Axion Dark Matter}
\begin{figure}[htbp]
\centering
\includegraphics[width=100mm]{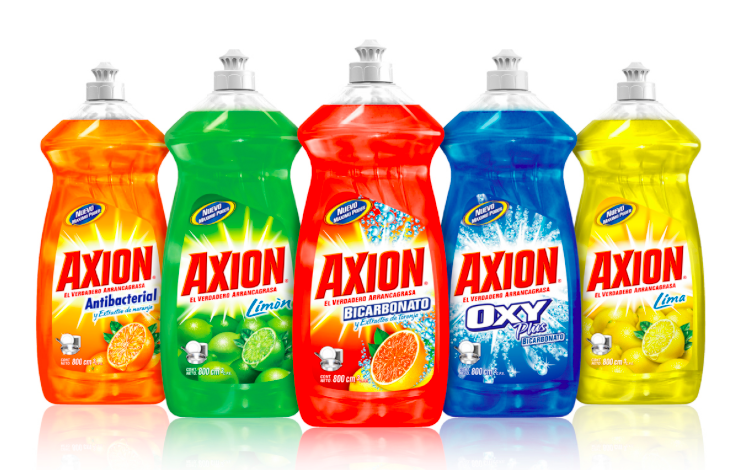}
\caption[Axion]{\it Picture of the detergent in which Wilczek was inspired to name the axion. Surely, the marketing department of the detergent brand never thought that their product would be part of the history of high energy physics.}
 \label{fig:axion}
\end{figure}

In 1977 Roberto Peccei and Helen Quinn proposed an elegant mechanism to solve the strong CP problem\footnote{See Sect.~\ref{sec:strongcpproblem} for more details.} \cite{PhysRevLett.38.1440}. This mechanism assumes the existence of a new symmetry spontaneously broken. After the SSB of the Peccei-Quinn symmetry, a new light particle appears, the so-called \textit{axion}\footnote{The particle was predicted at the same time, independently, by Wilczek and Weinberg. Wilczek was the one who baptised the particle with the name of \textit{axion}, inspired in a detergent brand (see Fig.~\ref{fig:axion}), while Weinberg called it \textit{Higglet}. The name that Wilczek gave to the particle became so popular that even Weinberg agreed to adopt it. The origin of the \textit{joke} is that the axion is a pseudoscalar particle, consequently, the symmetry broken is an axial symmetry.} \cite{PhysRevLett.40.279,PhysRevLett.40.223}. QCD non-perturbative effects generate a potential for the axion, giving mass for this particle. The mechanism did not predict the mass of this new light boson, that depends on the scale at which the Peccei-Quinn symmetry is broken. 

Axions were very popular in the scientific community since they may solve at the same time both, the strong CP problem and the DM problem. The way to produce the current DM abundance is not related with the thermal mechanism. In this case, it is assumed that the Dark Matter axions were produced in the early Universe as a result of coherent oscillations of the axion field. These oscillations generate bosonic condensates that today would be measured as CDM. The couplings between the Dark Matter axions and the other particles are model dependent and are generally assumed quite small. Nowadays, experimental bounds constrain the original Axion as a DM candidate. If the SSB of the PQ symmetry takes place after inflation, the misalignment angle is fixed, $\theta_{\text{CP}} \backsimeq \pi^{2}/3$, and the bounds over the mass exclude the axion DM.
However, particles that produce the DM abundance through the same mechanism are very dear to the scientific community. The name of these particles is \textit{Axion Like Particles}, ALPs.


Currently, the two most accepted benchmark realizations of the Peccei-Quinn mechanism are the KSVZ \footnote{Kim-Shifman-Vainshtein-Zakharov.} \cite{PhysRevLett.43.103,SHIFMAN1980493} and DFSZ\footnote{Dine-Fischler-Srednicki-Zhitnitsky.} \cite{DINE1981199,Zhitnitsky:1980tq} models. The feeble interaction between the axion DM field and the SM particles is also a consequence of the axion small masses, since mass and coupling are inversely proportional:
\be
m_a \backsimeq m_\pi \frac{f_\pi}{f_a} \, ,
\ee
where $m_\pi$ and $f_\pi$ are the pion mass and decay constant, respectively.

The number of experiments that try to find evidences of the existence of axions is enormous. Several experiments base their search in the Primakoff effect\footnote{The Primakoff effect is the resonant production of neutral mesons via high-energy photons interacting with a nucleus.} \cite{1974PhRvL..33.1400B} such as ADMX \cite{2010PhRvL.104d1301A}, HAYSTAC \cite{Brubaker_2017}, CULTASK \cite{Petrakou_2017} and ORGAN \cite{mcallister2017organ}. Other experiments, as PVLAS, search changes of the polarized light in a magnetic field \cite{1986PhLB..175..359M}. The mentioned experiments are only an infinitesimal example of the large experimental landscape. For more information about the detection and the astrophysical implications of axion DM see Refs.~\cite{Borsanyi:2016ksw,Wantz:2009it,Kim:2008hd}.

\subsection{Primordial Black Holes (PBHs)}

Primordial Black Holes (PBHs) were firstly proposed in the 1970's in Refs.~\cite{10.1093/mnras/168.2.399,1975ApJ...201....1C,10.1093/mnras/152.1.75}. While the standard Black Holes are the consequence of the gravitational collapse of a star, the PBHs were originated due to the extreme density of the Universe at the beginning of its expansion. Since PBHs were formed in the very first moments of the Universe, before the BBN, bounds on baryonic matter do not apply to them, became PBHs in a good DM candidate \cite{Chapline:1975ojl,Carr:2016drx}.

In order to obtain the correct relic abundance of the DM it is necessary that the PBHs survive until today. As the Primordial Black Holes are not stable\footnote{PBHs can evaporate through Hawking radiation \cite{Hawking:1971tu,1974Natur.248...30H,Hawking:1974sw}.}, a lower bound on their mass exists. If we assume that all DM abundance is due to PBHs, this lower bound is  $m_{\text{PBH}} > 3.5 \times 10^{-17} \, M_\odot\ $ \cite{PhysRevD.81.104019}. The idea of PBHs as DM has been revived with the detection of gravitational waves by LIGO\cite{Abbott_2016} since these observations can be explained with two coalescing Primordial Black Holes\cite{Bird_2016}.
\chapter{Evolution of the Universe: A Thermodynamic Description }
\label{sec:termo}

In order to understand any form of matter that surrounds us today, we need to ask ourselves which has been its evolution starting from the first moments of the Universe. This is obviously a problem of many bodies that must be statistically analysed. At the beginning of the 20th century it was thought that the Universe was practically empty, except for slight singularities (galaxies, planets, ...) that were completely lost in the immensity of space-time. In the middle of the century, the Cosmic Microwave Background was accidentally discovered. Nowadays, we know that the radiation from the CMB, measured at $T \backsimeq 2.725$ K \cite{Aghanim:2018eyx}, is the echo of the first moments after the Big Bang. The existence of a Cosmic Microwave Background is one of the great predictions of cosmological models based on the Big Bang hypothesis, according to which the original Universe was a plasma, at very high temperature, formed by baryons, electrons and photons. As the plasma cooled down, due to the adiabatic expansion of the Universe, the baryons and electrons recombined to form atoms, thus decoupling the photons in equilibrium.

The cooling of the Universe caused the different particles, that populated that hot and inert Universe, to decouple thermodynamically from the plasma until finally a small fraction remained, the CMB that we observe today, and slowly dilutes. 
Since the primordial Universe can be described as a plasma in thermodynamic equilibrium with good accuracy, developing any evolution model will involve understanding statistical thermodynamics.

\section{Equilibrium description}

\subsection{Fundamental Thermodynamic Variables}

Due to the asymptotic decrease in the strong interaction at high energies/temperatures, we can consider the plasma that formed the primordial Universe as a set of ideal gases in equilibrium with $g$ internal degrees of freedom. The number density $n$, the energy density $\rho$ and the pressure $p$ of this fluid can be written based on its distribution function in the phase space:
\bea
\left \{
\begin{array}{llll}
n &\equiv& \dfrac{g}{(2\pi)^3} \displaystyle \int_{-\infty}^{\infty} d^3 p \, f(\vec{p},t) = \dfrac{g}{2\pi^2} \int_{m}^{\infty} d E \, E \, (E^2 - m^2)^{1/2} \, f(E,T) \, \nonumber , \\
\rho &\equiv& \dfrac{g}{(2\pi)^3} \displaystyle \int_{-\infty}^{\infty} d^3 p \, E(\vec{p}) \, f(\vec{p},t) = \dfrac{g}{2\pi^2} \int_{m}^{\infty} d E \, E^2 (E^2 - m^2)^{1/2} \, f(E,T) \, \nonumber , \\
p &\equiv& \dfrac{g}{(2\pi)^3} \displaystyle \int_{-\infty}^{\infty} d^3 p \,\dfrac{|\vec{p}|^2}{3E(\vec{p})} \, f(\vec{p},t) = \dfrac{g}{6\pi^2} \int_{m}^{\infty} d E \, (E^2 - m^2)^{3/2} \, f(E,T) \, , \nonumber
\end{array}
\right .
\eea
\be
\,
\label{densidades_y_tal}
\ee
where the distribution function is Fermi-Dirac (FD) or Bose-Einstein (BE), depending on whether we are working with fermions or bosons:
\be
f(\vec{p}) = \dfrac{1}{e^{(E-\mu)/T} \pm 1} \, ,
\ee
where $\mu$ is the chemical potential of the species. The value of the sign in the denominator corresponds to $-1$ for the BE case and $+1$ for FD statistics, respectively. In the above expressions $E =\sqrt{|\vec{p}|^2 + m^2}$ represents the energy of a particle with momentum $p$ and mass $m$. If the species are in chemical equilibrium under the interaction $i + j \longleftrightarrow a + b$, the different chemical potentials associated with the species are related:
\be
\mu_i + \mu_j = \mu_a + \mu_b \, .
\ee

The different thermodynamic quantities described in Eq.~\ref{densidades_y_tal} have simple limits when $\mu/T \ll 1$. On the one hand, the different approximations for the $T\gg m$ case are given by:
\bea
\rho &=& \left \{
\begin{array}{llll}
g\dfrac{7}{8}\dfrac{\pi^2}{30} T^4 & \text{Fermions} \, , \nonumber \\[8pt] 
g\dfrac{\pi^2}{30} T^4 & \text{Bosons} \, ,
\end{array}
\right . \\
n &=& \left \{
\begin{array}{llll}
g\dfrac{3}{4}\dfrac{\zeta (3)}{\pi^2} T^4 & \text{Fermions}, \\[8pt] 
g\dfrac{\zeta (3)}{\pi^2} T^4 & \text{Bosons}.
\end{array}
\right .
\\[8pt]
p &=& \, \, \,\, \, \, \, \rho/3 \, , \nonumber
\label{aproximaciones_T_pequena}
\eea
where $\zeta (3) \backsimeq 1.202 $. On the other hand, for the $T \ll m$ case the Maxwell-Boltzmann distribution is a good approach for both, fermions and bosons. In this case, the energy density and the pressure take the following form:
\bea
\left \{
\begin{array}{llll}
n &=& g \left( \dfrac{m \, T}{2\pi} \right)^{3/2} e^{-(m-\mu)/T} \, , \\[8pt]
\rho &=& \left(\dfrac{3}{2} T+ m\right)\, n \, ,  \\[8pt]
p &=& n \, T \, .
\end{array}
\right .
\eea
In general, the average energy per particle can be obtained as $\langle E \rangle \equiv \rho/n$.

\subsection{Energy Density of the Universe}

The contribution of non-relativistic species to the total energy density is negligible with respect to the relativistic one. As a consequence, during the radiation dominated era the total energy density can be approximated as the radiation energy density, composed by the contribution of all relativistic particles:
\be
\rho \backsimeq \rho_R = \dfrac{\pi^2}{30}g_{\star}T^4 \, ,
\label{densidad_de_energia}
\ee
where
\be
 g_{\star}\equiv\sum_{i=\text{bosons}}{g_i\left(\dfrac{T_i}{T}\right)^4} \ + \ \dfrac{7}{8}\sum_{i=\text{fermions}}{g_i\left(\dfrac{T_i}{T}\right)^4} \, 
 \label{grados_de_libertad_energia}
\ee
is the effective number of relativistic degrees of freedom of the relativistic species, $g_i$ describes the degrees of freedom of each particle, $T_i$ its temperature and $T$ the temperature of the thermal bath, which coincides with the temperature of the photons. 

In general, relativistic species in thermal equilibrium with the photons have $T_i = T \gg m_i$. However, when the temperature of the thermal bath drops below the particle mass $m_i$, that specie becomes non-relativistic and must be removed from Eq.~\ref{grados_de_libertad_energia}. In the epoch where the temperature of the Universe was larger than the top mass $m_t$, all species were relativistic and $g_\star = 106.75$, its maximum value. Throughout the evolution of the Universe, the temperature decreases and the different particles become non-relativistic, decreasing the total number of relativistic degrees of freedom, until the current value:
\be
g_{\star (\text{today})} =2+\dfrac{7}{8}\times2\times3\times\left(\dfrac{4}{11}\right)^{4/3}=3.36 \, .
\label{current_degrees_of_freedom}
\ee
This value, which remains invariant since $e^-e^+$ annihilation, takes into account the three neutrino species and photons, the only relativistic particles. Neutrinos decoupled from the thermal bath when $T \backsim 1 \, \text{MeV}$, which led to a slightly cooler temperature from then of $T_\nu = (4/11)^{1/3} T_\gamma$ \cite{PhysRevD.98.030001}.
Under the hypothesis that Eq.~\ref{densidad_de_energia} represents a good approximation to the energy density of the Universe and that large-scale space-time is flat, Friedman's equations (Eq.~\ref{FriedmannEquations}) lead to an expression for the Hubble parameter (Eq.~\ref{Hubble_parameter}) as a function of the equilibrium temperature of the Universe:
\be
H = \sqrt{\dfrac{8\pi}{3}\dfrac{\rho}{M_{\text{P}}}} = \sqrt{\dfrac{4 \, \pi^3}{45}} \sqrt{g_\star} \dfrac{T^2}{M_{\text{P}}} \, ,
\label{Hubble_parameter_termo}
\ee
where $M_{\text{P}} = 1.22 \times 10^{19} \, \text{GeV}$ is the Planck mass.

\subsection{Entropy Conservation in the Universe}

An analysis of the evolution of any species of particle throughout the expansion of the Universe can be, in principle, complicated. In order to simplify the calculations, it is convenient to work with quantities that are conserved. Within the scope of equilibrium thermodynamics, the most commonly used conserved quantity in a \textit{comoving}\footnote{The comoving variables are defined in such a way that they are independent of the expansion of the Universe.} volume is the entropy, described by the second principle of thermodynamics:
\be
\label{segundo_principio}
T \, dS=dU+p \, dV=d(\rho V)+p \, dV=d[(\rho+p)V]-V \, dp \, ,
\ee
where $V$ is the comoving volume. Using the relation between the pressure and the temperature, $dp/dT = (p + \rho)/T$, Eq.~\ref{segundo_principio} is directly integrable:
\be
 S/V = \dfrac{p + \rho}{T}  \equiv \mathfrak{s} \, ,
\ee
where the entropy density $\mathfrak{s}$ is conserved through the expansion of the Universe, $d\mathfrak{s}/dt = 0$.  The net transfer of energy in a closed system is null, which means that the total creation and destruction of particles is zero for a Universe in equilibrium. Using Eq.~\ref{aproximaciones_T_pequena} the entropy density can be written as
\be
\mathfrak{s} = g_{\star \, s} \dfrac{2 \pi^3}{45} T^3 \, ,
\label{entropy_density}
\ee
where
\be
g_{\star s}\equiv\sum_{i=\text{bosons}}{g_i\left(\dfrac{T_i}{T}\right)^3} \ + \ \dfrac{7}{8}\sum_{i=\text{fermions}}{g_i\left(\dfrac{T_i}{T}\right)^3} \, 
\ee
is the effective number of degrees of freedom in entropy.

\begin{figure}[htbp]
\centering
\includegraphics[width=120mm]{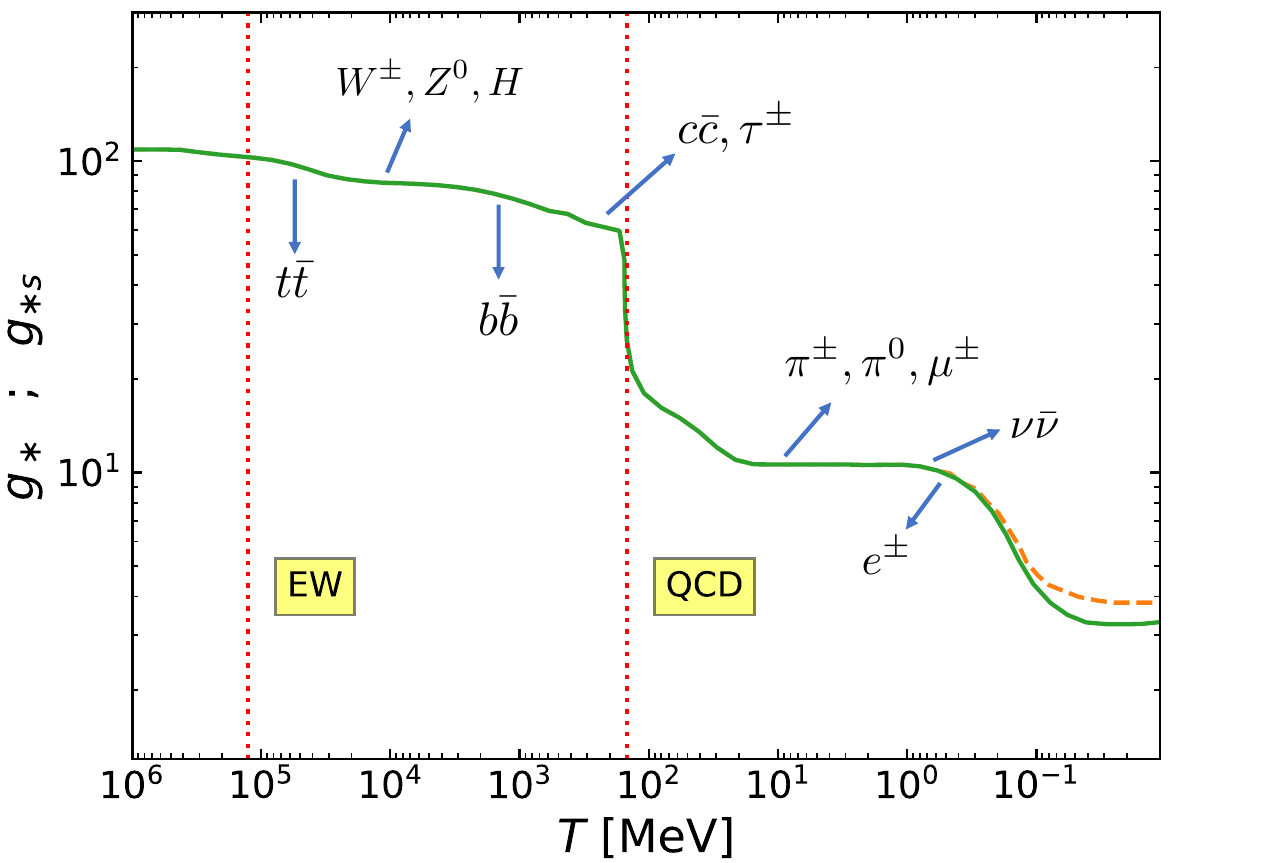}
\caption[Degrees of freedom in energy $g_\star$ and entropy $g_{\star s}$.]{\it Evolution of the relativistic degrees of freedom in density $g_\star$ (green solid line) and in entropy $g_{\star s}$ (orange dashed line). The red dotted lines show the temperature of the EW and QCD transition. The arrows indicate the moment when each species becomes non-relativistic.}
 \label{fig:degrees_of_freedom}
\end{figure}

Before neutrino decoupling $g_\star = g_{\star s}$ (all relativistic species were in the thermal bath). When $T \backsim 1 \, \text{ MeV}$, before nucleosynthesis, neutrinos decouple from the thermal bath, remaining constant its comoving temperature. At $T \sim 0.5\, \text{ MeV}$, photons are not energetic enough to create $e^\pm$ pairs anymore. Thus, electrons and positrons annihilate, slightly heating the thermal bath and increasing the temperature of the photons. Thenceforth, the number of relativistic degrees of freedom in entropy becomes
\be
g_{\star s (\text{today)}}=2+\dfrac{7}{8}\times2\times3\times\left(\dfrac{4}{11}\right) = 3.91 \, .
\ee
Since then, both $g_\star$ and $g_{\star s}$ remain constant, although differing, since the plasma has been heated up while neutrinos have not.
Fig.~\ref{fig:degrees_of_freedom} shows the evolution of $g_{\star}$ and $g_{\star s}$ as a function of the temperature. As can be seen in the Figure, the difference between the degrees of freedom in energy and entropy is only important  after the neutrino decoupling. This is due to the fact that neutrinos are the unique species that remains relativistic after its decoupling from the primordial plasma.

\section{Beyond the Equilibrium Description}

\subsection{The Idea of Thermal Decoupling}

At the beginning of time, the most part of the constituents of the Universe were in thermal equilibrium. For this reason, a description based on thermodynamic equilibrium is a good approximation of the early thermal history of the Universe. However, if the Universe were actually in complete thermal equilibrium, its current appearance would be that of a gas in equilibrium at the CMB temperature, which is not true. The Universe as we observe it today is the result of a multitude of processes out of equilibrium. A deep knowledge of these processes is the key to understand the evolution of the different particle species that were decoupled from the thermal bath, leaving a little background known today as relic abundances. There are different examples of decoupling from this thermal equilibrium, such as the case of neutrino decoupling, the background radiation, etc.

The problem is therefore reduced to analysing what has been the evolution of the abundance of these particles throughout the history of the Universe and what has been the remnant that they have left. The first task is to understand what equilibrium plasma decoupling actually means. Let's suppose the following $2 \rightarrow 2$ reaction:
\be
\chi \, \bar{\chi} \longleftrightarrow \psi \bar{\psi} \, ,
\label{doble_reaccion}
\ee
where $\chi$ represents the particle that will be decoupled (WIMP DM particles, for example) and $\psi$ are the rest of the particles of the primordial plasma. As long as the $\chi$ particle is in thermal equilibrium, the reaction given by Eq.~\ref{doble_reaccion} occurs. 
The reaction is possible in both directions while the temperature is high enough for the less massive particles to be annihilated giving rise to the more massive ones. The net destruction of $\chi$ particles is then null. As the Universe expands, however, the temperature drops until the process can only occur in one direction: there is destruction of $\chi$ particles, but there is no creation anymore:
\be
\chi \, \bar{\chi} \longrightarrow \psi \bar{\psi} \, .
\label{reaccion_solo_un_sentido}
\ee
At that moment, we say that $\chi$ is decoupled from the bath.

The criterion to determine if some kind of particles is coupled or decoupled to the primordial plasma involves the comparison of the interaction rate of the particle, usually called $\Gamma$, with the expansion rate of the Universe (Hubble parameter):
\bea
\left \{
\begin{array}{llll}
\Gamma & > & H  \,\,\,\,\,\, \text{(coupled)} \, , \\
\Gamma & \lesssim & H  \,\,\,\,\,\, \text{(decoupled)} \, .
\end{array}
\right .
\eea
The interaction rate is determined by all reactions that keep the species in thermal equilibrium. 

After the decoupling, the amount of $\chi$ particles falls to the point where the annihilation practically stops. The rest of the thermal bath particles will follow the equilibrium distribution, while the $\chi$ species will follow a new distribution function. How is the form of this new distribution function? To determine this it is necessary to understand the Boltzmann equation.

\subsection{Boltzmann Equation}
\label{sec:Boltzeq}

The evolution of the particle number densities depends on the evolution of the distribution function $f(p^{\mu}, x^{\mu})$ of $\chi$ species in phase space, but modelling this mathematically is tricky. Liouville's theorem\footnote{See Ref.~\cite{goldstein2002classical} for a modern description of the theorem.}  \cite{liouville1838note} tell us that the volume of the phase space of a distribution remains constant during the evolution of each particle of the system, as long as the system is collisionless. The theorem can be written in terms of the so-called Liouville operator (or Liouvillian):
\be
\hat{L}[f] = 0 \, .
\label{Liouville_theorem}
\ee
The general covariant form of this operator is given by \cite{Kolb:1990vq}:
\be
\hat{L} = p^\alpha \dfrac{\partial}{\partial x^\alpha} - \Gamma^\alpha_{\beta \gamma} p^\beta p^\gamma \dfrac{\partial}{\partial p^{\alpha}} \, .
\ee
All the gravitational effects of the problem then come from the affine connection of the metric. For the FLRW model, the phase space density is homogeneous and isotropic: this means that $f = f(|\vec{p}|,t)$ (or equivalently $f = f(E,t)$). The Liouville operator in this model takes the following form
\be
\hat{L}[f(E,t)] = E \dfrac{\partial \, f}{\partial \, t} - \dfrac{\dot{a}}{a} |\vec{p}|^2 \dfrac{\partial \, f}{\partial \, E} \, ,
\ee
where $a$ is the scale factor of the FLRW metric.

In order to describe a system where $\chi$ species interacts with the rest of the particles of the SM it is necessary to modify Eq.~\ref{Liouville_theorem} adding the \textit{collision operator}\footnote{For a derivation of the collision operator in quantum field theory see Ref.~\cite{collisionoperator}.} $\hat{C}$: 
\be
\hat{L}[f] = \hat{C}[f] \, ,
\label{BE_abstracta}
\ee
that receives the name of Boltzmann Equation\footnote{The equation was proposed in 1872 by Ludwig Boltzmann in the context of kinetic theory of gases \cite{Boltzmann1970WeitereS}.} and determines the evolution of the distribution function $f(p^{\mu}, x^{\mu})$ of any species of particles.
Using the definition of the number density in terms of the phase space density Eq.~\ref{densidades_y_tal}, and integration Eq.~\ref{BE_abstracta}, it is easy to obtain:
\be
\dfrac{d n_{\chi}}{d t} + 3 \dfrac{\dot{R}}{R} n_{\chi} = \dfrac{g}{(2 \pi)^3} \int \hat{C}[f] \dfrac{d^3 p}{E} \, ,
\label{BE_unpocomenosabstracta}
\ee
where $n_{\chi}$ refers to the numerical density of the $\chi$ particle.

In order to solve the equation, the collision term can be derived assuming that the colliding particles are not connected before the collision (the so-called \textit{Stosszahlansatz} or \textit{molecular chaos hypothesis}) \cite{jeans_2009,2006cond.mat..1566M,Boltzmann1970WeitereS}. Within this hypothesis, Eq.~\ref{BE_unpocomenosabstracta} can be written as:
\bea
\dfrac{g}{(2\pi)^3} \int \hat{C}[f] \dfrac{d^3 p_{\chi}}{E_{\chi}} &=& - \int d\Pi_{\chi} d\Pi_{a} d\Pi_{b} d\Pi_{j} d\Pi_{i} \nonumber \\
&\times & (2\pi)^4 \delta^4 (p_{\chi} + p_a + p_b - p_i - p_j) \nonumber \\
&\times & \left[ |\mathcal{M}|^2_{\chi + a + b \rightarrow i + j} f_a f_b f_{\chi} (1 \pm f_i)(1 \pm f_j) \right.\nonumber \\
& - & \left. |\mathcal{M}|^2_{ i + j \rightarrow \chi + a + b} f_i f_j  (1 \pm f_a) (1 \pm f_b) (1 \pm f_{\chi})  \right] \, , \nonumber \\
\, \, 
\label{colisionador_complicado}
\eea
having used the relativistic kinetic theory (see Ref.~\cite{Sarbach:2013fya}). In Eq.~\ref{colisionador_complicado} $f_i$, $f_j$, $f_a$ and $f_b$ are the phase space densities of species $i$, $j$, $a$, $b$; $f_{\chi}$ represents the phase space density of $\chi$ (the species that we try to analyse); $\pm$ changes for bosons ($+$) and for fermions ($-$). Finally, the integration measure is:
\be
d\Pi \equiv g\dfrac{1}{(2\pi)^3} \dfrac{d^3 p}{2E} \, ,
\ee
where $g$ counts the internal degrees of freedom. 
For simplicity, Eq.~\ref{colisionador_complicado} is particularized for $\chi \, + \, a \, + \, b \, \longleftrightarrow \, i \, + \, j$ case. However, it can be generalized to any number of colliding species.

Two well-motivated approximations can be used in order to simplify Eq.~\ref{colisionador_complicado}. The first assumption is CP invariance, that implies
\be
|\mathcal{M}|^2_{ i + j \rightarrow \chi + a + b} = |\mathcal{M}|^2_{\chi + a + b \rightarrow i + j} \equiv |\mathcal{M}|^2 \, .
\ee
The second one is to use the Maxwell-Boltzmann statistics for all species, instead than Fermi-Dirac for fermions or Bose-Einstein for bosons. In absence of Bose condensation or Fermi degeneracy, $1  \pm f \backsimeq 1$, $f_i(E_i) = e^{-(E_i - \mu_i)/T}$ can be used for all species in thermal equilibrium. With these approximations, the Boltzmann Equation takes the form
\bea
\dot{n}_\chi + 3 H n_\chi &=& - \int d\Pi_{\chi} \, d\Pi_{a}\, d\Pi_{b}\, d\Pi_{j}\, d\Pi_{i}\, (2\pi)^4 |\mathcal{M}|^2 \nonumber \\
&\times & \delta^4 (p_i + p_j - p_\chi - p_a - p_b) \, [f_a f_b f_\chi - f_i f_j] \, , \nonumber \\
\, \,   
\label{BE_mas_agradecida}
\eea
where $H \equiv \dot{a}/a$ is the Hubble rate. Analysing the meaning of the different terms of Eq.~\ref{BE_mas_agradecida} one finds that while $3 H n_\chi$ is the dilution of the particle density as a consequence of the expansion of the Universe, the right hand side term represents the variation produced by the interactions with the rest of the particles of the plasma.

In the analysis of the Boltzmann Equation it is very common to translate $n_\chi$ into the \textit{yield}:
\be
Y \equiv \dfrac{n_{\chi}}{\mathfrak{s}} \, .
\ee
This quantity takes into account the expansion of the Universe and remains constant throughout its evolution if interactions are absent. As a consequence, the yield only variates with the collision term.
The evolution of the yield since the beginning of time is better expressed in terms of temperature rather than time. For this reason, it is common to use the dimensionless variable
\be
x \equiv m/T \, ,
\ee
where $m$ is some mass scale useful for our problem (typically the mass of the $\chi$ species). Under the assumption that the number of relativistic degrees of freedom in energy ($g_{\star}$) and entropy ($g_{\star s}$) are independent of time, time and $x$ can be related during the radiation dominated epoch as $dt/dx = 1/(Hx)$. Eventually, it is very common to define
\be 
H(m) = \sqrt{\dfrac{4 \, \pi^3}{45}} \sqrt{g_\star} \dfrac{m^2}{M_{\text{P}}} \, ,
\ee
related with the Hubble parameter as $H = H(m)/x^2$. 

Under the manipulations described above, it is easy to obtain the more usual form of the Boltzmann Equation:
\bea
\dfrac{dY}{dx} &=& - \dfrac{x}{H(m) \, \mathfrak{s}} \int d\Pi_{\chi} d\Pi_{a} d\Pi_{b} d\Pi_{j} d\Pi_{i} (2\pi)^4 |\mathcal{M}|^2 \nonumber \\
&\times & \delta^4 (p_i + p_j - p_\chi - p_a - p_b)[f_a f_b f_\chi - f_i f_j] \, . \nonumber \\
\, \,   
\label{BE_casi_definitiva}
\eea

\section{Abundance analysis of the out of equilibrium species}

\subsection{Integrated Boltzmann Equation}
\label{sec:int_boltz_eq}

The general case of the Boltzmann Equation has been described in Sect.~\ref{sec:Boltzeq}. In this section we study the relic abundance generated by a stable or long-lived particle, the relevant case for the works that compose this Thesis. We can separate the analysis depending on the nature of the interaction: on the one hand, if the particles are stable, only processes $2 \, \rightarrow \, 2$, such as Eq.~\ref{doble_reaccion}, change the number of $\chi$ and $\bar{\chi}$ in a comoving volume. On the other hand, if the particles are unstable, other processes must be considered ($1 \rightarrow 2$, the different decays of $\chi$). The description performed in this section follows Ref.~\cite{Kolb:1990vq}.

First, we consider a $\chi \bar{\chi} \rightarrow \psi \bar{\psi}$ process, where $\psi$ and $\bar{\psi}$ particles represent some SM specie in thermal equilibrium. The distribution functions of these bath particles are given by the equilibrium distribution:
\bea
\left \{
\begin{array}{llll}
f_\psi &=&  e^{-E_{\psi}/T} \, , \\
f_{\bar{\psi}} &=& e^{-E_{\bar{\psi}}/T} \, .
\end{array}
\right .
\eea
The $\delta$-function in Eq.~\ref{BE_casi_definitiva} implies:
\be
E_{\chi} + E_{\bar{\chi}} = E_{\psi} + E_{\bar{\psi}} \, .
\ee
Using this information, it is easy to obtain:
\be
f_\psi  f_{\bar{\psi}} = e^{-(E_{\psi} + E_{\bar{\psi}})/T} = e^{-(E_{\chi} + E_{\bar{\chi}})/T}  = f^{\text{eq}}_\chi  f^{\text{eq}}_{\bar{\chi}} \, .
\ee
This information allows to simplify Eq.~\ref{BE_casi_definitiva}, obtaining $\left[f_\chi  f_{\bar{\chi}} - f_\psi  f_{\bar{\psi}} \right] = \left[f_\chi  f_{\bar{\chi}} - f^{\text{eq}}_\chi  f^{\text{eq}}_{\bar{\chi}} \right]$. Defining the thermal average annihilation cross-section for $2 \rightarrow 2$ processes:
\bea
\langle \sigma v \rangle &\equiv & (n_{\chi}^{\text{eq}})^{-2} \int d \Pi_\chi \Pi_{\bar{\chi}} \Pi_\psi \Pi_{\bar{\psi}} (2\pi)^4 \,  \nonumber \\ 
&\times & \delta^4(p_\chi + p_{\bar{\chi}} - p_\psi - p_{\bar{\psi}}) |\mathcal{M}|^2 e^{-E_{\chi}/T} e^{-E_{\bar{\chi}}/T} \, , \nonumber \\
\, \, 
\label{thermal_average_feo}
\eea
the Boltzmann Equation takes the form
\be
\dfrac{d Y}{dx} = \dfrac{-x \langle \sigma v \rangle \, \mathfrak{s}}{H(m)} \left(Y^2 - Y^2_{\text{eq}} \right) \, , 
\label{BE_buena}
\ee
where $Y = n_{\chi}/\mathfrak{s} = n_{\bar{\chi}}/\mathfrak{s}$ is the yield of $\chi$ and $\bar{\chi}$ particles while $Y_{\text{eq}} = n^{\text{eq}}_{\chi}/\mathfrak{s} = n^{\text{eq}}_{\bar{\chi}}/\mathfrak{s}$ is the equilibrium yield. Eventually, in order to obtain the total abundance, it is necessary to sum over all possible annihilation processes. To compute the evolution of the yield, it is necessary to know the abundance of all of the species of the Universe, the so-called equilibrium abundance \cite{Gondolo:1990dk}:
\be
Y_{\text{eq}} = \dfrac{45}{4\pi^4}\dfrac{x^2}{g_{\star s}}K_2(x) \, ,
\ee
where $K_2(x)$ is the second modified Bessel function of the second kind, which can be calculated using the following integral \cite{abramowitz+stegun}:
\be
K_n(y) = \frac{\sqrt{\pi}}{\left(n-1/2 \right)!} \left(y/2 \right)^n \int_1^\infty dt \, e^{-y \, t}\left(t^2-1\right)^{n-1/2} \, .
\ee

There are cases in which processes $1 \, \rightarrow \, 2$  have an important relevance in the evolution of the abundance. In those cases, the Boltzmann Equation must be modified as
\be
\dfrac{d Y}{d x} = -\dfrac{x \, \langle \Gamma \rangle}{H(m)} (Y - Y_{\text{eq}}) \, ,
\label{BE_copleta_con_decay}
\ee
where $\langle \Gamma \rangle$ represents the thermally averaged decay rate.
In the most general case, both terms are relevant. The Boltzmann Equation can be written then as: 
\be
\dfrac{d Y}{d x} = \dfrac{-x \left[\langle \sigma v \rangle \, \mathfrak{s} + \langle \Gamma \rangle\right] }{H(m)}\left(Y^2 - Y^2_{\text{eq}} \right) \, . 
\label{BE_buena_completa}
\ee

\subsection{Thermally-Average of Physical Observables}

Eq.~\ref{BE_buena} allows obtaining the abundance of some species out of the thermodynamic equilibrium. In order to obtain the yield, it is necessary to evaluate the thermal-averaged annihilation cross-section $\langle \sigma v \rangle$ and decay rate  $\langle \Gamma \rangle$. For the $\langle \sigma v \rangle$ case, the first task is to understand what exactly is $v$. In the non-relativistic case, $v$ is the relative velocity between the two initial particles, defined as $|v_1 - v_2|$. In the relativistic scenario (the most general case) the relative velocity is non-Lorentz invariant. Instead of the classical relative velocity expression, the so-called M{\o}ller velocity must be used \cite{moller1945general}
\be
v_{\text{M{\o}l}} = \left[|\vec{v_1} - \vec{v_2}|^2 - |\vec{v_1} \times \vec{v_2}|^2\right]^{1/2} \, .
\ee
Using this expression, the thermal-averaged annihilation cross-section can be written as:
\be
\label{thermal_average}
\langle \sigma v \rangle = \dfrac{1}{8 m^4 T K_2^2 (m/T)} \int_{4m^2}^{\infty} ds (s-4m^2) \sigma \sqrt{s} K_1 (\sqrt{s}/T)
\ee
where $K_1(y)$ and $K_2(y)$ are the modified Bessel functions of the second kind.
On the other hand, the thermally-averaged decay rate $\langle \Gamma \rangle$ can be written as
\be
\langle \Gamma \rangle = \Gamma \dfrac{K_1(x)}{K_2(x)} \, .
\ee

\section{Freeze-out: WIMP Dark Matter}
\label{Sec:freezeout}

As it has been commented in Sect.~\ref{sec:wimps}, the WIMP paradigm assumes that the DM was in thermal equilibrium with the rest of the particles of the SM in the early times of the Universe. This kind of DM was first studied by Benjamin W. Lee and Steven Weinberg \cite{PhysRevLett.39.165}. Since then, several studies have been done on this scenario. This section aims to understanding how the abundance of the DM is generated for WIMP particles using the Eq.~\ref{BE_buena}. The final form of the Boltzmann Equation for $2\rightarrow 2$ processes presented in Sect.~\ref{sec:int_boltz_eq} is written as a function of the Hubble rate and the entropy density. Replacing these two functions, the Boltzmann Equation takes the form:
\be
\dfrac{d Y}{d x} =- \dfrac{\lambda }{x^2} \langle \sigma v \rangle  \left(Y^2 - Y^2_{\text{eq}} \right) \, ,
\label{BE_buena_simplificada}
\ee
where
\be
\label{lambda_definida}
\lambda \equiv \sqrt{\dfrac{\pi}{45}} \, \dfrac{g_{\star s}}{\sqrt{g_{\star}}} \, M_{\text{P}} \, m_{\text{DM}} \, .
\ee

When $x=0$ the WIMP scenario assumes $Y = Y_{\text{eq}}$ (the DM is in thermal equilibrium with the SM species). The expansion of the Universe decreases the rate of the interactions, that for the particular $2 \rightarrow 2$ case varies as
\be
\Gamma_{\text{an}} = n_{\text{eq}} \langle \sigma v \rangle \, .
\ee 
When $\Gamma_{\text{an}}  \backsimeq H$, the DM species decouples from the primordial plasma. This occurs at $x=x_\text{fo}$, the so-called \textit{freeze-out}. Under this hypothesis, it is easy to get an approximated value for $x_\text{fo}$
\be
 x_\text{fo} \backsimeq \log \left[ \sqrt{\dfrac{45}{32\pi^6}}M_{\text{P}} \, m_{\text{DM}} \sqrt{\dfrac{x_\text{fo}}{g_{\star}}} \langle \sigma v \rangle_\text{fo}  \right] \, ,
\ee 
where $g_{\star}$ must be evaluated at the freeze-out and $\langle \sigma v \rangle_\text{fo} \equiv \langle \sigma v \rangle\rvert_{x=x_\text{fo}} $. The usual values of $x_\text{fo}$ in the WIMP scenario are $x_\text{fo} \backsim 20 \, - \, 25$, practically regardless of the DM mass in the GeV-TeV region. After the decoupling, the rate of the interactions becomes negligible, \textit{freezing} the abundance. 
It is possible to take into account two approximations about the evolution of the yield. On the one hand, the DM before the freeze-out is in thermal equilibrium with the primordial plasma. Therefore, when $x \leq x_\text{fo}$, the DM yield is equal to the equilibrium yield:
\be
Y(x) = Y_{\text{eq}}(x).
\ee
On the other hand, after the decoupling, the rate of the interactions decreases to practically zero. This fact implies that the yield remains constant
\be
Y(x) = Y(x_\text{fo}),
\ee 
when $x > x_\text{fo}$.

The background temperature today is $T_\infty = 2.725 \, \pm \, 0.001$ K \cite{PhysRevD.98.030001} while the observed abundance is given by $Y_{\infty} = Y_{x \rightarrow \infty}  \backsimeq Y(x \gg x_\text{fo})$. The yield is related with the relic density via \cite{Gelmini:2010zh}
\be
\Omega_{\text{DM}} h^2 = 2.755 \times 10^8 \dfrac{m_{\text{DM}}}{\text{GeV}} Y_{\infty} \, .
\ee
As we have commented in Tab.~\ref{tabla_cosmo_param}, the value of the relic abundance that we observe nowadays is $\Omega_{\text{DM}} h^2 = 0.1121 \,  \pm \, 0.0056 $ \cite{Aghanim:2018eyx}.

\begin{figure}[htbp]
\centering
\includegraphics[width=140mm]{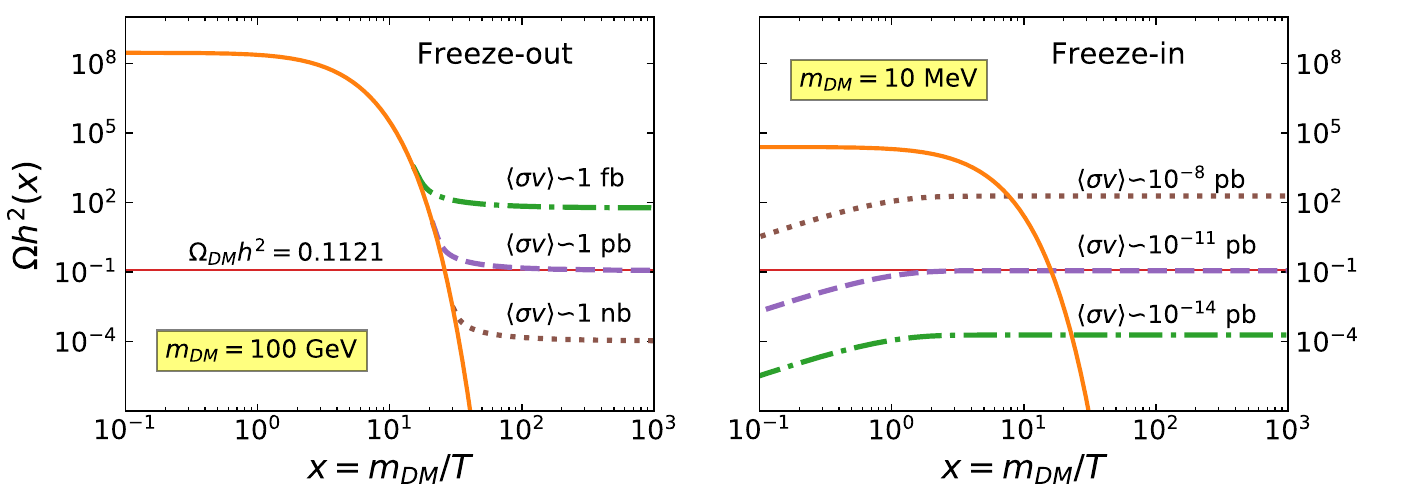}
\caption[Freeze-out and Freeze-in examples]{\it Different examples of the two thermal production mechanisms described in this Chapter. The plots show the abundance $\Omega h^2$ as a function of $x=m_{\text{DM}}/T$ for two representatives values of the DM mass. In both plots the red solid horizontal line shows the current DM abundance. Left plot: solution of the Boltzmann Equation \ref{BE_buena_simplificada} for different values of the thermal-averaged annihilation cross-section in the freeze-out regime $Y(x_0) = Y_{\text{eq}}(x_0)$. The orange solid line represents the abundance associated to the equilibrium distribution for a DM particles with $m_{\text{DM}} = 100 \, \text{GeV}$. The correct relic abundance is reached for $\langle \sigma v \rangle \sim 1 \, \text{pb}$. Right plot: solution for the freeze-in regime, $Y(x_0) = 0$. In this case, the orange solid line shows the abundance produced by the equilibrium distribution for $m_\text{DM} = 10 \, MeV$.}
 \label{fig:freeze_out_freeze_in}
\end{figure}

In order to solve Eq.~\ref{BE_buena_simplificada}, it is necessary to use different numerical techniques. The annihilation cross-section is, in general, too complicated to obtain an exact solution of the Boltzmann Equation. However, the special conditions of the evolution of the DM abundance in the WIMP scenario allow for an analytical approach to be found. As shown in the right panel of Fig.~\ref{fig:freeze_out_freeze_in}, after the freeze-out the DM yield remains constant, while the equilibrium yield falls. Therefore, it is reasonable to neglect $Y_{\text{eq}}$ for $x>x_\text{fo}$. Moreover, $Y = Y_{\text{eq}}$ before the freeze out. Under both assumptions
\bea
\dfrac{1}{Y_{\infty}} = \dfrac{1}{Y_{\text{fo}}} + \int^{\infty}_{x_\text{fo}} \dfrac{dx}{x^2} \, \lambda \, \langle \sigma v \rangle \, .
\eea
The dependence of $\lambda$ with $x$ comes from the variation of $g_{\star \, s}$ and $g_\star$ with the temperature. Taking $g_{\star \, s}$ and $g_\star$ at the value of $T_\text{fo}$ and neglecting $1/Y_\text{fo}$,
\bea
Y_{\infty} = \left(\lambda_\text{fo} \, \int^{\infty}_{x_\text{fo}}  \dfrac{dx}{x^2} \, \langle \sigma v \rangle\right)^{-1} \ \, ,
\label{approximacion_yield}
\eea
where $\lambda_\text{fo} \equiv \lambda \rvert_{x=x_\text{fo}}$.

The relative DM velocity in the WIMP scenario is small, fact which allows to write the cross-section in terms of the relative velocity between the two DM particles of the process. Under this assumption, we can expand the cross-section times $v$ as a power series of $v$:
\be
 \sigma v  \backsimeq a + b \, v^2 + c \, v^4 + \mathcal{O}(v^6) \, ,
\label{sigmavapprox}
\ee
where $v \backsimeq \sqrt{s/m_{\text{DM}}^2 - 4}$.
The different terms of the expansion represent the $s$-wave, $p$-wave and the $d$-wave contributions, respectively. From Eq.~\ref{thermal_average}, we thermally average the above expression \cite{Choi:2017mkk}
\be
 \langle \sigma v \rangle = \dfrac{x^{3/2}}{2\sqrt{\pi}}\int_0^{\infty} dv \, v^2 (\sigma v) e^{-xv^2/4} \backsimeq a + \dfrac{6 \, b}{x} + \dfrac{15 \, c}{x^2} + \mathcal{O}(1/x^3) \, 
 \label{thermal_simplificada_velocidades}
\ee

In the particular case where DM annihilation takes place in s-wave, the thermal-averaged cross-section remains constant and Eq.~\ref{approximacion_yield} gives a trivial solution for the DM yield:
\be
Y_{\infty} = \dfrac{x_\text{fo}}{\langle \sigma v \rangle \, \lambda_{\text{fo}}} \, .
\ee
Thanks to the relation between the yield and the relic density, it is easy to obtain:
\be
\label{aproximacion_omega}
\Omega_{\text{DM}} h^2 = \dfrac{1.04 \times 10^9 \, x_\text{fo}}{ \sqrt{g_{\star s}} \, M_{\text{P}} \, \langle \sigma v \rangle} \, \text{GeV}^{-1} \, .
\ee
Eq.~\ref{aproximacion_omega} assumes that the relativistic degrees of freedom in entropy and energy are equal for the typical decoupling temperatures in the WIMP scenario, $g_{\star} = g_{\star s} \backsimeq 80 - 100$. The exact value of $x_\text{fo}$ depends on the mass of the DM particle. However, $x_\text{fo} \backsim \, 20 - 30$ in the mass range for which it is possible to describe the DM relic abundance through freeze-out. Therefore, in that range, the relic abundance only depends on the annihilation cross-section, and not directly on the mass. 

In general, if $m_{\text{DM}} \in [10^{-1},10^4] \, \text{GeV}$, the value of $\langle \sigma v \rangle$ to obtain the correct relic density must be\footnote{The conversion factors between the cross-section units are $ 1 \, \text{GeV}^{-2} = 3.89 \times 10^8 \, \text{pb} = 1.17 \times 10^{-17} \text{cm}^3/\text{s}.$} $\langle \sigma v \rangle \sim 2 \times 10^{-26} cm^3/s$. Only small variations of $\langle \sigma v \rangle$ occurs in this mass range \cite{Steigman:2012nb}. However, the approximation described in Eq.~\ref{thermal_simplificada_velocidades} is not always valid. There are situations, close to a resonance for instance, where is more convenient solve Eq.~\ref{BE_buena_simplificada} numerically.

Left panel of Fig.~\ref{fig:freeze_out_freeze_in} shows the numerical solution of Eq.~\ref{BE_buena_simplificada} for different values of the thermal-averaged annihilation cross-section. Independently of the DM mass value, the current value of the DM abundance is reached for $\langle \sigma v \rangle \backsimeq 2 \times 10^{-26} cm^3/s \backsim 1 \, pb$. This value is pretty close to the typical electroweak interaction values: this fact receives the name of \textit{WIMP miracle}.

\section{Freeze-in: FIMP Dark Matter}
\label{Sec:freezein}

Sect.~\ref{Sec:freezeout} describes the case where the DM and the SM particles were in thermal equilibrium in the early Universe. However, when the visible and DM sectors interact with small couplings, $\backsim \, \mathcal{O}(10^{-7})$ \cite{Bernal:2017kxu}, the interaction rate is too small to reach thermal equilibrium. Therefore, the freeze-out mechanism cannot take place. In this particular case, the abundance of the DM in the early times was negligible: 
\be
Y(x_0) \backsimeq 0.
\ee
As long as the temperature is high enough, though, the interactions with the SM increases the yield. When the temperature of the Universe decreases, the possibility of generating more DM particles is reduced. As a consequence, the DM \textit{freezes-in} and the yield remains constant until today.
This kind of DM receives the name of FIMP\footnote{Despite the name was proposed in 2009, the idea was first studied in the late 1990's in Ref.~\cite{Chung:1998rq}.} (Feebly Interacting Massive Particles) \cite{Hall:2009bx}.
As we commented in Sect.~\ref{Sec:freezeout}, the freeze-out always occur for $x = m_{\text{DM}}/T \backsim \, 20 - 25$. This fact allows finding a typical value of the thermal-averaged cross-section to obtain the correct yield. Unlike WIMP, on the other hand, the FIMP scenario is highly dependent on initial conditions. Therefore, it is not possible to find a model-independent cross-section that reproduces the current relic abundance. 

In the freeze-in scenario, the term $Y^2$ in Eq.~\ref{BE_buena} can always be neglected with respect to the equilibrium one, because the DM never reaches the thermal equilibrium with the primordial bath. As a consequence, the DM abundance before the freeze-in is always smaller than the equilibrium abundance. The Boltzmann Equation for $2 \rightarrow 2$ processes can then be simplified as
\be
\dfrac{dY}{dx} = \dfrac{\lambda }{x^2} \langle \sigma v \rangle  \left(Y^2 - Y^2_{\text{eq}} \right) \backsimeq - \dfrac{\lambda \, \langle \sigma v \rangle}{x^2} Y_{\text{eq}}^2 \, .
\label{BEFIMP}
\ee
Unlike the freeze-out scenario, where the abundance of the DM decreases with the temperature, the yield of the FIMP increases through the evolution of the thermal history of the Universe, until the freeze-in. The difference between both frameworks produces a minus sign in Eq.~\ref{BEFIMP} with respect to Eq.~\ref{BE_buena}.
The right panel of Fig.~\ref{fig:freeze_out_freeze_in} shows the solution of the Boltzmann Equation in the freeze-in scenario for different constant values of the thermal average annihilation cross-section. In contrast to the freeze-out framework, in this case, the interaction must be much smaller to reach the correct relic abundance. In order to solve Eq.~\ref{BEFIMP} we take  $g_\star = g_{\star s} = 106.75$; this is a direct consequence of the ultra-relativistic nature of  DM species in the FIMP regime.

The dependence from the initial conditions makes useful to analyse the main aspects of the freeze-in Eq.~\ref{BE_buena} in terms of the temperature, instead of \footnote{Remember that there is a $-1$ factor between the Boltzmann Equation in terms of $T$ and $x$, $dx = -(m_{\text{DM}}/T^2) \, dT$.} $x=m_{\text{DM}}/T$.
Therefore, the Boltzmann Equation can be written as:
\be
\dfrac{dY}{dT} = - \dfrac{\gamma}{H \, \mathfrak{s} \, T} \left[ \left( \dfrac{Y}{Y_{\text{eq}}} \right)^2 - 1 \right] \backsimeq \dfrac{\gamma}{H \, \mathfrak{s} \, T}  \, ,
\label{FIMP_BE}
\ee
where $\gamma$ is the interaction rate density, defined for $a \rightarrow i + j $ processes as: 
\be
\gamma_{1 \rightarrow 2 }(T) =\dfrac{m_a^2 T}{2 \pi^2} K_1\left(\dfrac{m_a}{T}\right) \, ;
     \label{interaction_rate_density_1}
\ee
and, for $a + b \rightarrow i + j$ as:
\be
\gamma_{2 \rightarrow 2 }(T) = \dfrac{T}{64 \, \pi^4} \int_{s_{\text{min}}}^{\infty} ds \sqrt{s} \sigma_{R}(s) K_1(\sqrt{s}/T)  \, ,
\label{interaction_rate_density_2}
\ee
with $s_{\text{min}} = \text{Max}\left[ (m_a + m_b)^2 , (m_i + m_j)^2 \right]$. The reduced cross-section\footnote{The reduced cross-section represents the cross-section without the flux factors.} $\sigma_R(s)$ is related to the total annihilation cross-section $\sigma(s)$ via the K\" all\' en function\footnote{Defined as $\lambda (s,m_a^2,m_b^2) = \left[s-(m_a + m_b)^2\right]\left[s - (m_a - m_b)^2\right]$.}: 
\be
\sigma_R (s) = \dfrac{2 \lambda(s,m_a^2,m_b^2)}{s} \sigma (s) \, .
\ee
Eq.~\ref{interaction_rate_density_1} and Eq.~\ref{interaction_rate_density_2} show the interaction rate density for the two kind of processes that can contribute to the DM production in this scenario.

It is easy to integrate Eq.~\ref{FIMP_BE},
\be
Y(T) = \left(\dfrac{45}{4\pi^3} \right)^{3/2} \dfrac{2 \, M_{\text{P}}}{g_{\star s} \sqrt{g_\star}} \, \int^{T_{\text{rh}}}_{T} \dfrac{\gamma_{2 \rightarrow 2 }(T)}{T^6} \, ,
\ee
where $T_{\text{rh}}$ is the \textit{reheating temperature} which, in the approximation of a sudden decay of the \textit{inflaton}\footnote{Hypothetical scalar field responsible of the inflation in the very early universe \cite{Guth:1980zm}.}, corresponds to the maximal temperature reached by the primordial thermal bath. For the previous analysis to be valid, the DM has to be out of chemical equilibrium with the SM bath. One needs to guarantee, therefore, that the interaction rate density is $\gamma \ll  n_{\text{eq}} H$, which translates into a bound over the reheating temperature.
\chapter{Dark Matter Searches}
\label{sec:Detection}

In Chapter \ref{sec:DM} we explained the properties and characteristics of a viable Dark Matter candidate. In Chapter \ref{sec:termo} we analysed the WIMP and FIMP scenarios, explaining how the observed DM abundance is generated in the early Universe.
\begin{figure}[htbp]
\centering
\includegraphics[width=80mm]{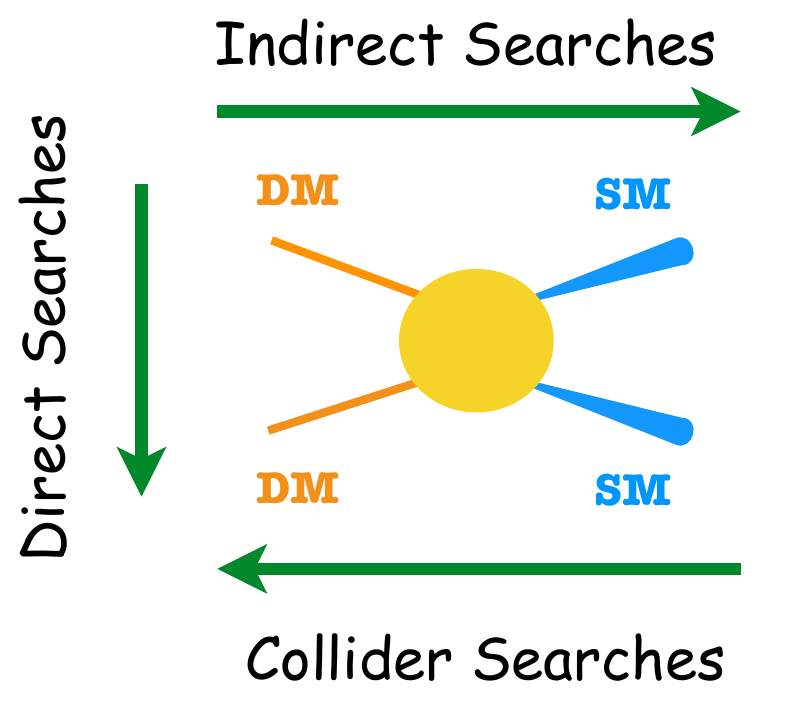}
\caption[Dark Matter detection techniques.]{\it Schematic representation of the different techniques of DM detection. Image taken from \cite{calore_ISAPP_2017}.}
 \label{fig:DM_searches}
\end{figure}
However, how can DM particles be detected? In the current particle physics landscape, it is possible to group the detection experiments into three categories: DM production at hadron colliders, such as the LHC \cite{Kahlhoefer:2017dnp}; Direct Detection (DD) of DM-nucleus scattering processes in ultra-sensitive low-background experiments \cite{Schumann:2019eaa}; and eventually, Indirect Detection (ID), or the detection of particles generated in Dark Matter annihilation processes \cite{Gaskins:2016cha}. Fig.~\ref{fig:DM_searches} shows a schematic representation of the three detection techniques. 
Although FIMPs can have similar properties to WIMPs, its coupling to the SM is much more suppressed, hence makes their detection more difficult.
Nevertheless, the detection techniques in both cases are the same.

Several experiments are currently trying to detect DM, and identify its nature and interactions beyond gravity. In this Chapter we analyse the current DM detection landscape, focusing on WIMP Dark Matter searches.

\section{Direct Detection}
\label{Sec:DD}

Nowadays, Direct Detection experiments represents one of the most promising detection techniques of BSM physics. The idea of the DD was first proposed by Mark W. Goodman and Edward Witten \cite{Goodman:1984dc}. Since the Dark Matter must be electrically neutral, the detection with electromagnetic techniques is impossible. However, the possibility of elastic scattering between the DM and atomic nucleis exist. As the Milky Way is surrounded by a Dark Matter halo, the knowledge about its different astrophysical properties allows us to predict the interaction rate of these DM particles with the detectors located on the Earth.

\subsection{Basic Ideas}

The first derivation of the different formulas presented in this section can be found in Ref.~\cite{Lewin:1995rx}. The following discussion is based on Refs.~\cite{Cerdeno:2010jj,Lin:2019uvt}. The most relevant quantity in DD experiments is the \textit{differential rate unit} (DRU) that represents the differential event rate, calculated per counts, kg, day and keV:
\be
\frac{dR}{dE_\text{NR}} = \frac{\rho_0}{m_\text{N} m_{\text{DM}}} \int_{v > v_{\text{min}}} v f(v) \frac{d\sigma}{dE_\text{NR}} (v,E_\text{NR}) dv \, ,
\label{Nuclear_recoil}
\ee
where $m_N$ is the nucleon mass, $E_{\text{NR}}$ is the nuclear recoil energy and $\sigma$ represents the DM-nucleon scattering cross-section. 

Typically, the DM Direct Detection experiments assumes that DM is distributed in an isotropic singular isothermal sphere, $\rho (r) \propto r^{-2}$. The local DM density is then $\rho_\odot\ = \rho\rvert_{r=R_\odot\ }$, where $R_\odot\ = 8.0 \pm 0.5 \, \text{Kpc}$ \cite{Reid:1993fx} is the approximate distance of the Sun from the Galactic Center. The most common value used in DD experiments for local DM density is given by\footnote{Note, however, that the most recent measurement finds $\rho_\odot = 0.46 \, \text{GeV}/\text{cm}^3$ \cite{Sivertsson:2017rkp}.} $\rho_\odot\ = 0.3 \, \text{GeV}/\text{cm}^3$ \cite{Read:2014qva}.

It is common to assume an isotropic and gaussian velocity distribution\footnote{Usually called Maxwellian.}
\be
f(\vec{v}) =\frac{1}{\sqrt{2\pi} \sigma_v} \, e^{|\vec{v}|^2/(2\sigma_v^2)} \, ,
\ee
where $\sigma_v$ represents the velocity dispersion in the DM \textit{gas}. This approximation is called \textit{Standard Halo Profile} and is supported by N-body simulations \cite{Kuhlen:2012ft}. The velocity dispersion is related to the total circular velocity of the galaxy by $\sigma_v = \sqrt{3/2} \, v_c$, where $v_c=220 \pm 20 \, \text{km}/\text{s}$ \cite{Kerr:1986hz}.

The integral is over all velocities above the minimal velocity required to induce a nuclear recoil. This velocity can be calculated with simple kinematics:
\be
v_\text{min} = \sqrt{\frac{m_\text{N} E_\text{NR}}{2\mu_{\text{DM-N}}^2}} \, ,
\ee
where $\mu_{\text{DM-N}} \equiv m_{N} m_\text{DM}/(m_{N} + m_\text{DM})$ is the reduced mass of the DM and nucleus system. When the velocity is larger than the escape velocity, $v>v_{\text{esc}} = 544 \, \text{km}/\text{s}$ \cite{Smith:2006ym}, the Dark Matter particles escapes from the Dark Matter halo. Therefore, integrating Eq.~\ref{Nuclear_recoil} up to the escape velocity is a good approximation.

The total event rate, calculated per kilogram and per day, can be obtained integrating Eq.~\ref{Nuclear_recoil} in the range of the possible nuclear recoil energies,
\be
R = \int_{E_\text{NR,low}}^{E_\text{NR,high}} dE_\text{NR} \epsilon(E_\text{NR}) \frac{dR}{dE_\text{NR}} \, ,
\label{total_rate}
\ee
where $\epsilon(E_\text{NR})$ represents the efficiency of the detector. The maximal recoil energy is constraint by the kinematics:
\be
E_\text{NR, high} = \frac{2 \mu_{\text{DM-N}} v_\text{esc}^2}{m_\text{N}} \, ,
\ee
while the $E_\text{NR, low}$ represents the threshold of the detector.

\subsection{DM-Nucleus Cross-Section}

Eq.~\ref{total_rate} gives the rate of the interaction per day and per kilogram of DM particles with the detector. All information about the interaction between the nucleus and the DM is given by the DM-nucleus cross-section,
\be
\frac{d\sigma}{dE_\text{NR}} = \left(\frac{d\sigma}{dE_\text{NR}}\right)_\text{SI} + \left(\frac{d\sigma}{dE_\text{NR}}\right)_\text{SD} \, ,
\ee
that consists of two contributions: Spin-Dependent (SD), the contributions that arise from the DM couplings to the quark axial-vector current, and the Spin-Independent (SI), that comes from the scalar and vector couplings in the Lagrangian.

The DM-nucleus cross-section depends on the DM-nucleon cross-section, that encodes the microscopic information of the collision. The small momentum transfer from the DM to the nucleus, $q = \sqrt{2m_\text{N}E_{\text{NR}}}$, allows us to obtain an expression that relates the microscopic and the macroscopic cross-sections.

\subsubsection{Spin-Dependent Cross-Section}

The SD cross-section depends on the spin of the DM and the angular momentum of the nucleus. For a fermionic\footnote{The expression for the spin-1 DM can be found in Ref.~\cite{Barger:2008qd}.} DM the expression is given by \cite{Cerdeno:2010jj}
\be
\label{spin_dependent_CS}
\left(\frac{d\sigma}{dE_\text{NR}}\right)_\text{SD} = \frac{16\, G_F^2\, m_\text{N}}{\pi v^2} \, \frac{J+1}{J} \left(a_p \langle S_p \rangle + a_n \langle S_n \rangle\right)^2 \frac{S(E_\text{NR})}{S(0)} \, ,
\ee
where $S(E_\text{NR})$ and $S(0)$ are the form factors, $\langle S_{n,p} \rangle$ are the expectation values of the spin content of the neutron and proton (that can be determined experimentally) and $J$ is the total angular momentum of the nucleus. The coefficients $a_p$ and $a_n$ are given by
\bea
\left \{
\begin{array}{lll}
     a_p = \sum_{q = u, d, s} \dfrac{\alpha_q^\text{A}}{\sqrt{2}G_\text{F}}\Delta_q^p \, , \\[8pt]
	 a_n = \sum_{q = u, d, s} \dfrac{\alpha_q^\text{A}}{\sqrt{2}G_\text{F}}\Delta_q^n \, .
\end{array}
\right .
\eea
The different $\alpha^\text{A}$ are the couplings of the DM to the axial-vector quark currents, which are given by the model. On the other hand, the $\Delta_q^{n,p}$  encode the information about the quark spin content of the nucleon and are proportional to $\left\langle N| \bar{q}\gamma_\mu \gamma_5 q |N \right\rangle$. These coefficients are usually calculated with two strategies: lattice QCD \cite{QCDSF:2011aa} and experimental nuclear physics techniques \cite{Alekseev:2010ub,Alekseev:2007vi,Aidala:2012mv}.
The values of $\Delta_q^{n,p}$ are summarized in Tab.~\ref{coeficientes_SD}.

\begin{table}
\centering
\begin{tabular}{c c c c}
    \hline
    \bfseries Nucleon  & \bfseries $\Delta_u$ & \bfseries $\Delta_d$ & \bfseries $\Delta_s$   \\ 
    \hline
     Neutrons & $-0.46(4)$ & $0.80(3)$ & $-0.12(8)$  \\
     Protons   & $0.80(3)$ & $-0.46(4)$ & $-0.12(8)$ \\
    \bottomrule
\end{tabular}
\caption[Matrix element of the axial-vector current in a nucleon.]{Matrix element of the axial-vector current in a nucleon. The first row represents $\Delta_q^n$ while the second represents $\Delta_q^p$. Data taken from \cite{Hill:2014yxa}.}
\label{coeficientes_SD}
\end{table}

\subsubsection{Spin-Independent Cross-Section}

In the zero-momentum transfer approximation \cite{Gresham:2014vja} Spin Independent contribution is independent of the DM and nucleus angular momentum. The expression is then given by:
\be
\left(\frac{d\sigma}{dE_\text{NR}}\right)_\text{SI} = \frac{2 m_\text{N}}{\pi v^2} \left( \left[Z \, f^p + (A - Z)\, f^n \, \right]^2 + \frac{B_\text{N}^2}{256} \right) F^2(E_\text{NR}) \, ,
\label{SI_ecuacion}
\ee
where $B_\text{N} \equiv \alpha_u^\text{V} (A + Z ) \, + \, \alpha_d^\text{V} (2A - Z)$ is the vector-vector contribution with $\alpha_{u,d}^\text{V}$ the vector-vector couplings between the DM and the $u$ and $d$ quarks, $(A,Z)$ the number of neutrons and protons of the nucleus and $F^2(E_\text{NR})$ another experimental form factor \cite{Engel:1991wq,Helm:1956zz}.

Finally, the $f^{p,n}$ quantities that appear in Eq.~\ref{SI_ecuacion} are
\be
\label{ecuacion_coeficientes_DD}
\frac{f^{p,n}}{m_{p,n}} = \sum_{q=u,d,s} \frac{\alpha_q^S}{m_q} f^p_{Tq} + \frac{2}{27} f^p_{TG}  \sum_{q=u,d,s} \frac{\alpha_q^S}{m_q} \, .
\ee
The scalar-scalar coupling between the DM and the quarks is given by $\alpha_q^\text{S}$. The coefficients $f^{p,n}_{Tq}$ encode the nucleon matrix elements and represent the contribution of each light quark to the nucleon. These coefficients are defined as:
\be
f^{p,n}_{Tq} = \frac{m_q}{m_{p,n}} \left\langle N| \bar{q} q |N \right\rangle \, ,
\ee
and must be calculated using Lattice QCD or experimentally, using measurements of the pion-nucleon sigma term \cite{Ellis:2008hf}. Finally, $f^{p,n}_{TG}$ represent the gluon contribution to the nucleon mass and is defined as
\be
f^{p,n}_{TG} = 1 \, - \sum_{q=u,d,s}  f^{p,n}_{Tq}.
\ee
These different constants are summarized in Tab.~\ref{coeficientes_SI}.

\begin{table}
\centering
\begin{tabular}{c c c c c}
    \hline
    \bfseries Nucleon & \bfseries $f_{TG}$ & \bfseries $f_{Tu}$ & \bfseries $f_{Td}$ & \bfseries $f_{Ts}$   \\ 
    \hline
     Neutrons & $0.910(20)$ & $0.013(3)$ & $0.040(10)$ & $0.037(17)$  \\
     Protons   & $0.917(19)$ & $0.018(5)$ & $0.027(7)$ & $0.037(17)$  \\
    \bottomrule
\end{tabular}
\caption[Contributions of the light quarks to the mass of the neutron and proton.]{Contributions of the light quarks to the mass of the neutron and proton. The numbers in parentheses are the one-sigma uncertainty. Data taken from \cite{Ellis:2018dmb}.}
\label{coeficientes_SI}
\end{table}

For a detailed explanation about the contributions of the light quarks to the mass and the matrix elements of the axial-vector currents see Refs.~\cite{Bishara:2017pfq, deAustri:2013saa}.

\subsection{Current Status of Direct Detection Landscape}

The search for Dark Matter has become one of the great milestones of high-energy physics. However, despite the efforts of many experimental groups, no conclusive direct detection of Dark Matter has ever been made, neither of WIMP particles nor of any other form of Dark Matter\footnote{There are some exceptions, such as the case of DAMA/LIBRA experiment, which obtained data compatible with the existence of WIMP particles at specific values of the mass, such as $m_\text{DM} = 7 - 12$ GeV \cite{Bernabei:2008yi,Bernabei:2010mq}. The current statistical significance of DAMA/LIBRA signal reaches the $12\sigma$ level. However, the annual modulation of the number of detection events found by DAMA/LIBRA is under debate since other experiments, as the experiments like LUX or Xenon1T, do not report any excess in that mass region.}. Therefore, currently we only have restrictive experimental bounds on theoretical models.

\begin{table}
\centering
\begin{tabular}{c c c c c}
    \hline
    \bfseries Experiment  & \bfseries Target & \bfseries Mass [Kg] & \bfseries Laboratory & \bfseries Ref.  \\ 
    \hline
     ANAIS-112 & NaI & $112$ & Canfranc & \cite{Amare:2019jul}  \\
     CDEX-10   & Ge & $10$ & CJPL  & \cite{Jiang:2018pic} \\
     CDMSLite  & Ge & $1.4$ & Soudan & \cite{Agnese:2017jvy}  \\
     COSINE-100  & NaI & $106$ & YangYang & \cite{Adhikari:2017esn} \\
     CRESST-II  & $\text{CaWO}_4$ & $5$ & LNGS & \cite{Angloher:2015ewa} \\
     CRESST-III  & $\text{CaWO}_4$ & $0.024$ & LNGS & \cite{Abdelhameed:2019hmk} \\
     DAMA/LIBRA-II  & NaI & $250$ & LNGS & \cite{Bernabei:2018yyw} \\
     Darkside-50  & Ar & $46$ & LNGS & \cite{PhysRevLett.121.081307} \\
     DEAP-3600 & Ar & $3300$ & SNOLAB & \cite{Ajaj:2019imk} \\
     DRIFT-II  & $\text{CF}_4$ & $0.14$ & Boulby & \cite{Battat:2016xxe} \\
     EDELWEISS  & Ge & $20$ & LSM & \cite{Hehn:2016nll} \\
     LUX  & Xe & $250$ & SURF & \cite{Akerib:2016vxi} \\
     NEWS-G & Ne & $0.283$ & SNOLAB & \cite{Arnaud:2017bjh} \\
     PandaX-II  & Xe & $580$ & CJPL & \cite{PhysRevLett.119.181302} \\
     PICASSO  & $\text{C}_4\text{F}_10$ & $3.0$ & SNOLAB & \cite{Behnke:2016lsk} \\
     PICO-60  & $\text{C}_3\text{F}_8$ & $52$ & SNOLAB & \cite{Amole:2019fdf} \\
     SENSEI  & Si & $9.5 \times 10^{-5}$ & FNAL & \cite{PhysRevLett.122.161801} \\
     SuperCDMS  & Si & $9.3 \times 10^{-4}$ & SNOLAB & \cite{PhysRevLett.121.051301} \\
     XENON-100  & Xe & $62$ & LNGS & \cite{PhysRevD.94.122001} \\
     XENON-1T  & Xe & $1995$ & LNGS & \cite{PhysRevLett.121.111302} \\
     XMASS  & Xe & $832$ & Kamioka & \cite{XMASS:2018bid} \\
    \bottomrule
\end{tabular}
\caption[Direct Detection experimental landscape.]{Current Direct Detection experimental landscape in alphabetic order. The table shows the target, mass in kg and the place of a great part of the current DD experiments. Not all current experiments are included. The different data are extracted from Ref.~\cite{Schumann:2019eaa}.}
\label{DD_experiments}
\end{table}

The first DD experiment started in 1987: Ultralow Background Germanium Spectrometer, with $0.72$ kg of high purity germanium crystal \cite{Ahlen:1987mn}. Since then, several experiments have appeared, improving the limits on DD. Nowadays, the landscape of DD is composed of a great number of experiments. The most common are the experiments that use noble gases, like xenon or argon, as a target. Tab.~\ref{DD_experiments} summarizes the most important of them, with its different properties.

\begin{figure}[htbp]
\centering
\includegraphics[width=130mm]{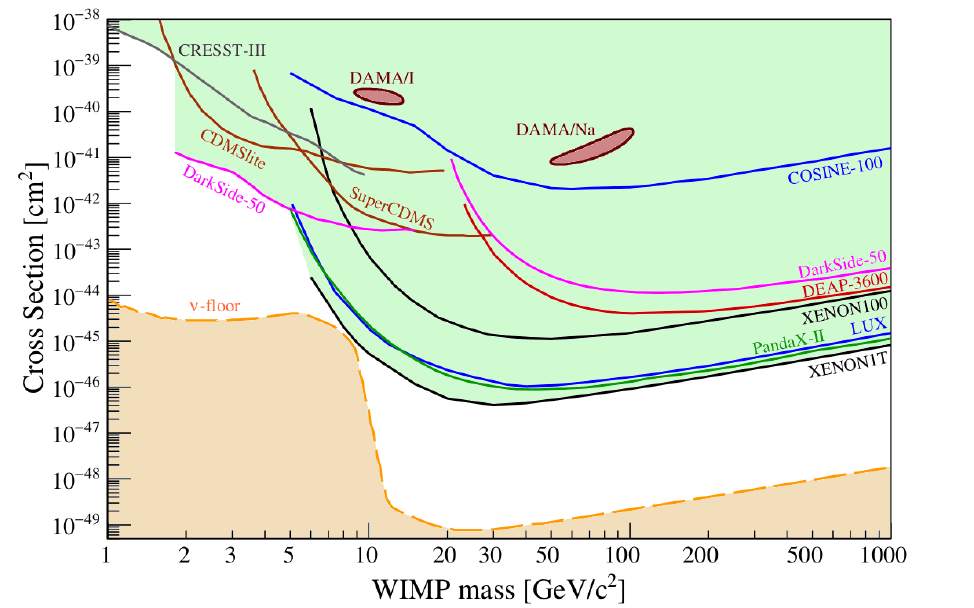}
\caption[Bounds from Dark Matter Direct Detection Spin Independent experiments]{\it Bounds from Dark Matter Direct Detection SI experiments. The space above the different lines is excluded at 90\% confidence level. The two contour red regions represent the DM observation reported by DAMA/LIBRA experiment. The yellow region represents the neutrino floor \cite{Billard:2013qya}, the parameter space region where the detectors should detect the coherent neutrino-nucleus scattering (CNNS). Image taken from Ref.~\cite{Schumann:2019eaa}.}
 \label{fig:direct_detection_bounds}
\end{figure}

The different DD experiments represent an important improvement in the detection of the Dark Matter particles, placing strong bounds. On most models, the strongest bounds come from the SI cross-section. Fig.~\ref{fig:direct_detection_bounds} shows some of this current limits. 

\section{Indirect Detection}
\label{Sec:ID}

Indirect Detection experiments try to observe the SM products of the annihilation of stable particles in the cosmic rays fluxes. In general, it is possible to distinguish between three kinds of detectable fluxes: charged particles, like electrons and positrons, protons and antiprotons, deuterium and antideuterium; photons and, finally, neutrino fluxes. Since the 1970's, several works appeared trying to find DM signatures. First publications are: in $\gamma$-rays \cite{1978ApJ223.1015G,1978ApJ223.1032S,Zeldovich:1980st,Ellis:1988qp}, in positrons fluxes \cite{Ellis:1988qp,PhysRevLett.53.624,Rudaz:1987ry,Stecker:1988fx}, in antiproton fluxes \cite{Ellis:1988qp,PhysRevLett.53.624,Rudaz:1987ry,Stecker:1988fx,PhysRevLett.55.2622} and in antideuterons fluxes \cite{Donato:1999gy,Baer:2005tw,Donato:2008yx}. There are several reviews about this topic. In this Thesis we have used Refs.~\cite{Cirelli:2010xx,Gaskins:2016cha}.

Information on stable particle fluxes reaching the Earth can be used to constrain DM models under specific conditions. In general, in all BSM models, the DM can be annihilated into SM particles, resulting, in its final states, in stable particles. If these processes are possible, the signature of the DM annihilations remain in the cosmic rays detected at the Earth. The ID tries to trace the footsteps of these DM annihilations in the stable particle fluxes detected in the experiments.
However, not all DM annihilations leave evidences in the cosmic rays. If the annihilation cross-section depends on the relative DM velocity, the contribution of these processes to the stable particle flux will be negligible, since the relative velocity of the DM particles today is small. This situation takes place when the angular momentum of the collision is $l>0$. According to the velocity dependence, the different annihilation cross-section terms receives the names summarized in Tab.~\ref{waves}. 

In general, ID is possible in processes that take place in \textit{s-wave}, where the annihilation cross-section is not suppressed by the DM velocity. However, the velocity suppression only affects the indirect signals today. This fact is compatible with the DM production in the early Universe. Indeed, the DM production takes place when DM is relativistic and, as a consequence, the velocity suppression does not prevent reaching the current abundance \cite{Gondolo:1990dk}. 

\begin{table}
\centering
\begin{tabular}{c c c}
    \hline
    \bfseries Name  & \bfseries $l$ & \bfseries  Velocity dependence of $\langle \sigma v \rangle $ \\ 
    \hline
     s-wave & $0$ & $-$ \\
     p-wave & $1$ & $\langle \sigma v \rangle \propto v^2$  \\
     d-wave & $2$ & $\langle \sigma v \rangle \propto v^4$ \\
     f-wave & $3$ & $\langle \sigma v \rangle \propto v^6$  \\
    \bottomrule
\end{tabular}
\caption[Velocity dependence of the cross-section according to the collision angular momentum.]{Velocity dependence of the cross-section according to the collision angular momentum $l$.}
\label{waves}
\end{table}  

\subsection{Hadrons, leptons and photons spectra}

\begin{figure}[htbp]
\centering
\includegraphics[width=130mm]{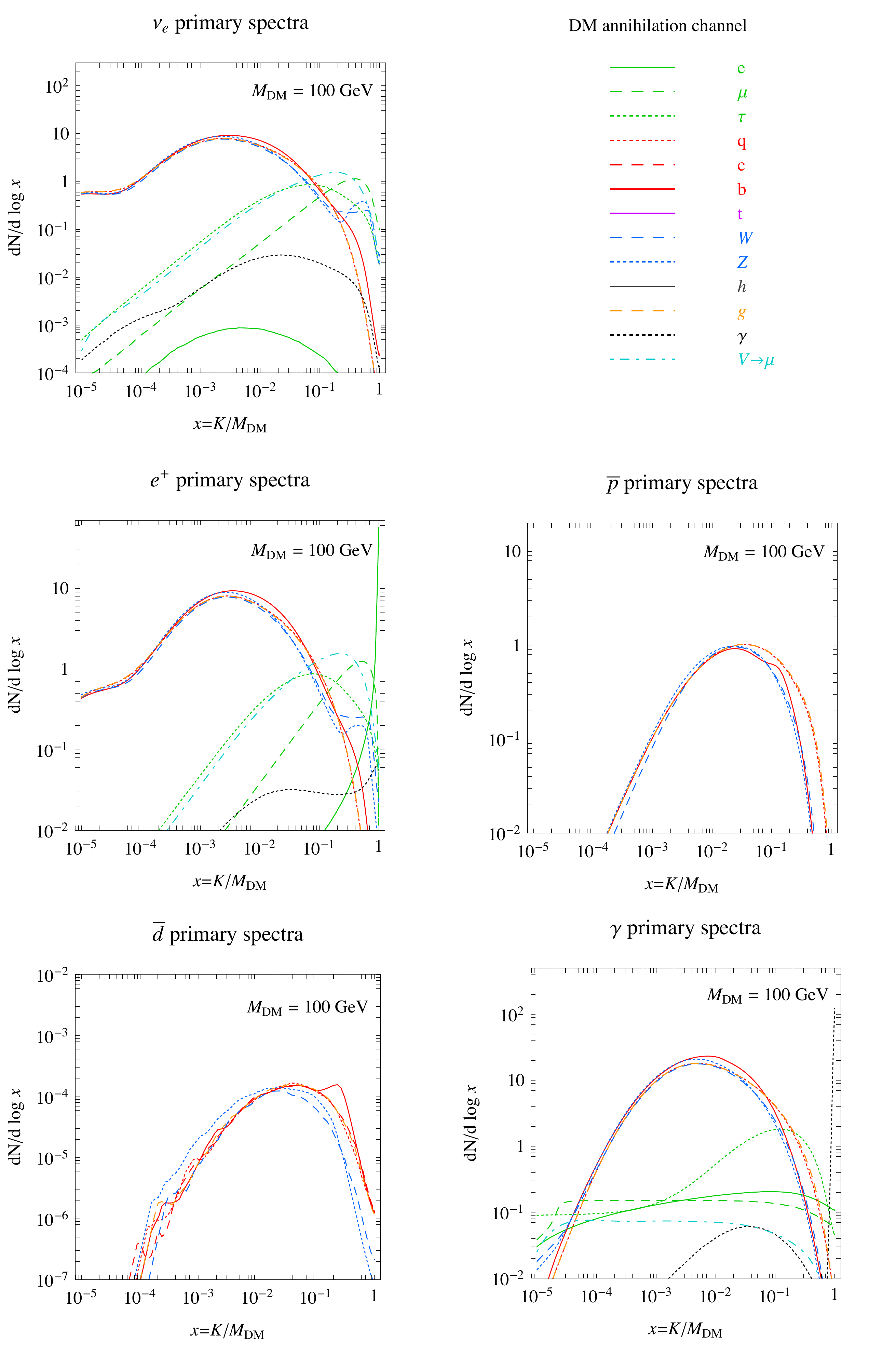}
\caption[SM spectra generated by DM ]{\it Different examples of SM particles fluxes (from top to bottom, left to right, photons, neutrinos, positrons, antideuterons and antiprotons, respectively) produced by annihilation of two DM particles with $m_\text{DM}  = 100 $ GeV. In all plots $K$ represents the kinetic energy of the final stable states. This examples are taken from Ref.~\cite{Cirelli:2010xx}.}
 \label{fig:espectros}
\end{figure}

For a given particle physics model, the spectrum of SM particles is not easily calculated. Nowadays, the most efficient way to obtain the different fluxes is to use a specific software, such as those in Refs.~\cite{Alloul_2014,Sj_strand_2008,Alwall_2014,Staub_2015,Porod_2003}. Fig.~\ref{fig:espectros} shows different examples of spectra generated by annihilation of DM into photons, neutrinos, positrons antideuterons and antiprotons.

In general, the different processes that generate the final SM particle spectrum do not occur close to the Earth, where the detection is produced. As a consequence, it is necessary to propagate the spectrum given by our BSM model. The propagation is strongly dependent on the particle properties and of the cosmic-ray model employed. Indeed, there are different propagation models for photons, neutrinos, positrons, antiprotons and antideuterons. A very useful package to this task is \textit{PPPC4DMID} and can be found in Ref.~\cite{Cirelli:2010xx}.

\subsection{Propagation models of charged particles}

In this section, we will provide a general benchmark for the propagation of the spectra of differently charged particles. The most common ones are the antiprotons and positrons (both cases will be analysed). For a complete description of deuterium propagation models see Refs.~\cite{Donato:2008yx,Brauninger:2009pe,Kadastik:2009ts}.

\subsubsection{Electrons and positrons}
\label{sec:posi}

The same formalism is used for electrons and positrons. Therefore in the following expressions we will not distinguish between them. The evolution of the electrons spectrum $f_e \equiv dN_e/dE$ along the galaxy obeys the diffusion loss equation
\be
\ -\nabla \left[ \mathcal{K}(\vec{x},E) \, \nabla f_e \right] - \frac{\partial}{\partial E} \left[ b(E) \, f_e \right] = Q(\vec{x},E) \, ,
\label{dif_elec}
\ee
where $Q(\vec{x},E)$ takes into account of all sources, $\mathcal{K}(\vec{x},E)$ is the diffusion coefficient function and $b(E)$ the energy loss coefficient function, that describes the energy lost by charged particles.
In general, the diffusion coefficient depends on the position. However, in order to obtain a semi-analytical solution of Eq.~\ref{dif_elec}, the spatial dependence is usually neglected in the literature: $\mathcal{K}(E) = \mathcal{K}_0 \epsilon^\delta$, where $\epsilon \equiv E/\text{GeV}$. In the same way, for high energy $b(\vec{x},E) \backsimeq b(E) \propto E^2$ \cite{Meade:2009iu}.

\begin{table}
\centering
\begin{tabular}{c c c}
    \hline
    \bfseries Model  & \bfseries $\delta$ & \bfseries  $\mathcal{K}_0 \, [\text{kpc}^2/\text{Myr}]$ \\ 
    \hline
     Min & $0.55$ & $0.00595$ \\
     Med & $0.70$ & $0.0112$ \\
     Max & $0.46$ & $0.0765$  \\
    \bottomrule
\end{tabular}
\caption[Propagation coefficients of electrons and positrons through the galaxy.]{Propagation Coefficients of electrons and positrons through the galaxy. The different data are extracted from Ref.~\cite{Delahaye:2007fr}.}
\label{positron_propagation_tabla}
\end{table}

The propagation model is defined by the constants $\mathcal{K}_0$ and $\delta$. Moreover, Eq.~\ref{dif_elec} is usually solved in a diffusion region defined by a cylinder that sandwiches the galactic plane. In Tab.~\ref{positron_propagation_tabla} we summarize the three most used models in the literature.
The electron/positron flux $\Phi_e$ produced by DM annihilation and decay can be obtained by solving Eq.~\ref{dif_elec}:
\bea
\left \{
\begin{array}{llll}
\dfrac{d\Phi_e}{dE}(E,\vec{x}) = \dfrac{v_e}{8 \pi b(E,\vec{x})} \left(\dfrac{\rho_\odot\ }{m_\text{DM}}\right)^2\displaystyle \sum_f \langle \sigma v \rangle_f \displaystyle \int_E^{m_\text{DM}} dE_s \dfrac{dN_e^f}{dE_s} I(E,E_s,\vec{x}) \, , \nonumber \\[8pt]
\dfrac{d\Phi_e}{dE}(E,\vec{x}) = \dfrac{v_e}{4 \pi b(E,\vec{x})} \left(\dfrac{\rho_\odot\ }{m_\text{DM}}\right)\displaystyle \sum_f \Gamma_f \displaystyle \int_E^{m_\text{DM}/2} dE_s \dfrac{dN_e^f}{dE_s} I(E,E_s,\vec{x}) \, , \nonumber \\
\end{array}
\right .\\
\eea
where $v_e$ is the velocity of the electrons, $E_s$ is the particle energy at the production point and $I(E,E_s,\vec{x})$ is the generalized halo function, that encodes all astrophysical information of the propagation. Both $b(E,\vec{x})$ and $I(E,E_s,\vec{x})$ can be calculated for the three models described in Tab.~\ref{positron_propagation_tabla} with the \textit{PPPC4DMID} package \cite{Cirelli:2010xx}.

\subsubsection{Protons and Antiprotons}
\label{sec:antiprot}

Protons and antiprotons are charged particles as positrons and electrons. Therefore, their propagation is defined by differential equation similar to that given Sect.~\ref{sec:posi}. However, it is necessary to include new terms and effects in the model. It is common to find in the literature the expression in cylindrical coordinates $(r,z)$, where $z$ is the distance from the Earth to the source. The equation is given by
\be
-\mathcal{K}(K) \cdot \nabla^2 f_p + \frac{\partial}{\partial z} \left[ \text{sign}(z) f_p V_\text{conv} \right] = Q - 2h	\delta(z) \Gamma_\text{ann} f_p
\label{dif_antipro}
\ee
where $f_p \equiv dN_p/dE$, $K$ is the kinetic energy of protons/antiprotons and $\mathcal{K} =\mathcal{K}_0 \beta \, (p/GeV)^\gamma $ is the diffusion function, with $p = \sqrt{K^2 + 2m_pK}$ the momentum and $\beta = v_p/c$ the velocity of the proton/antiproton. 

There are two extra terms in Eq.~\ref{dif_antipro} with respect to Eq.~\ref{dif_elec}. The first one, $V_\text{conv}$, is the convective wind, assumed to be constant and directed outward from the galactic plane. The value of $V_\text{conv}$, such as $\delta$ and $\mathcal{K}_0$, is fixed by the model. The second new term takes into account the annihilation of protons/antiprotons confined in the galactic plane, that has $h = 0.1 \, \text{kpc}$ of thickness (see Ref.~\cite{Hisano:2005ec} for more details). Tab.~\ref{antiproton_propagation_tabla} summarizes the three most common models for proton/antiproton propagation.

\begin{table}
\centering
\begin{tabular}{c c c c}
    \hline
    \bfseries Model  & \bfseries $\delta$ & \bfseries  $\mathcal{K}_0 \, [\text{kpc}^2/\text{Myr}]$ & \bfseries $V_\text{conv} \, [\text{km}/\text{s}]$  \\ 
    \hline
     Min & $0.85$ & $0.0016$ & $13.5$  \\
     Med & $0.70$ & $0.0112$ & $11$  \\
     Max & $0.46$ & $0.0765$ & $5$  \\
    \bottomrule
\end{tabular}
\caption[Propagation coefficients of protons and antiprotons through the galaxy.]{Propagation coefficients of protons and antiprotons through the galaxy. The different data are extracted from Ref.~\cite{Donato:2003xg}}
\label{antiproton_propagation_tabla}
\end{table}

Assuming steady state conditions, the first term in Eq.~\ref{dif_antipro} can be neglected, and the equation can be solved analytically \cite{Bottino:1998tw,Chardonnet:1996ca,Maurin:2001sj,Maurin:2002ua}. Then, the proton/antiproton flux $\Phi_p$ due to the DM annihilation and decay is given by
\bea
\left \{
\begin{array}{llll}
\dfrac{d\Phi_p}{dE}(K) &=& \dfrac{v_p}{8 \pi } \left(\dfrac{\rho_\odot\ }{m_\text{DM}}\right)^2 R(K)\displaystyle \sum_f \langle \sigma v \rangle_f \dfrac{dN_p^f}{dK}  \, , \\[8pt]
\dfrac{d\Phi_p}{dE}(K) &=& \dfrac{v_p}{4 \pi } \left(\dfrac{\rho_\odot\ }{m_\text{DM}}\right) R(K)\displaystyle \sum_f \Gamma_f \dfrac{dN_p^f}{dK}  \, , 
\end{array}
\right .
\eea
where $R(K)$ encodes all astrophysical information about the propagation. This function can be approximated with an accuracy better than 6\% as
\be
\log_{10}\left(\frac{K}{\text{Myr}} \right) = a_0 + a_1 \, \kappa + a_2 \, \kappa^2 + a_3 \, \kappa^3 + a_4 \, \kappa^4 + a_5 \, \kappa^5     \, ,
\ee
where $\kappa = \log_{10} \left(K/\text{GeV}\right)$. The $a_i$ coefficients depends on the propagation model (Min, Med, Max) and the DM density profile (the values can be found in Ref.~\cite{Cirelli:2010xx}).

Since the mass of the protons/antiprotons is larger than the electron/positron mass, it is necessary to take into account the effect of the solar modulation. A complete description of this effect in the cosmic rays can be found in Ref.~\cite{Gleeson:1968zza}.

\subsection{Propagation of Uncharged Particles}


Two fluxes of uncharged particles arrive at Earth: neutrinos and photons. Regarding the neutrino flux, the most significant contribution arriving at Earth is generated in the Sun (\textit{solar neutrinos}) or in the Earth's atmosphere (\textit{atmospheric neutrinos}). The weak interaction of the neutrinos with the rest of the particles makes easier their propagation and larger their mean path. However,  it is necessary to take into account different effects, such is the case for neutrino oscillations. A complete description of the subtleties of the propagations of neutrinos can be found in Ref.~\cite{Cirelli:2005gh}.

The other neutral particles that reach the earth are $\gamma$-rays. The differential photon flux produced by DM annihilations that arrives at Earth from a window with size $\Delta \Omega$, is given by \cite{Cirelli:2010xx}
\be
\frac{d\Phi_\gamma}{dE}(E)  = \frac{J}{8\pi m_\text{DM}^2} \sum_f \langle \sigma v \rangle_f \frac{N^f_\gamma}{dE} (E) \, ,
\ee
where 
\be
J = \int_{\Delta\Omega} d\Omega \int \rho^2(s) ds \, 
\ee
is called \textit{$J$-factor} and it encodes all astrophysical information.
In other words, the $J$-factor is the integration of the DM profile along the line of sight.
If the $\gamma$-rays are generated through DM decay, the flux takes the form
\be
\frac{d\Phi_\gamma}{dE}(E)  = \frac{J}{4\pi m_\text{DM}} \sum_f \Gamma_f \frac{N^f_\gamma}{dE} (E) \, ,
\ee
with
\be
J = \int_{\Delta\Omega} d\Omega \int \rho(s) ds \, .
\ee
\subsection{Experimental status of indirect detection: Landscape and limits}

The current landscape of ID experiments provides a good source of constraints to the BSM models that include Dark Matter candidates. In this Section, we try to give a general overview of the experimental status.

\subsubsection{$\gamma$-rays searches}
\label{sec:gammaraysearches}

The $\gamma$-ray search experiments represent the most promising source of bounds in ID. The observation of photons coming from Dwarf Spheroidal Galaxies can be used to set limits in different BSM models. DSphs are objects dominated by DM and, thanks to their high latitude, these astronomical objects suffer from low diffuse $\gamma$-ray emission.

In the last years, Fermi-LAT experiment\footnote{The Fermi Large Area Telescope.} has analysed the photon flux of 15 different dSphs. In general, the Fermi collaboration has studied photons with energies between 500 MeV and 500 GeV \cite{Fermi-LAT:2016uux,Ackermann:2015zua}. It is easy to analyse the bounds imposed on some BSM models by the dSphs using \textit{gamLike v.1.0} \cite{Workgroup:2017lvb}.

Although the dSphs are the strongest source of bounds, different advances are being made in the $\gamma$-rays coming from the GC and other galaxy groups \cite{Lisanti:2017qlb,Chang:2018bpt}.

\subsubsection{Charged particle searches}
\label{sec:charged}


Several experiments have reported the observation of fluxes for positrons and antiprotons. PAMELA has analysed the positron flux coming from the centre of our galaxy \cite{Adriani_2011}, whereas AMS-02 did the same analysis but additionally observed the antiproton flux \cite{PhysRevLett.117.091103,PhysRevLett.113.221102,PhysRevLett.113.121101}. Some DM models predict extra positrons and antiprotons that increase the fluxes predicted by the SM. The SM+BSM flux can be studied and compared using different backgrounds model, allowing to set bounds in specific regions of the parameter space. 
In the last years, an excess of $\backsimeq 10-20$ GeV cosmic-ray antiprotons has been reported by several authors in the data taken by AMS-02 experiment \cite{Cuoco:2016eej,Cui:2016ppb,Bringmann:2014lpa,Cirelli:2014lwa,Hooper:2014ysa}. This excess, with a $4.7\sigma$ of significance with respect to the background signal \cite{Cuoco:2016eej}, has been studied as a DM prove by several authors, some examples can be found in Refs.~\cite{Cholis:2019ejx,Cui:2018nlm}.

In general, the bounds imposed by charged particles are always worse than the bounds from $\gamma$-rays or Direct Detection. Their propagation models have many uncertainties and this makes difficult to set robust constraints.

\subsubsection{Neutrino searches}

Most of the neutrinos that reach the Earth are produced in the Sun or in the Earth's atmosphere. DM could be captured by the Sun and annihilate into neutrinos, which would then be detected by different neutrino experiments giving an excess with respect to solar neutrinos due to nuclear reactions in the Sun. However, this is not the only neutrino source: fluxes coming from the GC are looked for, too. Both neutrino fluxes can be used to constrain DM models.

The weak interaction of neutrinos hinders their detection. However, there are several neutrino experiments on Earth making possible the detection of these elusive particles. Nowadays, the two most important neutrino telescopes are KM3Net and IceCube\footnote{KM3Net is located $2.5$ km under the Mediterranean Sea off the coast of Toulon, France (in the same place where ANTARES was located). On the other hand, IceCube is located at the Amundsen-Scott South Pole Station in Antarctica, in the same location that its predecessor AMANDA. In order to suppress the \textit{atmospheric neutrino background}, the neutrino telescopes explore upward-going neutrinos. Therefore, while ANTARES explores the Southern Hemisphere, IceCube explores the Northern.}.

With Respect to the GC neutrino bounds, the small number of detections in Icecube and Antares makes the bound over DM models due to GC neutrino fluxes $\backsim 3$ order of magnitude worse than the bounds from $\gamma$-rays \cite{Aartsen:2017ulx,Albert:2016emp}. However, very competitive bounds from the solar neutrino searches are presented by both experiments \cite{Adrian-Martinez:2016gti,Aartsen:2016zhm}.

\subsection{Galactic Center $\gamma$-ray Excess (GCE)}
\label{sec:GCE}

The different fluxes explained in the previous sections describe measurements that can be explained using only SM particle. This fact set limits over the DM models. However, there is an unexpected signal detected in the $\gamma$-ray data reported by the Fermi-LAT collaboration from the center of the Milky Way, the so-called Galactic Center Excess (GCE). The distribution and morphology of this photon excess is compatible with the predictions about DM annihilation \cite{Goodenough:2009gk, Vitale:2009hr, Hooper:2010mq, Gordon:2013vta, Hooper:2011ti, Daylan:2014rsa, 2011PhLB..705..165B, Calore:2014xka, Abazajian:2014fta,Zhou:2014lva}. According to the last Fermi-LAT analysis, the GCE is peaked at $\backsim 3$ GeV.

The physical origin of the GCE is unclear. The DM explanation is not the only one, as the GCE could be caused by the emission of unresolved point sources \cite{Abazajian:2010zy,Bartels:2015aea,Lee:2015fea,Fermi-LAT:2017yoi,Caron:2017udl} or due to cosmic-ray particles injected in the galactic center region, interacting with the gas or radiation fields \cite{Cholis:2015dea}. In addition, the nature of the GCE seems different below and above $\backsim 10$ GeV. The high energy tail may be explained as an extension of the \textit{Fermi bubbles} observed at higher latitudes \cite{Caron:2017udl}, whereas the low energy excess might be produced by DM annihilation, unresolved \textit{Millisecond Pulsars}, or both.

It is true that the interpretation of the GCE as a signal of DM annihilation is not robust, but currently it can not be ruled out either.


\section{Collider Searches}
\label{Sec:Collider}

Sect.~\ref{Sec:DD} and Sect.~\ref{Sec:ID} give an overview about the different techniques of Direct and Indirect DM Detection. In order to complete the DM detection landscape it is necessary to talk about the DM production at colliders. The strongest current bounds come from the searches at LHC. In general, the DM signals at colliders consist on the detection of some missing energy or momentum in the collision. Several reviews can be visited by the reader to expand the brief summary made in this section, for instance Refs.~\cite{Boveia:2018yeb,Kahlhoefer:2017dnp,Abdallah:2015ter,Kahlhoefer:2015bea} 

We can distinguish two kind of models analysed at colliders: models where DM couple directly to SM particles and models where do not exist such direct couplings. In the first case, we can find different interesting channels to search for DM. On the one hand, channels related with the Higgs boson have been one of the most promising searches as a consequence of its special role in the electroweak interaction. The current bounds over the invisible decay $\text{Br}( H \rightarrow \text{inv})$ imposed by ATLAS and CMS can be found in Refs.~\cite{Aaboud:2019rtt,Sirunyan:2018owy} and constraints models where DM couple directly to the Higgs. The limits over the DM mass in this case are $m_{\text{DM}} \lesssim m_H/2$ . On the other hand, models where DM couple to the Z boson are constrained by the precise measurements in LEP \cite{ALEPH:2005ab}. Analogous to the Higgs case, the limits over this kind of models are $m_{\text{DM}} \lesssim m_Z/2$.
The second kind of models is composed by scenarios where DM do not couple directly to the SM particles. In this context, the dijet and dilepton searches \cite{Aaboud:2017buh,Aaij:2017rft,Harris:2011bh,Aaboud:2017yvp,Sirunyan:2016iap} play an important role when  DM interacts with quarks and leptons through BSM mediators. In these cases, strong experimental constraints apply \cite{Aaboud:2017buh,Aaij:2017rft,Harris:2011bh,Aaboud:2017yvp,Sirunyan:2016iap}.
Finally, the study of monojets has important implications in the DM collider searches. In some DM scenarios,  it is expected to produce DM at colliders together with QCD jets which set strong bounds,
for instance, on DM models with leptophobic and coloured mediators mediators as shown by ATLAS \cite{Aaboud:2017dor} and CMS \cite{Sirunyan:2017hci} experiments.

\chapter{Extra-Dimensions}
\label{sec:ED}

\section{Motivation}

To understand the original motivation for the extra dimensions it is necessary to go back to the second half of the 19th century. Between 1860 and 1870 James Clerk Maxwell published his work about the electromagnetic field \cite{Maxwell:1865zz}, which represented the unification of the electric and magnetic interactions into the same force, the electromagnetism. The unification of electromagnetism inspired many scientists to try to unify the two interactions that were known at that time: electromagnetism and gravity. In 1916 Einstein published his results on General Relativity \cite{Einstein:1916vd}, the gravitational interaction being fully described as a field theory. The first attempts at unifying electromagnetism and General Relativity came soon. In 1921 Theodor Kaluza presented an extension of the theory of General Relativity into five dimensions \cite{1921SPAW.......966K}, with a metric tensor of fifteen components. These fifteen components would be distributed as follows: ten would correspond to the classic 4D metric, explaining gravity; four would represent the potential vector of electromagnetism; finally, the last component would be an unidentified massless scalar field, usually called \textit{radion} or \textit{dilaton}. The equation of motion of the theory provides both the Einstein equations and the Maxwell equations, and identifies the electrical charge with the motion into the fifth dimension.

In 1926, Oskar Klein adds a quantum interpretation to the Kaluza theory\footnote{That same year quantum physics began to take its first steps with the publication of the Erwin Schr$\ddot{\text{o}}$dinger Equation, Ref.~\cite{PhysRev.28.1049}.} \cite{1926ZPhy...37..895K,1926Natur.118..516K}, imposing the quantization of linear momentum in the fifth dimension. Klein's quantum interpretation gives a solution to the invisibility of the extra-dimension: the new dimension is closed and periodic. Indeed, the characteristic radius of the fifth dimension estimated by Klein was $\backsim 10^{-30} \, \text{cm}$, which explains the non-observation of the extra-dimension.

The discovery of the weak and strong interactions, and the subsequent electroweak unification, made the original motivation of Kaluza and Klein's theory lost\footnote{See, however, the works of Refs.~\cite{Salam:1981xd,WETTERICH1985309}.}. Years later, in the 1970's, the emergence of string theories \cite{Danielsson_2000} revived the extra-dimensional theories in order to obtain a consistent quantum gravity theory. Since the 1990's, theories of extra-dimensions have received much more attention in the scientific community. The Universe being formed by more than 4 dimensions could, for example, give a solution to the hierarchy problem\footnote{For a description of the problem see Sect.~\ref{sec:hierarchyproblem}.}. Also, many extra-dimensional models present natural candidates for Dark Matter, such as the case of the lightest Kaluza-Klein state in Universal Extra Dimensions (UED) \cite{Appelquist:2000nn,Arun:2017zap}.
In the extra-dimensional theories it was assumed that the compactification radius of the extra-dimension was of Planck lenght. However,
In the 1990's Ignatius Antoniadis in Ref.~ \cite{Antoniadis:1990ew} and Arkani-Hamed, Dimopoulos, and Dvali in Refs.~\cite{Antoniadis:1997zg,ArkaniHamed:1998rs,Antoniadis:1998ig,ArkaniHamed:1998nn} proposed the Large Extra Dimensions (LED). In this scenario, the extra-dimension can be \textit{large} of order $TeV^{-1}$, provided that only gravity propagates along the new dimension. Sect.~\ref{Sec:LED} summarizes the fundamental characteristics of LED models.

The space-time described by LED assumes new \textit{flat} dimensions, that is, with the same structure as the other three spatial dimensions already known.
This is equivalent to neglect the curvature effects of the gravitational field over the new extra-dimension. The approximation is accurate when the tensions of the branes are small. However, new interesting phenomenology appears when this is not the case and its curvature becomes relevant. These are the so-called Warped Extra-Dimensions scenarios, also known as Randall-Sundrum models after the physicists who proposed them. We review them in Sect.~\ref{sc:warped}.

Ever since Lisa Randall and Raman Sundrum proposed their extra-dimensional model, this one and its variants have been the only models of Warped Extra-Dimensions until 2016, when Gian Giudice and Matthew McCullough proposed a new Warped Extra-Dimensional model, the Clockwork/Linear Dilaton (CW/LD) model \cite{Giudice:2016yja,Giudice:2017fmj}. 

In the next Sections we develop the basic concepts of the extra-dimensional models. Several reviews can be found to complete the information of this Chapter, for instance Refs.~\cite{Csaki:2018muy,Csaki:2004ay,Kribs:2006mq,Cheng:2010pt}.

\section{Kaluza-Klein Decomposition}

The Kaluza-Klein decomposition allows to write the extra-dimensional fields as the sum of a tower of 4D fields. In this section we show the example of the procedure for a scalar field in the 5-dimensional flat space. However, this decomposition is valid as long as we work with a separable metrics.

In the General Relativity 5-dimensional extension, the space-time metric can be written as $ds^2 = g^{(5)}_{MN}dx^Mdx^N$. 
In the rest of the Chapter we use Greek letters when we refer to the classical 4-dimensions $x^\mu = (x^0,x^1,x^2,x^3)$ and to denote the fifth-dimension we use $x^5 = y$. For the 5-dimensional index we use Latin capital letters $x^M = (x^0,x^1,x^2,x^3,y)$. Thereafter, the signature of the metric is understood to be $(1,-1,-1,-1,-1)$.

Let us now consider the specific case of a free real scalar field. The action for a 5-dimensional Minkowski metric can be written as
\be
\label{accion_5d_scalar}
S = \int d^4x \, dy \, \frac{1}{2} \, \left[(\partial_\mu \phi)^2 - (\partial_y \phi)^2 \right] \, .
\ee
The equation of motion is then given by $\partial^2_\mu \phi - \partial^2_y \phi = 0 $. Imposing the periodic boundary conditions in the extra-dimension, the equation accepts as a solution:
\be
\phi(x,y) = \frac{1}{\sqrt{2\pi r_c}} \sum_{n=0}^\infty \phi^{(n)}(x)\, e^{i\,n\,y/r_c}\, ,
\ee
where $r_c$ is the compactification radius of the extra-dimension. Using this expression in Eq.~\ref{accion_5d_scalar}, it is easy to obtain:
\be
\label{accion_5d_scalar_guay}
S = \int d^4x \, \left[ \, \sum_{n>0} \partial_\mu \phi^{(n)\dagger} \, \partial^\mu \phi^{(n)} - \frac{n^2}{r_c^2}|\phi^{(n)}|^2 \right] \, ,
\ee
where the 5D field can be written as a sum of infinite 4D massive fields with mass
\be
m_n = \frac{n}{r_c} \, .
\ee
If the 5D field has a mass parameter $m_0$, the mass spectrum is shifted as $m_n = m_0 + n/r_c$.
As we can see, the Kaluza-Klein decomposition is an expansion that transforms a 5D Lagrangian into a 4D Lagrangian with an infinite spectrum of 4D massive particles. 

\section{Large Extra-Dimensions (LED) }
\label{Sec:LED}

The most famous scenario of flat extra dimensions is called Large Extra-Dimensions \cite{Antoniadis:1990ew,Antoniadis:1997zg,ArkaniHamed:1998rs,Antoniadis:1998ig,ArkaniHamed:1998nn}. This model implements one of the fundamental concepts of the modern extra-dimensions, the so-called \textit{branes}. Branes are $(3+1)$-dimensional hypersurfaces that can trap fields on their surfaces. The presence of these hypersurfaces implies the existence of fields that only propagate on the brane (4-dimensional fields). In addition to the brane fields, there can also exist fields that freely propagate into the extra-dimensional space (the so-called \textit{bulk}).
\begin{figure}
\centering
\includegraphics[width=80mm]{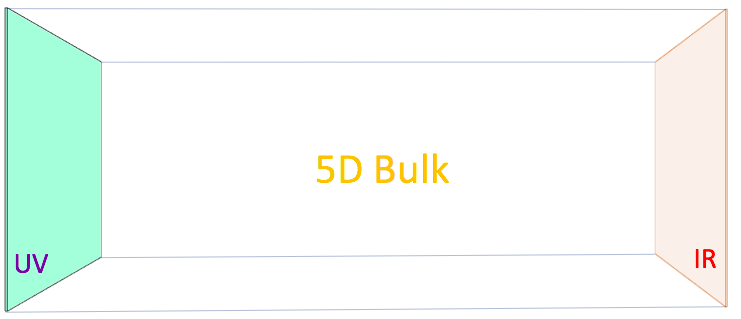}
\caption[Large Extra-Dimensions 5D space-time]{\it Representation of Large Extra-Dimensions 5D space-time. }
 \label{fig:LED}
\end{figure}
If the SM is confined in the brane and gravity freely propagates along the bulk, the gravitational interaction is diluted along  the extra-dimensional space. Therefore, while the fundamental scale of the higher dimensional gravity ($M_D$) can be $\mathcal{O}(1)$ TeV, the fundamental scale on the brane is $M_\text{P}$. The hierarchy problem then is only an effect of the existence of the extra-dimensions. Fig.~\ref{fig:LED} shows a pictorial representation of the Large Extra-Dimensions 5D space-time.

In order to obtain a relation between both fundamental scales, we assume that the metric of the higher $D$-dimensional space-time is given by:
\be
ds^2 = G_{MN}^{(D)}dx^Mdx^N.
\ee
The generalization of the Einstein-Hilbert action to more than 4 dimensions keeps the 4-dimensional structure:
\be
S_{n} = -M_{D}^{D-2} \int d^{D}x\, \sqrt{G^{(D)}} R^{(D)},
\ee
where $R^{(D)}$ is the Ricci tensor in $D=d+4$ dimensions. On the other hand, the usual 4-dimensional action is given by
\be
S_{4} = -M_{p}^{2} \int d^{4}x\, \sqrt{G^{(4)}} R^{(4)}.
\ee
To know how the classical 4-dimensional gravity is contained inside the higher dimensional metric (or equivalently, how the 4-dimensional graviton is contained in the $D$-dimensional metric) we can expand the 4-dimensional part of the metric:
\be
ds^2 = (\eta_{\mu \nu} + h_{\mu \nu})dx^\mu dx^\nu - r_c^2d\Omega_d^2,
\ee
where $r_c$ is related to the size of the extra-dimensions (the compactification radius) and $d\Omega_d$ is the line element of the flat extra-dimensional space. The perturbation $h_{\mu \nu}$ represents the 4-dimensional graviton in 5D.
Finally, the necessity to reproduce the Newton's law in four dimensions gives a relation between both fundamental scales:
\be
\label{relation_LED}
M_\text{P}^2 = M_{D}^{D-2} (2\pi r_c)^D.
\ee 

Stringent limits for LED models come from the deviations of Newton's law. If we assume $M_{D} \backsim 1$ TeV (value that solves the hierarchy problem), the distance scale $r_c$ where we found $\mathcal{O}(1)$ deviations order one is given by Eq.~\ref{relation_LED}. 
\begin{table}
\centering
\begin{tabular}{c c}
    \hline
    \bfseries Number of extra-dimensions & \bfseries $r_c$ [cm]  \\ 
    \hline
     $d=1$ & $10^{13}$ \\
     $d=2$ & $10^{-2}$ \\
     $d=3$ & $10^{-7}$ \\
     $d=4$ & $10^{-10}$ \\
     $d=5$ & $10^{-12}$ \\
     $d=6$ & $10^{-13}$ \\
    \bottomrule
\end{tabular}
\caption[LED bounds from deviations of the Newton's law.]{LED bounds from deviations of the Newton's law. $r_c$ represents the distance scales where we expect deviations order one.}
\label{limits_LED}
\end{table}
Tab.~\ref{limits_LED} shows the expected values for $r_c$ as a function of the number of dimensions. It is clear that the one extra-dimension case is totally ruled out because the scale is larger than the size of the Solar System! The effects of the deviation should have been observed in that case. On the other hand, for $d \geq 2$ the LED model  solves the hierarchy problem, being $r_c$ compatible with present bounds on deviations from the Newton's law\footnote{In addition to the limits on deviations from the Newton's law, supernovae and neutron stars are sources of bounds for LED models \cite{Hannestad:2003yd}.} \cite{Kapner:2006si}.


\section{Warped Extra-Dimensions}
\label{sc:warped}

Complementary to the flat case, Warped Extra-Dimensions was proposed, where the new dimensions are curved. This section summarizes the basic concepts of RS scenario, whereas for a complete mathematical description we address to Ref.~\cite{Folgado:2019sgz} (included in Part \ref{sec:papers} of this Thesis). For simplicity, we will only study the 5-dimensional case. However, the generalization to $D$-dimensional bulk can be found in several references (see, for instance, Refs.~\cite{Csaki:2018muy,Csaki:2004ay,Kribs:2006mq,Cheng:2010pt}).

\subsection{The Randall-Sundrum Background}

The first steps in these models were given by Lisa Randall and Raman Sundrum at the end of \footnote{An alternative form of the model was published by the same authors shortly after \cite{Randall:1999vf}.}  1990's \cite{Randall:1999ee}. The popular Randall-Sundrum scenario consider a non-factorizable 5-dimensional metric in
the form:
\be
ds^2 = e^{-2\sigma (y)}\eta_{\mu \nu} dx^\mu dx^\nu - r_c^2 dy^2 \, ,
\label{RS_metric}
\ee
where $\sigma (y) = kr_c|y|$ and the signature of the metric is $(+,-,-,-,-)$. In RS scenario $r_c$ is the compactification scale, as in LED, while $k \backsim \mathcal{O}(M_\text{P})$ is the curvature along the 5th-dimension.
We impose periodical boundary conditions over the extra dimension, $y = y + 2\pi$, and reflectivity $y = -y$. Therefore, the metric is defined in $0 \leq y \leq \pi$ region. The resulting space $S_1/\mathbb{Z}_2$ is called \textit{orbifold}.
We only consider a slice of the space-time between two branes located conventionally at the two fixed-points of this orbifold, $y = 0$ (the so-called UV-brane) and $y=\pi$ (the IR-brane), with compactification radius $r_c$. The 5-dimensional space-time is a slice of \textit{anti-de Sitter}\footnote{This mathematical space was proposed and studied by Willem de Sitter and Albert Einstein in the 1920's.} ($\text{AdS}_5$) space and the exponential factor that multiplies the $\mathcal{M}_4$ Minkowski 4-dimensional space-time is called \textit{warp factor}.
Planck mass in this scenario is related with the fundamental $M_5$ as
\be
\bar{M_\text{P}}^2 = \frac{M_5^3}{k} \left[ 1 - e^{-2 k \pi r_c} \right] \, ,
\ee
where $\bar{M_\text{P}} = M_\text{P}/\sqrt{8\pi}$ is the reduced Planck mass.
Unlike the flat case, in RS $M_\text{P}$ and the new fundamental mass parameter $M_5$ are the same order.
\begin{figure}
\centering
\includegraphics[width=80mm]{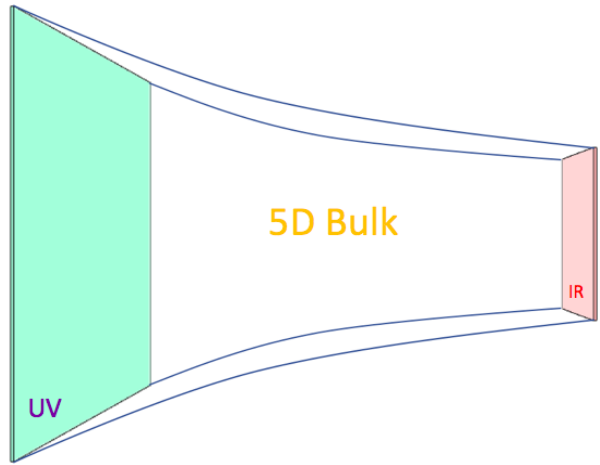}
\caption[Randall-Sundrum 5D space-time]{\it Representation of Randall-Sundrum 5D space-time. }
 \label{fig:RS}
\end{figure}
Fig.~\ref{fig:RS} shows how the extra-dimension changes along the 5-dimensional bulk. The difference between the fundamental masses of the SM and the Planck Mass is explained by the exponential growth between the IR and the UV branes. The hierarchy problem is then a consequence of the warping of the 5-dimensional space-time.

The original RS model assumes that all fields are confined on the IR-brane, being gravity the only field that can propagate freely along the bulk. While in the classical 4-dimensional space-time the scale of the interactions is the Planck mass, $\bar{M_\text{P}}^2$, in RS is given by
\be
\Lambda \equiv \bar{M_\text{P}} e^{-k \pi r_c} \, .
\label{lambda_RS}
\ee
Choosing $k$ and $r_c$ such that $\Lambda \ll \bar{M_\text{P}}$, the RS scenario can address the hierarchy problem (for $\sigma = kr_c \backsim 10$).

To study in RS scenario the gravitational interaction in the brane we expand the 4-dimensional component metric around the flat space metric:
\be
G^{(4)}_{\mu \nu} = e^{-2\sigma} (\eta_{\mu \nu} + \kappa_5 h_{\mu \nu}) \, ,
\ee
with $\kappa_5 = 2M_5^{-2/3}$. The 5-dimensional $h_{\mu \nu}$ field play the same role that in the classical space-time linearised gravity, the graviton. This field can be decomposed as a KK-tower of infinite 4-dimensional massive modes in the brane, usually called KK-gravitons.
Notice that in the 4-dimensional decomposition of a 5-dimensional metric, two other fields are
generally present: the graviphoton, $h_{\mu 5}$ and the graviscalar $h_{55}$. It has been shown elsewhere
\cite{Giudice:1998ck} that the graviphoton KK-modes are reabsorbed by the (massive) KK-gravitons.
On the other hand, the graviscalar field is relevant to stabilize the size of the extra-dimension and it will be discussed in Sect.~\ref{Sec:radionRS}. 

The mass spectrum of the KK-gravitons is given by:
\be
m_{n} = kx_n e^{-k\pi r_c} \, ,
\ee
where $x_n$ are the zeros of \footnote{$J_1$ is the first Bessel functions of the first kind. The first zero is $x_1 \approx 3.83$ while the rest can be approximated by $x_n \approx \pi (n + 1/4) + \mathcal{O}(n^{-1})$ \cite{davis1927}.} $J_1(x_n)$. Then, in RS the spacing between two consecutive KK-modes is 
$\Delta m \backsim k(x_{n-1} - x_n) e^{-k\pi r_c}$.
Notice that, for low $n$, the KK-graviton masses are not equally spaced. This is very different from LED where the spacing between the masses of two adjacent KK-modes is $1/r_c^2$. However, for large $n$, as a consequence of the $x_n$ structure, the spacing becomes approximately constant.

The strongest constraints in RS are given by the resonance searches at LHC, assuming that all fields are located in the IR-brane. Once a KK-graviton resonance is produced, we can study its decay modes in the narrow width approximation.
The KK-graviton decay channels that provide the most stringent bound
on $m_{1}$ and $\Lambda$ are $pp \rightarrow G_1 \rightarrow \gamma\gamma$ \cite{Aaboud:2017yyg} and $pp \rightarrow G_1 \rightarrow \ell\ell$ \cite{Aaboud:2017buh}. 

\subsection{Size Stabilization: The Goldberger-Wise Mechanism}
\label{Sec:radionRS}

Stabilizing the size of the extra-dimension to be $y = \pi r_c$ is a complicated task: bosonic quantum loops have a net effect on the border of the extra-dimension such that the extra-dimension
itself should shrink to a point (see, e.g., Refs.~\cite{Appelquist:1982zs,Appelquist:1983vs,deWit:1988xki}). This feature, in a flat extra-dimension, can only be compensated by fermionic quantum loops
and, usually, some supersymmetric framework is invoked to stabilize the radius of the extra-dimension (see, e.g., Ref.~\cite{Ponton:2001hq}). 
In Randall-Sundrum scenarios, on the other hand, a new mechanism has been considered: if we add a bulk scalar field $\Phi$
with a scalar potential $V(\Phi)$ and some {\rm ad hoc}  localized potential terms, $\delta (y=0) V_{\rm UV}(\Phi)$ and 
$\delta (y = \pi) V_{\rm IR} (\Phi)$, it is possible to generate an effective potential $V(\varphi)$ for the 4-dimensional field $\varphi = f_{\text{IR}} \, e^{- k \pi T }$, where $f_{\text{IR}}$ is the IR-brane tension. In order to have a stable background metric in Eq.~\ref{RS_metric} and $\langle T \rangle  = r_c$, the condition $f_{\text{IR}} = \sqrt{24 M_5^3/k}$ must be satisfied. The minimum of this potential can yield the desired value of $k r_c$ without extreme fine-tuning of the 
parameters \cite{Goldberger:1999wh,Goldberger:1999uk}.

As in the spectrum of the theory there is already a scalar field, the graviscalar $G^{(5)}_{55}$, the $\Phi$ field will generically
mix with it. The KK-tower of the graviscalar is absent from the low-energy spectrum, as they are eaten by the KK-tower 
of graviphotons to get a mass (due to the spontaneous breaking of translational invariance caused by the presence of one or more
branes). On the other hand, the KK-tower of the field $\Phi$ is present, but heavy (see Ref.~\cite{Goldberger:1999un}). 
The only light field present in the spectrum is a combination of the graviscalar zero-mode and the $\Phi$ zero-mode.
This field is usually called the {\em radion}, $r$. Its mass can be obtained from the effective potential $V (\varphi)$ and is given by
\be
m_\varphi^2 = \frac{k^2 v_v^2}{3 M_5^3} \, \epsilon^2 \, e^{-2 \pi k r_c} \, , 
\ee
where $v_v$ is the value of $\Phi$ at the IR-brane and 
\be
\epsilon =\frac{m^2}{4 k^2} \, ,
\ee
with $m$ the mass of the field $\Phi$. Quite generally, $\epsilon \ll 1$ and, therefore, the mass of the radion can be much smaller than the first KK-graviton mass. Notice that $m_r$ is, thus, a new free parameter of the RS model, in addition to $m_1$ and $\Lambda$ (or, alternatively, $M_5$ and $k$).

\subsection{AdS/CFT Correspondence and RS Model}

In the original Randall-Sundrum scenario (and its subsequent generalizations), the space-time is a slice of the AdS space. $\text{AdS}_{n}$ is a maximally symmetric \textit{Lorentzian manifold}\footnote{Mathematical space that are described by a \textit{Lorentzian metric}.} with the peculiarity that presents a constant negative scalar curvature (opposite to a de Sitter space, with positive curvature.). In 1998 the so-called AdS/CFT duality was conjectured, establishing a relationship between quantum gravity theories (like M-theory and string theory) defined in some $D$-dimensional AdS mathematical space with \textit{conformal field theories} (CFT) living on the boundary of such space. The idea was proposed by Juan Maldacena\footnote{Hitherto, in 2020, Maldacena's article is the most cited paper in high-energy physics with 16000 citations!} in Ref.~\cite{Maldacena:1997re}. However, some mathematical aspects were clarified by Steven Gubser, Igor Klebanov, Alexander Polyakov and Edward Witten in Refs.~\cite{Gubser:1998bc,Witten:1998qj}.
\begin{figure}
\centering
\includegraphics[width=100mm]{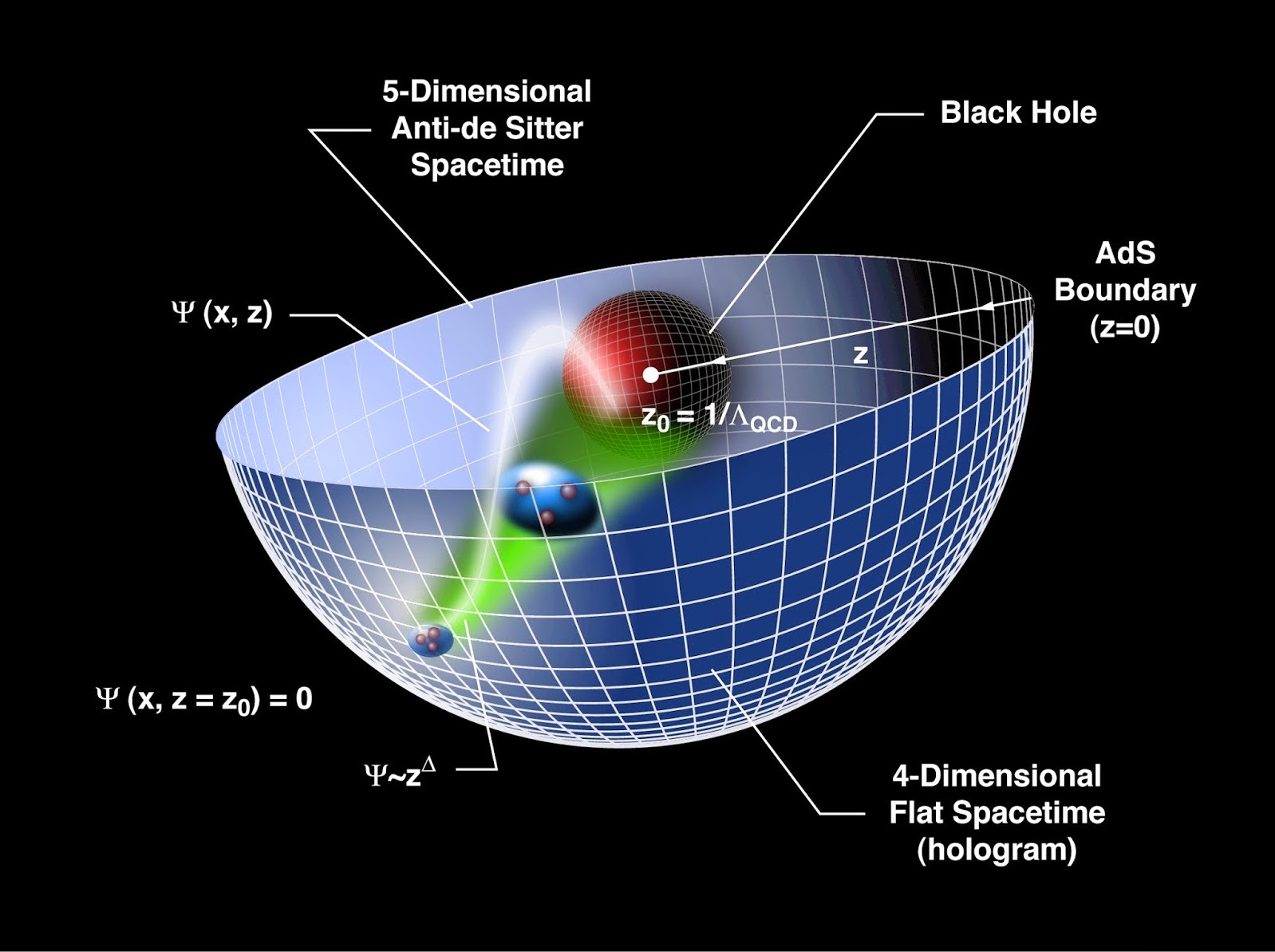}
\caption[ADS/CFT correspondence]{\it Artistic representation of ADS/CFT correspondence. Image taken from Ref.~\cite{Brodsky:2008pg}.}
 \label{fig:ADS_CFT}
\end{figure}
The AdS/CFT conjecture is also called \textit{holographic duality} because the CFT can be interpreted as a hologram that contains all physical information about the higher-dimensional quantum gravity theory. Fig.~\ref{fig:ADS_CFT} shows an artistic representation of this duality.

Since AdS/CFT duality was proposed, different authors have studied the implications of this conjecture in RS models. The idea was first explored in the non-compact Randall-Sundrum model\footnote{Usually called RS2, to distinguish it from original RS model, also called RS1.} \cite{Randall:1999vf} (some examples can be found in Refs.\cite{Gubser:1999vj,Giddings:2000mu,Giddings:2000ay,Verlinde:1999fy,Duff:2000mt}). Shortly after, the implications of the Maldacena's duality were studied in the original RS model (first publications in this direction include, for instance, Refs. \cite{ArkaniHamed:2000ds,Rattazzi:2000hs}).

A complete review about the ADS/CFT conjecture can be found in Ref.~\cite{Gherghetta:2010cj}.

\section{Clockwork/Linear Dilaton (CW/LD) Extra-Dimensions }
\label{sec:cw_ld}

In 2016 Clockwork/Linear Dilaton model was proposed by Gian Giudice and Matthew McCulloug \cite{Giudice:2016yja,Giudice:2017fmj}. In this extra-dimensional scenario a KK-graviton tower, with a spacing very similar to that of LED models, starts at a mass gap $k$ with respect to the zero-mode graviton. 
The fundamental gravitational scale $M_5$ can be as low as the TeV, where $k$ is typically chosen in the GeV to TeV range.
In this Section we have summarize the most relevant properties of CW/LD model. A more complete and technical review with all mathematical details can be found in Ref.~\cite{Folgado:2019gie}, included in Part \ref{sec:papers} of this Thesis.

Clockwork/Linear Dilaton scenario is defined by the metric:
\be
\label{eq:5dmetric}
ds^2 = e^{4/3 k r_c |y|} \left ( \eta_{\mu \nu} dx^{\mu}dx^{\nu} - r_c^2 \, dy^2 \right ) \, ,
\ee
where the signature of the metric is $(+,-,-,-,-)$.
This particular metric was first proposed in the context of {\em Linear Dilaton} (LD) 
models and {\em Little String Theory} (see, {\em e.g.} Refs.~\cite{Antoniadis:2011qw,Baryakhtar:2012wj,Cox:2012ee} and references therein). 
The metric in Eq.~(\ref{eq:5dmetric}) implies that the space-time is non-factorizable, as the length scales on our 4-dimensional space-time depending 
on the particular position in the extra-dimension due to the warping factor $e^{2/3 \, k r_c \, |y|}$. Notice, however, that in the limit $k \to 0$ the standard, 
factorizable, flat LED case is immediately recovered. 
As for the case of the Randall-Sundrum model, also in the CW/LD scenario the extra-dimension is 
compactified on a ${\cal S}_1/{\cal Z}_2$ orbifold (with $r_c$ the compactification radius), and two branes
are located at the fixed points of the orbifold, $y = 0$ (IR-brane) and at $y = \pi$ (UV-brane).
\begin{figure}
\centering
\includegraphics[width=80mm]{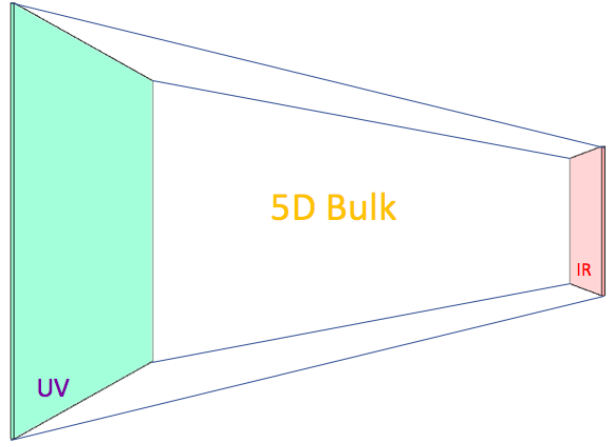}
\caption[Clockwork/Linear Dilaton 5D space-time]{\it Representation of Clockwork/Linear Dilaton 5D space-time. }
 \label{fig:CW_LD}
\end{figure}
Fig.~\ref{fig:CW_LD} shows the structure of the 5-dimensional CW/LD model. As in the RS case, the hierarchy problem is solved by the growth of the fundamental parameters along the bulk. However, there is a fundamental difference between these two models: the warping factor in Eq.~\ref{RS_metric} multiplies only the four dimensional components, whereas, in the CW/LD case it multiplies all the 5-dimensional metric. The growing then is different in the CW/LD respect to RS, giving a totally different phenomenology \cite{Giudice:2017fmj}. 

In the minimal scenario, Standard Model fields are located in one of the two branes (usually the IR-brane).
The scale $k$ (also called the \textit{clockwork spring}\footnote{A term inherited by its r\^ole in the discrete version of the Clockwork model \cite{Giudice:2016yja}.}) is the curvature along the 5th-dimension and it can be much smaller than the Planck scale. 
Being the relation between $\bar{M_\text{P}}$ and the fundamental gravitational scale $M_5$ in the CW/LD model: 
\be
\bar{M_\text{P}}^2 = \frac{M_5^3}{k} \left ( e^{2 \pi k r_c}  - 1 \right ) \, ,
\ee
it can be shown that, in order to solve or alleviate the hierarchy problem, $k$ and $r_c$ must satisfy the following relation:
\begin{equation}
k \, r_c = 10 + \frac{1}{2 \pi} \, \ln \left ( \frac{k}{\rm TeV} \right ) - \frac{3}{2 \pi} \, \ln \left ( \frac{M_5}{ 10 \, {\rm TeV}} \right ) \, .
\end{equation}
For $M_5 = 10$ TeV and $r_c$ saturating the present experimental bound on deviations from the Newton's law, $r_c \sim 100 \, \mu$m 
\cite{Adelberger:2009zz}, this relation implies that $k$ could be as small as $ k \sim 2$ eV, and KK-graviton modes would therefore 
be as light as the eV, also. This \textit{extreme} scenario does not differ much from the LED case, but for the important difference that the hierarchy
problem could be solved with just one extra-dimension (for LED models, in order to bring $M_5$ down to the TeV scale, an astronomical 
lenght $r_c$ is needed and, thus, viable hierarchy-solving LED models start with at least 2 extra-dimensions). In the phenomenological 
application of the CW/LD model in the literature, however, $k$ is typically chosen above the GeV-scale and, therefore, $r_c$ is accordingly
diminished so as to escape direct observation. Notice that, differently from the case of Warped Extra-Dimensions, where scales 
are all of the order of the Planck scale ($M_5, k \sim \bar{M_\text{P}}$) or within a few orders of magnitude, in the CW/LD scenario, both the fundamental
gravitational scale $M_5$ and the mass gap $k$ are much closer to the electro-weak scale $\Lambda_{\rm EW}$ than to the Planck scale, 
as in the LED model.

Expanding the metric at first order around its static solution, we have: 
\be
\label{eq:metricexpansion}
G^{(5)}_{MN} =  e^{2/3  S} \left(\eta_{MN} + \frac{2}{M_5^{2/3}} h_{MN}\right) \, ,
\ee
where $s=2 k r_c |y|$ is the dilaton field.
The 4-dimensional component of the 5-dimensional field $h_{MN}$ can be expanded in a Kaluza-Klein tower of 4-dimensional fields (4-dimensional massive gravitons)
with masses
\begin{equation}
m_0^2 = 0 \, ; \qquad m_n^2 = k^2 + \frac{n^2}{r_c^2} \, .
\end{equation}
Instead of $\bar{M_\text{P}}$, in CW/LD the scale of the gravitational interactions is enhanced (as it was for RS). Indeed, the scale of the interaction of this KK-gravitons with the particles located in the IR-brane can be $\mathcal{O}(\text{TeV})$. This scale is related with the fundamental parameters of the model as
\bea
\left \{
\begin{array}{llll}
\dfrac{1}{\Lambda_0} &=& \dfrac{1}{M_{\rm P}} \, , \nonumber \\[8pt]
\dfrac{1}{\Lambda_{n}} &=&  \dfrac{1}{\sqrt{M_5^3 \pi r_c}}  \left ( 1 + \dfrac{k^2 r_c^2}{n^2} \right )^{-1/2} 
=  \dfrac{1}{\sqrt{M_5^3 \pi r_c}} \left ( 1 - \dfrac{k^2}{m_n^2} \right )^{1/2}  \,  , \nonumber \\
\end{array}
\right .\\
\label{Lambda_graviton}
\eea
from which it is clear that the coupling between KK-graviton modes with $n \neq 0$ is suppressed by the effective scale $\Lambda_n$
and not by the Planck scale, differently from the LED case and similarly to the Randall-Sundrum one.
In the RS scenario this scale is a global parameter (equal for all KK-gravitons). However, in CW/LD each gravitons is coupled different to the brane particles.

Stabilization of the radius of the extra-dimension $r_c$ is always an issue.
In the CW/LD scenario, differently from the RS one, we can use the already present bulk dilaton field to stabilize the compactification radius. A complete description of the mechanism can be found in Ref.~\cite{Folgado:2019gie}, included in Part \ref{sec:papers} of this Thesis.

As a final comment, In CW/LD scenario the graviton resonances are close enough to considerate a continuum spectrum. This fact allows to constrain the model using non-resonant searches at LHC in $G_1 \rightarrow \gamma\gamma$ and $G_1 \rightarrow \ell\ell$ channel. \cite{Aaboud:2017yyg,Aaboud:2017buh,Sirunyan:2018wnk}.

\chapter{Summary of the Results}
\label{sec:summary}
In Chapters \ref{sec:SM} to \ref{sec:ED} a summary of the most relevant aspects of the Dark Matter and Extra-Dimensions has been made. The aim of the introduction is to offer the tools needed to understand the different models that compose the original works of this Thesis.
In this Chapter, on the other hand, we summarize the basic ideas and results of the four papers that constitute the second part of the Thesis. Technical details can be found in the complete articles that are collected in Part \ref{sec:papers}.

\section{Probing the Sterile Neutrino Portal with $\gamma$-rays}

One of the most important open problems in high-energy physics is Dark Matter, but, as we commented in Sect.~\ref{sec:openproblems}, it is not the only one. Among the various problems that currently exist in the Standard Model, one of them is the neutrino masses: the model predicts zero mass for them. However, neutrino oscillations was suggested more than half a century ago as a distinctive signature of neutrino masses. This interesting effect, experimentally detected in 1998 \cite{Fukuda:1998mi}, consists of a quantum-mechanical oscillation in the leptonic flavor. The phenomenon has deep implications: the effect can only happen if at least one of the three SM neutrino is massive\footnote{Despite that the phenomenon could be explained with only one massive neutrino, the observation of the effect in both atmospheric and solar neutrinos needs at least two neutrinos to be explained \cite{Donini:2011jh}.}. However, the mass of these particles must be much smaller than the masses of all the other SM particles in order to escape observation.
This fact favoured the development of models where the mass of the neutrinos is generated by the so-called \textit{seesaw mechanisms}. In Sect.~\ref{sec:neutrinomasses} a small review about  this topic can be found.

The attempt to solve both the Dark Matter and the neutrino mass problems, at the same time\footnote{The most economical scenario, namely that the sterile neutrinos constitute the DM \cite{Dodelson:1993je}, has been thoroughly studied \cite{Adhikari:2016bei}.} led to the development of models with a \textit{sterile neutrino portal to dark matter}. This scenario has been studied by several authors, setting limits on it using Direct Detection \cite{Escudero:2016tzx,Macias:2015cna,Escudero:2016ksa} and Indirect Detection \cite{Tang:2015coo,Batell:2017rol,Campos:2017odj} experiments. This model is interesting from the point of view of Indirect Detection for several reasons (see Sects.~\ref{Sec:DD} and \ref{Sec:ID} for details on DD and ID): on the one hand, Direct Detection does not happen at the lowest order in perturbation theory. As a consequence, the limits on the model due to Direct Detection experiments are worse than in other models. On the other hand, the mixing of sterile neutrinos with active neutrinos causes Dark Matter annihilations to produce photons and charged particles, as a result of several decays. All this makes it the perfect candidate to be studied from the point of view of Indirect Detection, as we have done in Ref.~\cite{Folgado:2018qlv}.

We analysed a particular model in which, besides the sterile neutrinos, the SM is extended by a dark sector that contains a scalar field 
$\phi$ and a fermion $\Psi$. These fields are both singlets of the SM gauge group but charged under a dark sector symmetry group, $ G_{\text{dark}}$, such that the combination $ \overline{\Psi} \phi$ is a singlet of this hidden symmetry. 

\begin{figure}[t]
\centering
\includegraphics[width=120mm]{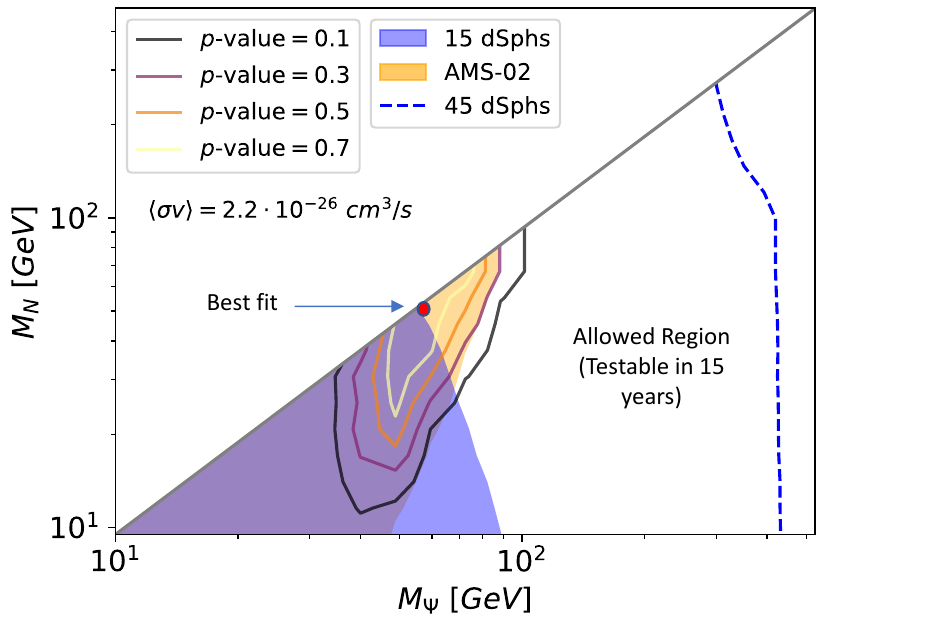}
\caption[Results from \textit{Probing the sterile neutrino portal to Dark Matter with $\gamma$-rays}.]{\it Limits over the \textit{sterile neutrino portal to Dark Matter} model in the sterile neutrino and DM masses space ($M_N, M_\psi$). The yellow region shows the antiproton limits, whereas the blue region are the dSphs limits. The different contours represent the region where the GCE can be fitted with its respective $p$-value (with increasing $p$-value going from outer to inner contours). Finally, the blue-dashed line shows our prediction about the foreseen future limit from the dSphs in the next 15 years of the Fermi-LAT experiment.}
 \label{fig:resultados_paper_1_ingles}
\end{figure}

The lightest of the two dark particles ($\phi$ or $\Psi$) turns out to be stable if all SM particles, as well as the sterile neutrinos, are singlets of $G_{\text{dark}}$, irrespective of the nature of the dark group. As a consequence, the stable particle is a good DM candidate. We assume for simplicity that the dark symmetry $G_{\text{dark}}$ is a global symmetry at low energies, although we do not expect significant changes in our analysis if it was local.

The most relevant terms in the Lagrangian are given by:
\bea
\mathcal{L} &\supset & \mu_H^2 H^\dagger H - \lambda_H (H^\dagger H)^2  -
\mu_\phi^2 \phi^\dagger \phi - \lambda_\phi (\phi^\dagger \phi)^2 - 
\lambda_{H\phi} (H^\dagger H) \, (\phi^\dagger \phi)   
\nonumber \\
&-&\left[ \phi \overline{\Psi}(  \lambda_a +  \lambda_p \gamma_5) N 
 + Y \overline {L}_L H N_{R}  + {\rm h.c.}  \right]  \, . \nonumber \\
 \,
\eea
The Yukawa couplings $Y$ between the right-handed fermions $N_R$ and the SM leptons lead to masses for the active neutrinos after electroweak symmetry breaking, via type-I seesaw mechanism. Although two sterile neutrinos are required to generate the neutrino masses observed in oscillations, at least, in our analysis we consider that only one species is lighter than the DM and therefore relevant for the determination of its relic abundance and indirect searches. The results can be easily extended to the case of two or more sterile neutrinos lighter than the DM. Assuming that the DM  is described by the fermionic field $\Psi$ (the analysis would be similar for Dark Matter being represented by $\phi$) the masses of the model fulfill the relation $m_N < m_\Psi < m_\phi$.

Fig.~\ref{fig:resultados_paper_1_ingles} shows the final results of our analysis. Fixing the mass of the scalar mediator field such as to obtain the correct relic abundance via the freeze-out mechanism (this means $\langle \sigma \, v \rangle \sim 2 \times 10^{-26} cm^3/s$), the figure shows the different limits from photons and antiprotons  in the parameter space $(M_N,M_\Psi)$. As it has been commented in Sect.~\ref{sec:GCE}, the Fermi-LAT experiment has reported a Galaxy-Center $\gamma$-ray Excess (GCE). The studied model predicts a photon excess that can be compatible with the GCE in a small region of the parameter space $(M_N,M_\Psi)$. In our analysis we assume that there are two distinct sources for the GCE: one astrophysical, responsible for the high energy tail of the $\gamma$-ray spectrum, and DM annihilation, that we considered the only source of the low energy GCE,
\be
\Phi = \Phi_{\text{astro}} + \Phi_{\text{DM}} \, .
\ee
Notice that this astrophysical contribution to the flux is always needed to fit the GCE, independently of the DM model considered.
The contour areas in Fig.~\ref{fig:resultados_paper_1_ingles} show the region where this fit is possible with different $p$-values where the outer contours have a lower $p$-value than the inner contours). However, the extra photons predicted by the model must also be compatible with the rest of  measurements made on the different photon fluxes. Specifically, the same experiment performs measurements on the $\gamma$-rays from 15 different Dwarf Spheroidal Galaxies\footnote{See Sect.~\ref{sec:gammaraysearches} for more information about this measurement.}. The dark blue-shaded region shows the area of the parameter space where the results obtained are not compatible with these measurements at 90 \% C.L.

On the other hand, the model also predicts an increase of antiproton flux. This increase has been compared with the antiproton flux from the galactic center measured by the AMS-02 experiment\footnote{See Sect.~\ref{sec:charged} for a description of the experimental results and Sect.~\ref{sec:antiprot} for the details of the antiprotons propagation along the galaxy.}, observing that there are areas in which the predictions of the model would not be compatible with the experimental measurements at 95 \% C.L. (light yellow-shaded area in the Figure). 
However, notice that the antiproton limits are less robust than the dSphs bounds, due to the large astrophysical uncertainties in the propagation models of charged particles.
Finally, an analysis of the possible impact of future Indirect Detection experiments on the model has also been carried out. Particularly, an improvement in the dSphs data taken by Fermi-LAT is expected and could set strong bounds on the studied model (blue-dashed line). As a final comment, DM models in general, would only marginally solve the GCE, but in our case, the model could be fully tested in the next decade.

\section{Gravity-mediated Scalar Dark Matter in RS}
\label{sec:paper2ing}

All the evidence we have today about the existence of Dark Matter is only related to gravitational interaction. This leads us to think about the possibility that Dark Matter particles may only interact gravitationally. In this case, DM would be undetectable by current and future particle physics experiments and it could not be a WIMP, since the gravitational interaction is too weak to produce the observed dark matter abundance through the freeze-out mechanism. However, what would it happen if we lived in more than 4 dimensions? This is the idea that inspired Ref.~\cite{Folgado:2019sgz}.

In this work we explored the possibility to obtain the current DM abundance, under the assumption that is composed by WIMP scalar particles, only through gravitational interaction and assuming a 5-dimensional RS space-time. In the described scenario, Dark Matter and the Standard Model live confined in the TeV-brane. Both types of matter interact through gravity, which propagates in the 5-dimensional bulk, and is described in the effective 4-dimensional theory as a tower of massive gravitons (Kaluza-Klein modes).

The model is described using four physical parameters: the scale of the interaction of 4-dimensional massive gravitons with matter, $\Lambda$; the mass of the first graviton of the 4-dimensional KK-tower, $ m_{1} $; the Dark Matter mass, $ m_{\text {DM}} $; and, the radion mass, $ m_r $. 
Our analysis shows that when $m_r < m_\text{DM}$ and, therefore, the annihilation channel into radions is open, the results obtained are largely independent of the particular value of $m_r$. Regarding the virtual radion-exchange annihilation cross-section into SM particles, it only becomes relevant close to the resonance, $m_\text{DM} \backsim m_r/2$. Thus, for the study of the phenomenology we fix the radion mass and focus on the remaining parameters.

\begin{figure}[t]
\centering
\includegraphics[width=150mm]{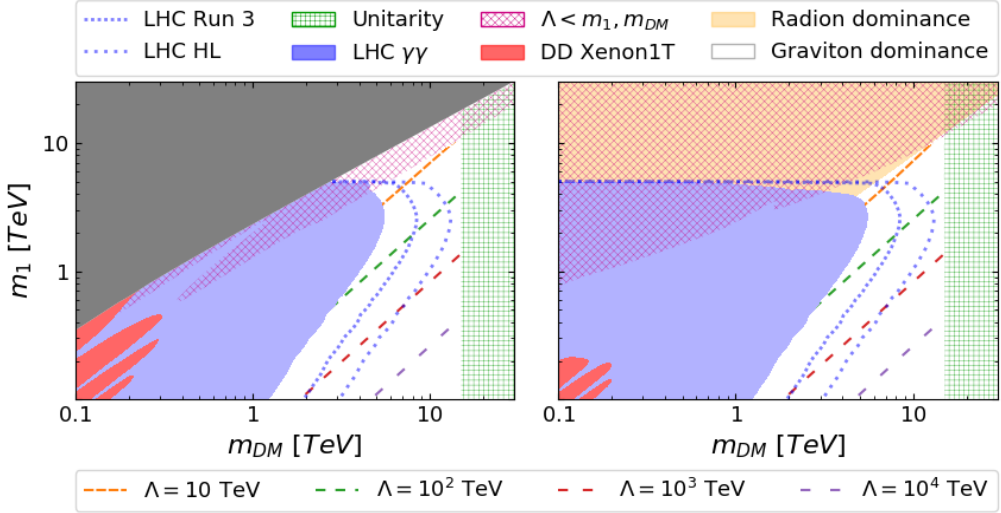}
\caption[Results from \textit{Gravity-mediated Scalar Dark Matter in Warped Extra-Dimensions}.]{\it Region of the $(m_{\text{DM}}, m_{1})$ plane for which 
$\left\langle \sigma v \right\rangle = \left\langle \sigma_{\rm fo} v \right\rangle$. 
Left panel: the radion and the extra-dimension stabilization mechanism play no role 
in DM phenomenology.
Right panel: the extra-dimension length is stabilized with  the Goldberger-Wise mechanism, 
with radion mass $m_r = 100$ GeV.
In both panels, the grey area represents the part of the parameter
space where it is impossible to achieve the correct relic abundance; 
the red-meshed area is the region for 
which the low-energy RS effective theory is untrustable, as $\Lambda < m_{1}$; 
the wiggled red area in the lower left corner is the region excluded by DD experiments; 
the blue area is excluded by resonant KK-graviton searches at the LHC with 36 fb$^{-1}$ at $\sqrt{s} = 13$ TeV; 
the dotted blue lines represent the expected LHC exclusion bounds at the end of the Run-3 (with $\sim 300$ fb$^{-1}$)
and at  the HL-LHC (with $\sim 3000$ fb$^{-1}$); eventually, the green-meshed area on the right is
the region where the theoretical unitarity constraints are not fulfilled.
In the left panel, the allowed region is represented by the white area, for which $\left\langle \sigma_{\rm fo} v \right\rangle$
is obtained through on-shell KK-graviton production. In the right panel, in addition to the white area, 
within the tiny orange region  $\left\langle \sigma_{\rm fo} v \right\rangle$
is obtained through on-shell radion production.
The dashed lines depicted in the white region represent the values of $\Lambda$ needed to obtain the correct relic abundance.
}
\label{fig:resultados_paper_2_ingles}
\end{figure}

The method followed for the analysis of the model has been the following: we have first computed the relevant annihilation cross-sections for DM into SM particles and KK-gravitons; then, we have studied a
two-dimensional grid with different values of the mass parameters $ (m_{1}, m_\text{DM}) $; for each point on this grid, we have searched for the $ \Lambda $ value to obtain the current DM abundance (for which $\langle \sigma v \rangle \backsimeq \langle \sigma v \rangle_\text{fo} = 2 \times 10^{-26} \, \text{cm}^3/\text{s}$). In this way, for each point the three free parameters ($m_{1}$, $m_\text{DM}$, $\Lambda$) are fully defined, which allows us to establish different theoretical and experimental limits on them. 

Fig.~\ref{fig:resultados_paper_2_ingles} shows the final results of the phenomenological analysis of the model. Following the above strategy, on the left panel the case without radion has been explored, assuming that some alternative method could be found to stabilize the radius of the fifth dimension. In comparison, on the right panel it has been considered that the mass of the radion is $ m_r = 100 $ GeV (it is important to remember that the phenomenology is not affected by the value of this mass). The dark gray-shaded area is the region where it is not possible to obtain the current DM abundance for any value of $ \Lambda $, meanwhile the orange area represents the parameter space region where the abundance is achieved thanks to the contributions of radionic interaction channels. The green-meshed area is the region where we found unitarity problems\footnote{Dark Matter particles have a small relative velocity, so that  $s \backsimeq m_{\text{DM}}^2 $. Since to obtain the correct relic abundance $\sigma = \sigma_{\text{fo}}$ is needed, then the unitarity limit becomes a restriction directly on the DM mass, $m_{\text{DM}}^2 \lesssim 1/\sigma_{\text{fo}}$. Therefore, in the mass plane ($m_\text{DM}$, $m_1$) this bound appear as a vertical line.}, $\sigma > 1/s$. In addition to this limit, there is another theoretical constraint: if $ \Lambda < m_{\text {DM}}, m_{1} $ the effective theory that describes the interaction of these quantum fields is not valid (as they should have been integrated out). This occurs in the red-meshed region.

So far, we have summarized the different limits to the model from theoretical reasons. Now we turn to the experimental bounds. The current Direct Detection experiments and the resonance searches in the ATLAS and CMS experiments at the LHC can provide much more information to our analysis. The red areas show the points where the cross-section of DM-nucleon interaction is already excluded by Xenon1T Direct Detection experiment, while  the blue area is the one excluded by the resonance searches (KK gravitons searches, in our case) at the LHC. More concretely, the strongest bound comes from searches at the LHC with 36 fb$^{-1}$ at $\sqrt{s} = 13$ TeV in the $\gamma \gamma$ channel. The two dotted lines show our prospect for the LHC-Run-3 (with $\sim 300$ fb$^{-1}$) and the HL-LHC (with $\sim 3000$ fb$^{-1}$).

The results of this work have been very rich: although similar analysis had already been carried out in the Randall-Sundrum scenario, this is the first paper that takes into account the Dark Matter annihilation channels directly into KK-gravitons in such high regions of mass space (various TeV). Without this annihilation channel, it is not possible to obtain the correct DM relic abundance in this RS scenario.
 Likewise, a new diagram totally forgotten in the literature has been studied: the annihilation into gravitons without a mediator, coming from the second order expansion of the interaction Lagrangian. Apart from that, it should be noted that this analysis has only been carried out for scalar Dark Matter. However, in Ref.~\cite{Folgado:2020vjb}, which is currently in publication process, the fermionic and vector Dark Matter cases are analysed. This new study shows that fermionic DM is disfavoured respect to the scalar and vector ones. The reason is that the dominant process (the annihilation directly into gravitons) is more suppress in that case.

\section{Gravity-mediated Dark Matter in CW/LD}

After the analysis of the implications of purely gravitational WIMP Dark Matter in the Randall-Sundrum scenario, the question of what would occur in the recent Clockwork/Linear Dilaton model almost naturally arises. This idea inspired Ref.~\cite{Folgado:2019gie}. CW/LD scenario displays more technical complications than RS: the KK-tower of massive gravitons in this case has a very small separation that makes more complicated the numerical analysis of its phenomenology. A brief review about CW/LD extra-dimensions can be found in Sect.~\ref{sec:cw_ld}.

The strategy to analyse the model is the same that we used in the RS case. The main difference with the previous model is the parameters chosen to study the phenomenology. In contrast with RS, in CW/LD the couplings of the massive 4-dimensional gravitons to the rest of the particles are not universal, but depend on the order $n$ of the KK mode. Therefore, it is more useful to characterize the model in terms of $ M_5 $ instead of the effective coupling $\Lambda_n$, that depends on the particular KK-mode studied. In addition to that, the mass of the first graviton coincides with the value of the curvature along the fifth dimension, $ m_{1} = k $.
In the original RS scenario a stabilization mechanism was absent, and a new scalar field is necessary to stabilize the fifth dimension. On the contrary, in CW/LD the 5-dimensional dilaton field takes this role. Unlike RS, where the radion mass is a new parameter, in this scenario the mass of the radion is also determined by $k$. However, there are several ways to stabilize the size of the extra-dimension with the dilaton field. The minimal case assumes that the tension of the 4-dimensional branes is infinite. This framework receives the name of \textit{rigid limit} and it is the assumed case in this work. Currently, we are working on the implications of the phenomenology out of the rigid limit \cite{TFM_Jesus}. There is another important difference between both frameworks relevant for the phenomenological study: in CW/LD the complete dilaton KK-tower is relevant. In RS the KK tower of the Goldberger-Wise scalar field was present, but heavy \cite{Goldberger:1999un}. As a consequence, the only light field present in the spectrum in that case was the radion.

\begin{figure}[htbp]
\centering
\includegraphics[width=150mm]{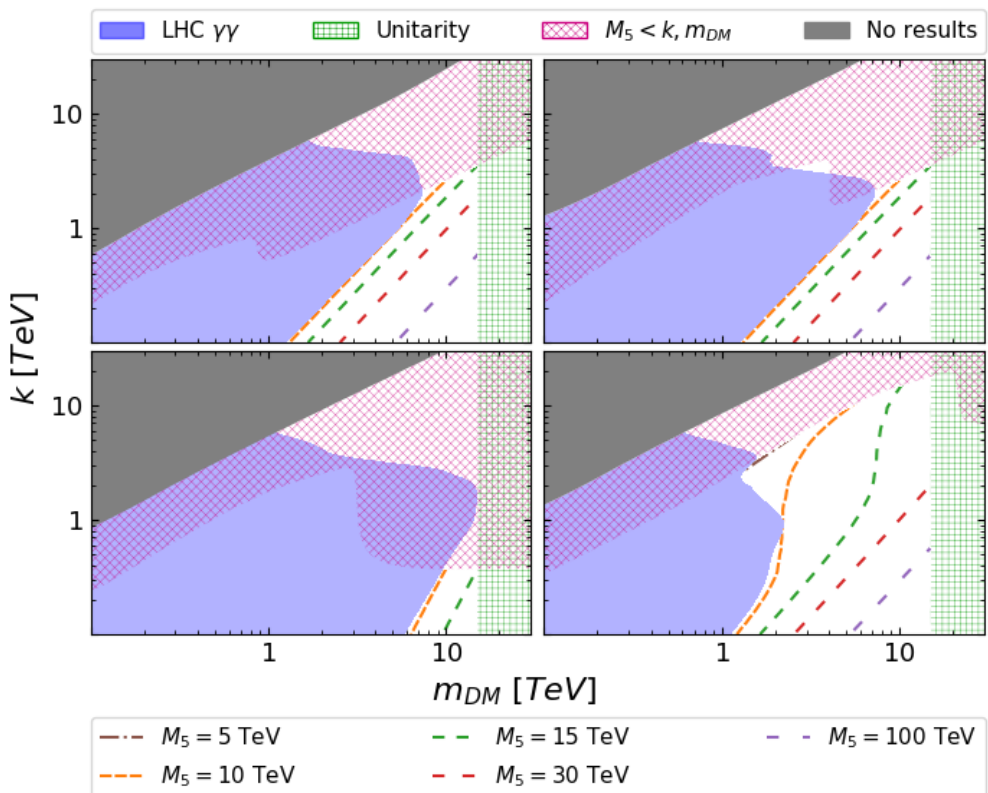}
\caption[Results from \textit{Gravity-mediated Dark Matter in CW/LD Extra-Dimensions}.]{\it Region of the $(m_{\rm DM}, k)$ plane for which $\left\langle \sigma v \right\rangle = \left\langle \sigma_{\rm fo} v \right\rangle$. 
Upper left panel: scalar DM (unstabilized extra-dimension);
Upper right panel:  scalar DM (stabilized extra-dimension in the rigid limit); 
Lower left panel: fermion DM (stabilized extra-dimension in the rigid limit);
Lower right panel: vector DM (stabilized extra-dimension in the rigid limit).
In all panels, the grey-shaded area represents the part of the parameter space for which it is impossible to achieve the correct relic abundance; 
the red diagonally-meshed area is the region for which the low-energy CW/LD effective theory is untrustable, as $M_5 < k, m_\text{DM}$; 
the blue-shaded area is excluded by non-resonant searches at the LHC with 36 fb$^{-1}$ at $\sqrt{s} = 13$ TeV \cite{Giudice:2017fmj}; 
eventually, the green vertically-meshed area on the right is the region where the theoretical unitarity constraints are not fulfilled, 
$m_{\rm DM} \gtrsim 1/\sqrt{\sigma_{\rm fo}}$.
In all panels, the white area represents the region of the parameter space for which the correct relic abundance is achieved
(either through direct KK-graviton and/or radion/KK-dilaton production, as in the case of scalar DM, or through virtual KK-graviton exchange,
as for fermion and vector DM) and not excluded by experimental bounds and theoretical constraints.
The dashed lines depicted in the white region represent the values of $M_5$ needed to obtain the correct relic abundance.
}
 \label{fig:resultados_paper_3_ingles}
\end{figure}

Fig.~\ref{fig:resultados_paper_3_ingles} shows the results obtained for this scenario, following the same strategy outlined in Sect.~\ref{sec:paper2ing}. As in the RS case, $ M_5 $ has been fixed to set the current abundance of Dark Matter for each point in the parameter space ($ m_{\text{DM}}, k $). The different limits studied are the same as in the RS case: the red-meshed region shows the area where the effective field theory is untrustable, $M_5 < m_{\text{DM}}, m_{G_1}$; the green-meshed region represents the area where $\sigma < 1/m_{\text{DM}}^2$ and, therefore, suffers from unitarity problems; eventually, the blue-shaded area represents the limits imposed by the LHC. As a consequence of the small separation between the KK-gravitons, the strongest bound imposed by the LHC comes from non-resonant searches in $\gamma \gamma$ channel. Finally, it should be noted that in the CW/LD case the limits imposed by the Direct Detection of Dark Matter exclude very small DM masses and, as a consequence, they do not appear in the Figure.

In this case, three possible Dark Matter particles spin have been analysed: scalar, fermion and vector. The two upper plots correspond to the scalar case without taking into account the radion and the dilaton-tower (left) and taking it into account (right). This is the only case where radion and dilatons play an important role in the phenomenology of the model and therefore it is worth showing what their impact is on the final results. The lower panels correspond to the fermionic case (left) and the vector case (right). In both cases the radion and the dilatons do not play any role. The Figure shows that the fermionic case is disfavoured with respect to the other two: the non-resonant searches at LHC impose strong limits in this case. This fact is because in the fermionic case the dominant channel, the annihilation of Dark Matter directly into KK gravitons, is suppressed.

As a final comment to gravitational-interacting DM in RS and/or CW/LD scenarios, we can say that in both cases a viable region of the parameter space exists, for the DM masses in the range $[1,10]$ TeV approximately and for $m_1$ smaller than $\backsim 3$ TeV, $\backsim 400$ GeV and $\backsim 10$ TeV for the scalar, fermionic and vectorial cases, respectively. Most of the allowed region could be tested by the LHC Run-3 or its high luminosity upgrade. Notice that in the allowed region typically the scale of new physics (either $\Lambda$ or $M_5$) is a bit too large to solve the hierarchy problem.

\section{Kaluza-Klein FIMP Dark Matter in RS}

In the three models analysed before it has been considered that the DM is composed by WIMP particles. However, FIMP Dark Matter\footnote{Described in Sect~\ref{sec:FIMP_intro}.} brings interesting properties for the purely gravitational case. In the last work included in this Thesis we explore the possibility to obtain the DM abundance using gravitational interaction and FIMP particles in the RS scenario \cite{Bernal:2020fvw} (an extension to the CW/LD is in progress). The FIMP case raises very different mathematical and numerical difficulties from the WIMP case: due to the feeble interaction that displays these kind of particles, the mechanism to obtain the DM abundance for FIMP particles is the freeze-in\footnote{For a complete description of the freeze-in mechanism see Sect.~\ref{Sec:freezein}.}, instead of the freeze-out. Indeed, in the WIMP DM case, the abundance is always obtained for $\left\langle \sigma v \right\rangle = \left\langle \sigma_{\rm fo} v \right\rangle$ for DM masses in the  GeV-TeV range. However, in the FIMP scenario the strong dependence of the evolution with the initial conditions makes necessary to solve the Boltzmann Equation, Eq.~\ref{BEFIMP}, for each point of the parameter space.

\begin{figure}[t]
\begin{center}
\includegraphics[width=115mm]{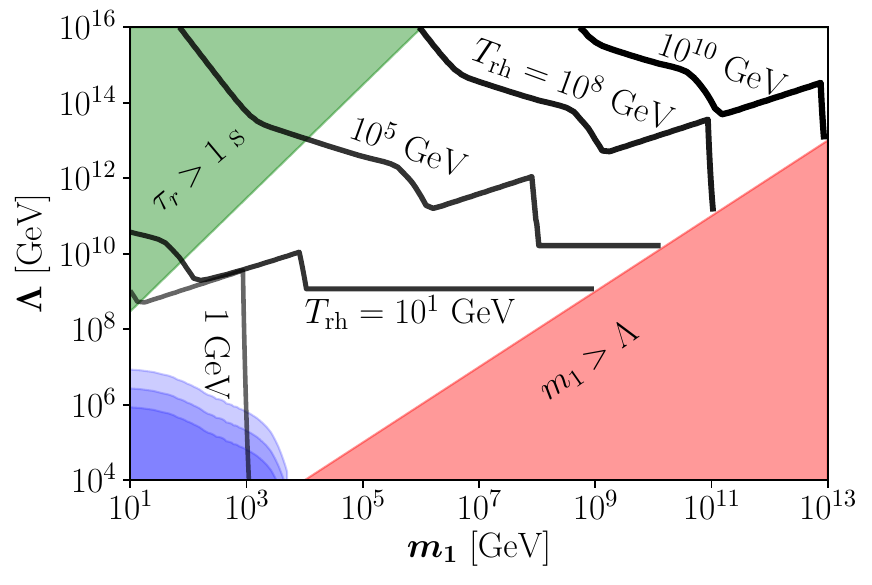}
	\caption[Results from \textit{Kaluza-Klein FIMP Dark Matter in Warped Extra-Dimensions}.]{Parameter space required to reproduce the observed DM abundance for $m_\text{DM}=1$~MeV and $m_r=m_1/10^3$, for several values of the reheating temperature $T_\text{rh}$.
	The blue areas are excluded by resonant searches at LHC and represent the current bound and our prospects for the LHC Run-3 and the High-Luminosity LHC in the $\gamma \, \gamma$ channel~\cite{Aaboud:2017buh,Aaboud:2017yyg}.
	The upper left green corner corresponds to radion lifetimes longer than 1~s.
	In the lower right red area ($m_1>\Lambda$) the EFT approach breaks down.
}
	\label{fig:resultados_paper_4_ingles}
\end{center}
\end{figure}

In the FIMP case, the abundance also has a strong dependence on a new parameter: the highest temperature of the universe, the so-called \textit{reheating temperature} $T_{\text{rh}}$. Due to the complexity of the parameter space, the analysis in this scenario has been performed for a specific value of the Dark Matter mass: $ m_{\textit{DM}} = 1 $ MeV. The values of $T_\text{rh}$ needed to obtain the observed DM relic abundance are shown in Fig. \ref{fig:resultados_paper_4_ingles}. The blue-shaded region shows the experimental limits imposed by resonance searches in $pp \rightarrow G_1 \rightarrow \gamma \, \gamma$ channel at the LHC  (and the two expected bounds from the Run-3 and the HL-LHC). The red-shaded area represents the region where the EFT approach breaks down. On the other hand, the upper left green corner corresponds to radion lifetimes higher than $1$ s, potentially problematic for BBN (all the KK-graviton states are heavier than the radion and therefore will have naturally shorter lifetimes).

In contrast with the WIMP case, this work shows that the RS model with FIMP is much less constrained, because in order to obtain the correct DM relic abundance via freeze-out $\Lambda$ can not be larger that  $10^4$ TeV and $m_1 < 10 $ TeV, while the allowed range of these parameters when the DM abundance is set via freeze-in expands over several orders of magnitude. On the other hand, in such regions the model does not solve at all the hierarchy problem.



\setcounter{part}{1}

\part{Scientific Research}\label{sec:papers}\thispagestyle{empty}
\renewcommand{\headrulewidth}{0.4pt}


\phantomsection\addcontentsline{toc}{section}{Probing the sterile neutrino portal to Dark Matter with $\gamma$ rays}
\thispagestyle{empty}
\renewcommand{\headrulewidth}{0.4pt}
\includepdf[pages=-,scale=0.9,pagecommand={\pagestyle{fancy}}]{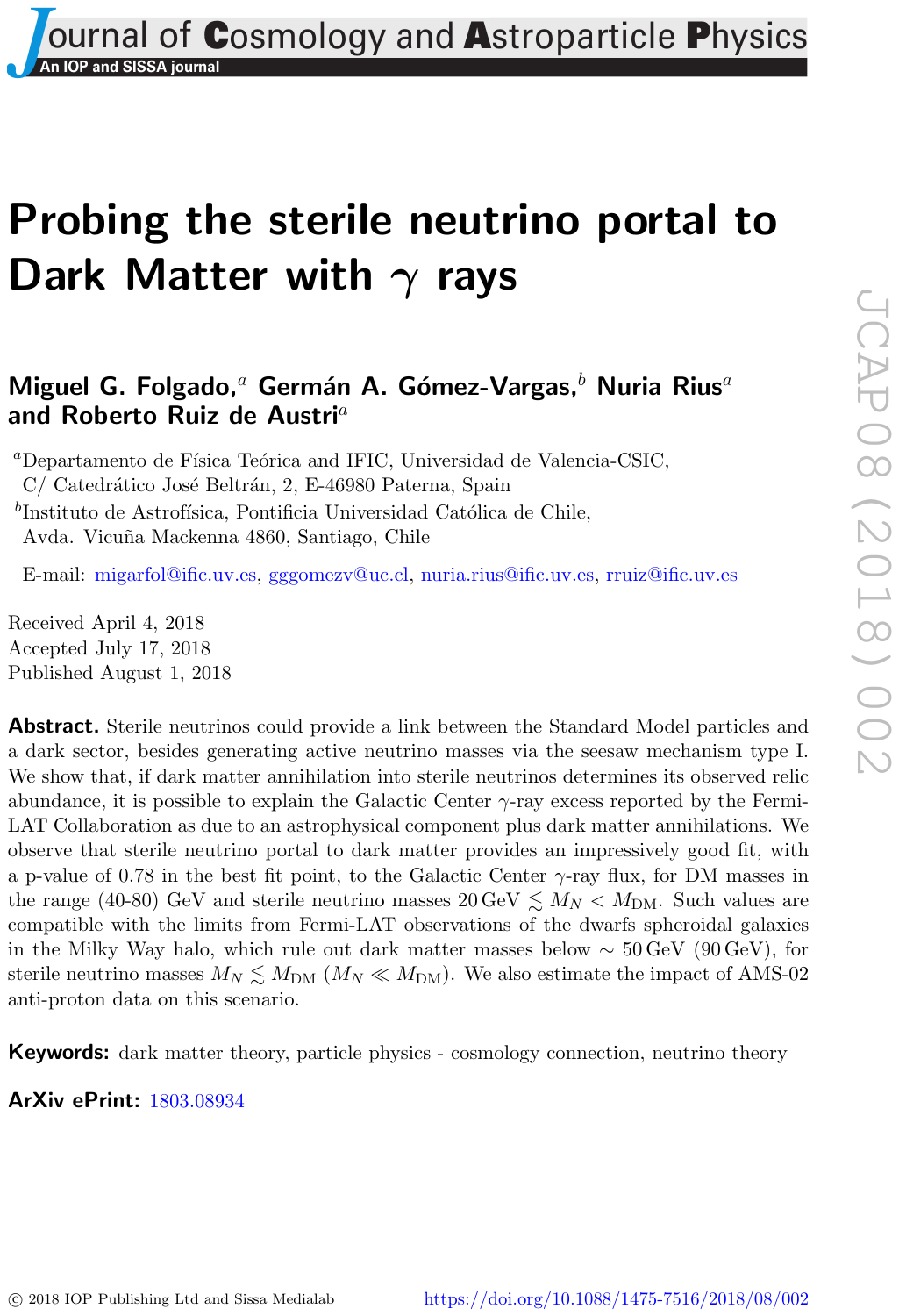}

\phantomsection\addcontentsline{toc}{section}{Gravity-mediated Scalar Dark Matter in Warped Extra-\\Dimensions}
\thispagestyle{empty}
\renewcommand{\headrulewidth}{0.4pt}
\includepdf[pages=-,scale=0.9,pagecommand={\pagestyle{fancy}}]{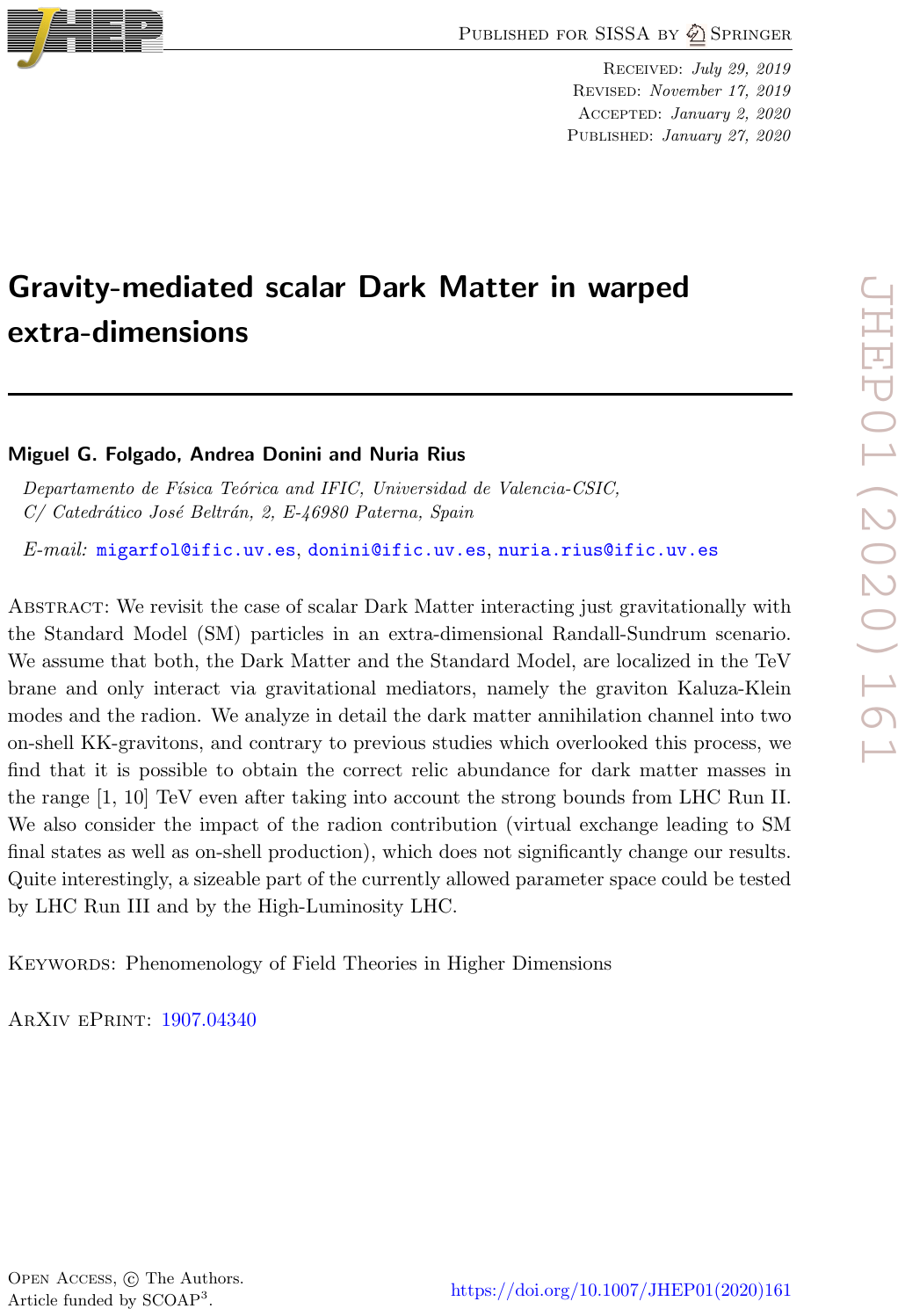}

\phantomsection\addcontentsline{toc}{section}{Gravity-mediated Dark Matter in Clockwork/Linear Dilaton Extra-Dimensions}
\thispagestyle{empty}
\renewcommand{\headrulewidth}{0.4pt}
\includepdf[pages=-,scale=0.9,pagecommand={\pagestyle{fancy}}]{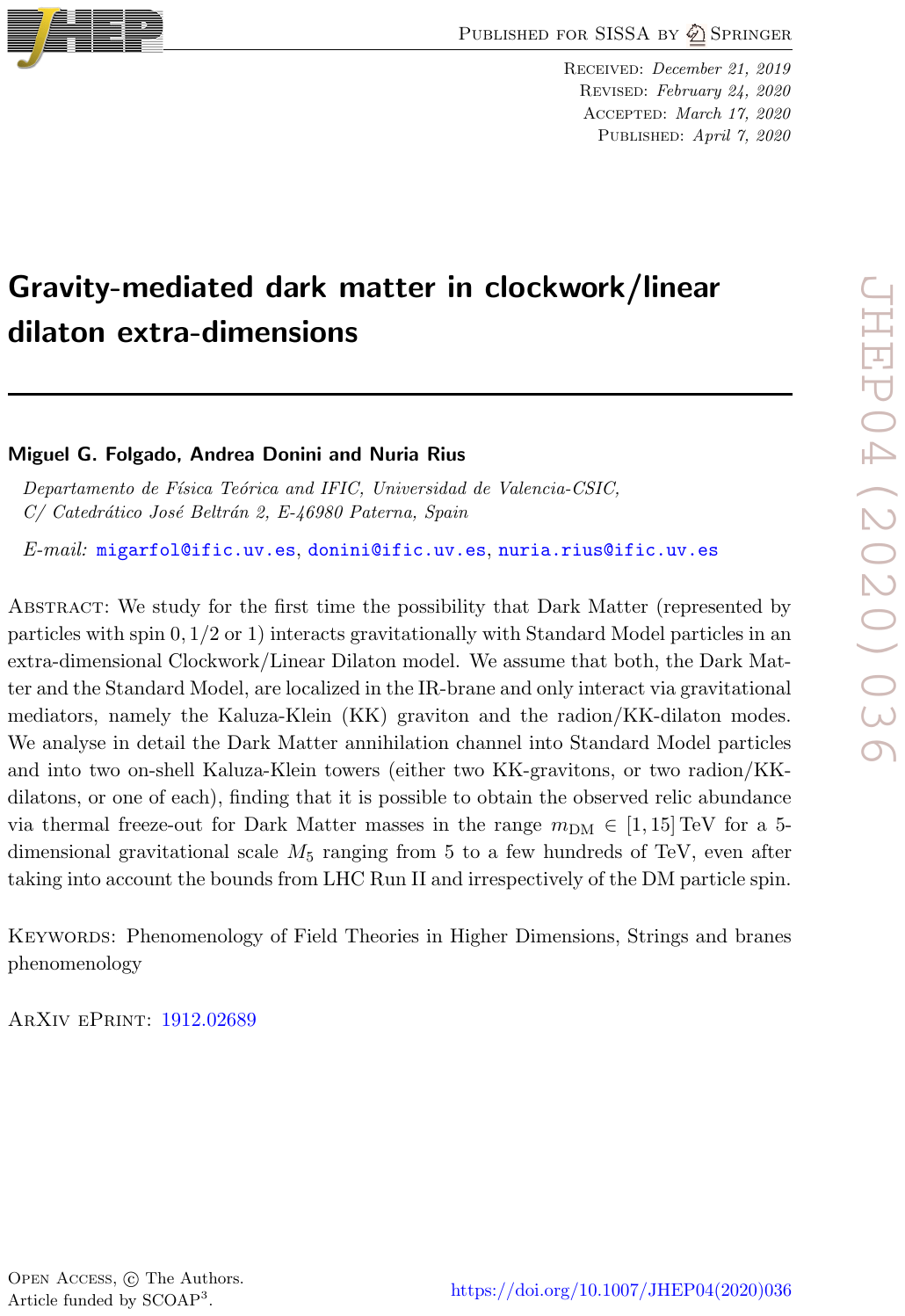}

\phantomsection\addcontentsline{toc}{section}{Kaluza-Klein FIMP Dark Matter in Warped Extra-Dimensions}
\thispagestyle{empty}
\renewcommand{\headrulewidth}{0.4pt}
\includepdf[pages=-,scale=0.9,pagecommand={\pagestyle{fancy}}]{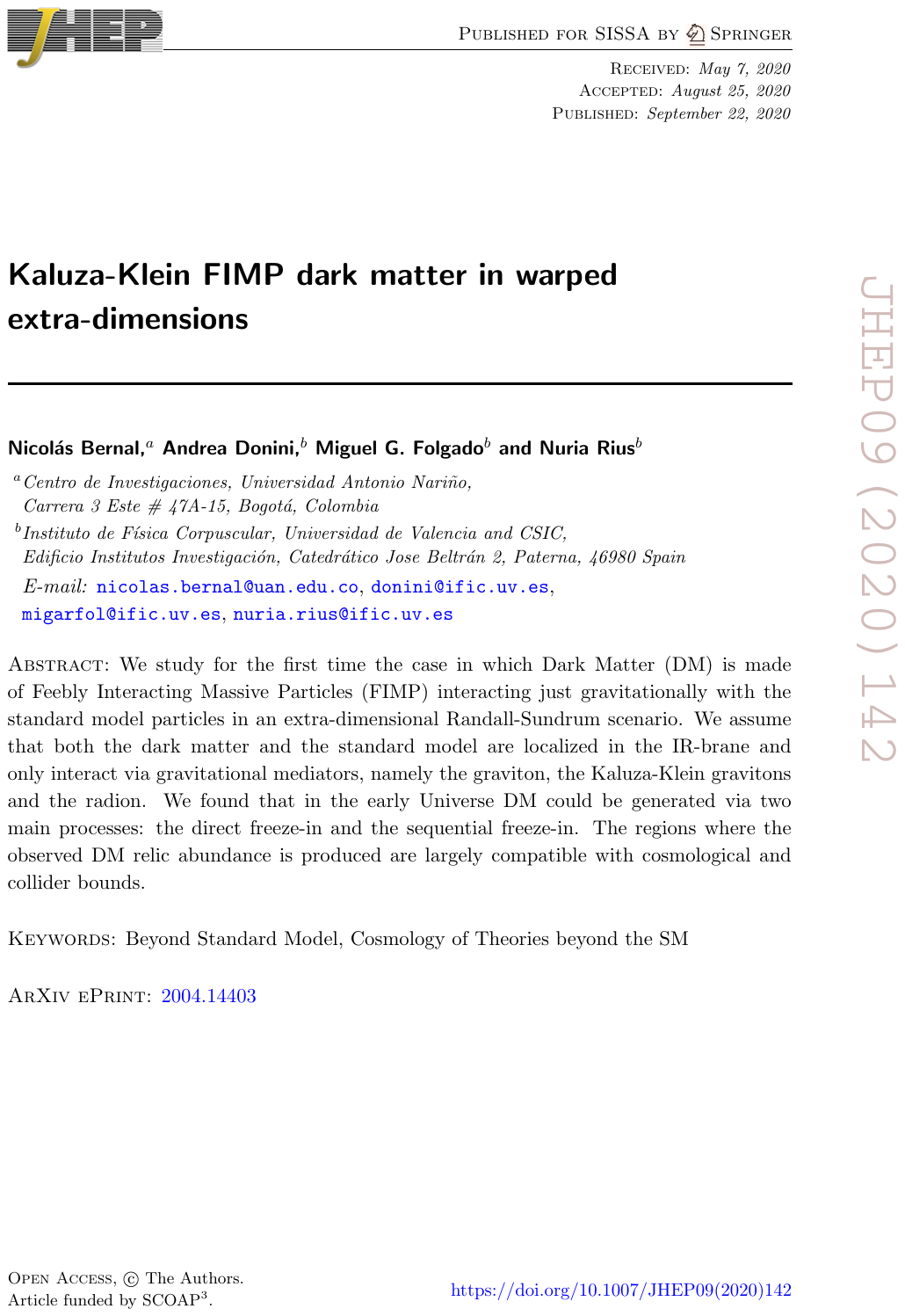}



\part{Bibliography}\label{sec:bib}\thispagestyle{empty}
\renewcommand{\headrulewidth}{0.4pt}

\cleardoublepage

\phantomsection

\renewcommand{\headrulewidth}{0.5pt}

\lhead[{\bfseries \thepage}]{The Bibliography}
\rhead[{The Bibliography}]{\bfseries \thepage}

\bibliographystyle{jhep}
\bibliography{biblio}

\cleardoublepage

\end{document}